\documentclass[showpacs,amsmath,amssymb]{revtex4}
\usepackage{bm}

\begin{document}

\title{Second quantization method in the presence of bound
states of particles}
\author{Sergey V. Peletminskii}
\email{spelet@kipt.kharkov.ua}
\author{Yuriy V. Slyusarenko}
\email{slusarenko@kipt.kharkov.ua} \affiliation{ National Science
Center "Kharkov Institute of Physics
and Technology"\\
Akademicheskaya Str.,1, Kharkov, 61108, Ukraine}

%\date{November 12, 2003}
\date{\today}

\begin{abstract}
We develop an approximate second quantization method for
describing the many-particle systems in the presence of bound
states of particles at low energies (the kinetic energy of
particles is small in comparison to the binding energy of compound
particles). In this approximation the compound and elementary
particles are considered on an equal basis. This means that
creation and annihilation operators of compound particles can be
introduced. The Hamiltonians, which specify the interactions
between compound and elementary particles and between compound
particles themselves are found in terms of the interaction
amplitudes for elementary particles. The nonrelativistic quantum
electrodynamics is developed for systems containing both
elementary and compound particles. Some applications of this
theory are considered.
% \verb+\pacs{#1}+command.
\end{abstract}

\pacs{03.65.Ca, 05.30.-d, 12.20.-m}

\keywords{Second quantization method, bound states, operator,
state vector, Wick's theorem, normal product, spontaneous
radiation, van der Waals forces}

%Use showkeys class option if keyword display desired

\maketitle

\newpage
\section{Introduction}

The second quantization is the effective method, which is used for
description of physical processes in quantum theory of
many-particle systems. The key role of this method consists in
introduction of creation and annihilation operators of particles
in a certain quantum state. The operators of physical quantities
are constructed in terms of creation and annihilation operators.
Such a description of quantum many-particle systems implies the
particles to be elementary (not consisting of other particles).
Moreover, it is absolutely accurate despite of the possible
existence of compound particles. Since the interactions between
particles may lead to formation of bound states, the standard
second quantization method becomes too cumbersome. For this reason
the construction of approximate quantum-mechanical theory for
many-particle systems consisting of elementary particles and their
bound states represents an actual problem. In this theory it is
necessary to introduce the creation and annihilation operators of
bound states as the operators of elementary objects (not
compound). In addition we should preserve the required information
concerning internal degrees of freedom for bound states. Such an
approach may be actually realized in the approximation in which
the binding energy of compound particles is much greater in
comparison to their kinetic energy.

The problems leading to the necessity of accounting for compound
particles along with elementary ones are typical in studying of
the interaction of irradiation with matter (the matter consists of
neutral atoms or molecules). The atoms or molecules may be both in
the ground and excited states when it is necessary to take into
account their internal structure retaining the convenience and
simplicity of the second quantization method. Such a situation
should take place in describing the experiments on laser cooling
of atoms${}^{1-3}$ or experiments on observation of Bose-Einstein
condensates in magnetic traps, where a condensate temperature is
reached by laser cooling of gases of alkali atoms${}^{4}$. The
similar problem of accounting for the bound states occurs also in
quantum chemistry, for example, in studying of chemical reactions.

Thus, the construction of the quantum theory of many-particle
systems in the presence of bound states of particles within the
second quantization method represents not only academic but also
applied interest. The formulation of the quantum theory itself is
to be done so that the operators of physical quantities are
constructed through the creation and annihilation operators of
elementary particles and their bound states.

In this paper we demonstrate a possibility of construction of the
above-mentioned quantum-mechanical theory by considering a system
of two kinds of charged fermions. In this theory the compound
particles (atoms) consist of two fermions of various kinds. The
choice of such model is not associated with principle difficulties
and makes it possible to simplify calculations and to obtain the
visual results.

As applications of the developed theory we study a spontaneous
radiation of two-component excited atom and obtain the expression
for its probability (see, e.g.,${}^{5}$). We also investigate the
attractive forces acting between neutral atoms at long distances
(the van der Waals forces) and derive the expression for the
potential of the van der Waals forces${}^{6}$. Finally we consider
a question concerning the scattering of photons and elementary
particles by bound states${}^{7}$.

\section{Creation and annihilation operators of bound states of particles}

Consider a system consisting of two kinds of fermions with masses
$m_{1}$ and $m_{2}$. As was mentioned in the Introduction this
case is more visual in order to demonstrate the procedure for
obtaining the operators of basic physical quantities within the
second quantization method in the presence of bound states of
particles. We shall see that the developed methodology does not
contain any difficulties for its generalization (the bound states
of more than two particles, the presence of bosons in the system,
the bound states of bosons and boson-fermion bound states). For
simplicity we do not take into account a spin variable (this can
be done without any difficulties).

Let ${\hat\psi}_{1}({\mathbf x})$, ${\hat\psi}_{2}({\mathbf x})$ be
the annihilation operators of two kinds of fermions at the point
${\mathbf x}$,
\[
{\hat\psi}_{1}({\mathbf x})|0\rangle = {\hat\psi}_{2}({\mathbf
x})|0\rangle = 0,
\]
where $|0\rangle$ is a vacuum state vector. Then the state vectors
\begin{equation}
|{\mathbf x}_{1},\ldots ,{\mathbf x}_{n},{\mathbf y}_{1},\ldots
,{\mathbf y}_{m}\rangle = {\hat\psi}_{1}^{+}({\mathbf x}_{1})\ldots
{\hat\psi}_{1}^{+}({\mathbf x}_{n}) {\hat\psi}_{2}^{+}({\mathbf
y}_{1}) \ldots{\hat\psi}_{2}^{+}({\mathbf y}_{m})|0\rangle,
\label{1}
\end{equation}
($n,m=0,1,2,\ldots$) form a basis in the space of states $H$. In
these states the particles are at the certain points ${\mathbf
x}_{1}, \ldots ,{\mathbf x}_{n}; {\mathbf y}_{1},\ldots ,{\mathbf
y}_{m} \in R$ of the coordinate space. The state vectors (\ref{1})
satisfy the orthogonality and normalization relations and form a
complete set of state vectors.

We assume that the particles of two different kinds form a bound
state specified by the wave function
\begin{equation}
\varphi_{\alpha}({\mathbf x}_{1},{\mathbf x}_{2},{\mathbf
x})=\varphi_{\alpha}({\mathbf x}_{1}-{\mathbf x}_{2})\delta({\mathbf
x}-{\mathbf X}), \qquad {\mathbf X}={m_{1}{\mathbf
x}_{1}+m_{2}{\mathbf x}_{2}\over {m_{1}+m_{2}}}, \label{1'}
\end{equation}
where ${\mathbf x}$ is the space coordinate and $\alpha$ are the
quantum numbers of the bound state (atom) (we suppose that the
particles of the same kind do not form the bound states). The
corresponding state vector has the form
\[
|\alpha,{\mathbf x}\rangle=\int d{\mathbf x}_{1}d{\mathbf x}_{2}
\varphi_{\alpha}({\mathbf x}_{1}-{\mathbf x}_{2})\delta({\mathbf
x}-{\mathbf X}) \hat{\psi}_{1}^{+}({\mathbf
x}_{1})\hat{\psi}_{2}^{+}({\mathbf x}_{2})|0\rangle.
\]
For this reason the operators
\begin{equation}
{\hat{\varphi}}_{\alpha}^{+}({\mathbf x})=\int d{\mathbf
x}_{1}d{\mathbf x}_{2} \varphi_{\alpha}({\mathbf x}_{1}-{\mathbf
x}_{2})\delta({\mathbf x}-{\mathbf X}) \hat{\psi}_{1}^{+}({\mathbf
x}_{1})\hat{\psi}_{2}^{+}({\mathbf x}_{2}) \label{2}
\end{equation}
we shall call as the creation operators of the bound states
(atoms), so that
\[
{\hat{\varphi}}_{\alpha}^{+}({\mathbf x})|0\rangle=|\alpha,{\mathbf
x}\rangle, \qquad {\hat{\varphi}}_{\alpha}({\mathbf x})|0\rangle =0.
\]
If the atom has a certain momentum, then its state vector is given
by
\[
\vert\alpha,{\mathbf p}\rangle={1\over \sqrt{\mathcal V}}\int
d{\mathbf x}_{1}d{\mathbf x}_{2}\varphi_{\alpha}({\mathbf
x}_{1}-{\mathbf x}_{2})e^{i{\mathbf p}{\mathbf
X}}\hat{\psi}_{1}^{+}({\mathbf x}_{1})\hat{\psi}_{2}^{+}({\mathbf
x}_{2})|0\rangle,
\]
where ${\mathcal V}$ is the volume of the system. The corresponding
creation operator $\hat{\varphi}_{\alpha}^{+}({\mathbf p})$ of the
atom in the state with momentum ${\mathbf p}$ is defined by
\[
\vert\alpha,{\mathbf p}\rangle = \hat{\varphi}_{\alpha}^{+}({\mathbf
p})|0\rangle, \quad \hat{\varphi}_{\alpha}^{+}({\mathbf x})={1\over
\sqrt{\mathcal V}}\sum_{\mathbf
p}\hat{\varphi}_{\alpha}^{+}({\mathbf p})e^{-i{\mathbf p}{\mathbf
x}}.
\]
Taking into account that
\[
\int d{\mathbf y}_{1}\varphi_{\alpha}^{*}({\mathbf y}_{1}-{\mathbf
y}_{2})\varphi_{\beta}({\mathbf y}_{1}-{\mathbf
y}_{2})=\delta_{\alpha\beta},\label{3}
\]
it is easy to find the following commutation relations:
\begin{equation*}
[{\hat{\varphi}}_{\alpha}({\mathbf x}),{\hat{\varphi}}_{\alpha
'}^{+}({\mathbf x}')]=\delta_{\alpha\alpha '}\delta({\mathbf
x}-{\mathbf x}') + \hat{\chi}_{\alpha\alpha '}({\mathbf x},{\mathbf
x}'), \qquad [{\hat{\varphi}}_{\alpha}({\mathbf
x}),{\hat{\varphi}}_{\alpha '}({\mathbf x}')]=0,\label{4}
\end{equation*}
where
\begin{eqnarray*}
\hat{\chi}_{\alpha\alpha '}({\mathbf x},{\mathbf x}') &=& \int
d{\mathbf y} d{\mathbf y}'\varphi^{*}_{\alpha}({\mathbf
y})\varphi_{\alpha\prime}({\mathbf y}')\biggl\{{\hat
\psi}^{+}_{1}\left({\mathbf x}+{m_{2}\over M}{\mathbf
y}\right){\hat\psi}_{1}\left({\mathbf x}'+{m_{2}\over M}{\mathbf
y}'\right)\delta\left({\mathbf
y}-{\mathbf y}' - {m_{1}\over M}({\mathbf x}-{\mathbf x}')\right)\\
&&+{\hat \psi}^{+}_{2}\left({\mathbf x}'-{m_{1}\over M}{\mathbf
y}'\right){\hat \psi}_{2}\left({\mathbf x}-{m_{1}\over M}{\mathbf
y}\right)\delta\left({\mathbf y}-{\mathbf y}' + {m_{2}\over
M}({\mathbf x}-{\mathbf x}')\right)\biggr\}, \qquad M=m_{1}+m_{2},
\end{eqnarray*}
moreover, $\hat{\chi}_{\alpha\alpha '}({\mathbf x},{\mathbf
x}')|0\rangle =0$. The vectors
\begin{equation}
\vert\underbrace{{\mathbf x}_{1},\ldots}_{n},\underbrace{{\mathbf
y}_{1},\ldots}_{m},\underbrace{{\mathbf
z}_{1},\ldots}_{l}\rangle\equiv
\prod_{i=1}^{n}\hat{\psi}_{1}^{+}({\mathbf x}_{i})
\prod_{k=1}^{m}\hat{\psi}_{2}^{+}({\mathbf
y}_{k})\prod_{j=1}^{l}{\hat{\varphi}}_{\alpha_{j}}^{+}({\mathbf
z}_{j})|0\rangle \label{5}
\end{equation}
have an obvious physical meaning under the following conditions:
\begin{align}
|{\mathbf x}_{i}-{\mathbf x}_{j}| &\gtrsim a, \quad |{\mathbf
y}_{i}-{\mathbf
y}_{j}| \gtrsim a, \quad |{\mathbf z}_{i}-{\mathbf z}_{j}| \gtrsim a, \nonumber \\
|{\mathbf x}_{i}-{\mathbf y}_{j}| &\gtrsim a, \quad |{\mathbf
x}_{i}-{\mathbf z}_{j}| \gtrsim a, \quad |{\mathbf y}_{i}-{\mathbf
z}_{j}| \gtrsim a, \label{6}
\end{align}
(${\mathbf x}, {\mathbf y}, {\mathbf z}\in R_{a}$, $a>>r_{0}$,
$r_{0}$ is the radius of the bound state; the definition of $a$ see
below). In this case the elementary particles and their bound states
are at the certain space points.

Notice that the state vectors (\ref{5}) do not form a basis in the
Gilbert space $H$ if (\ref{6}) are valid. However, their linear
span, which is a totality of the following vectors
\begin{equation}
\sum_{n,m,l}\underbrace{\int d{\mathbf x}_{1}\ldots\int d{\mathbf
y}_{1}\ldots \int d{\mathbf
z}_{1}\ldots}_{R_{a}}C(\underbrace{{\mathbf x}_{1},\ldots}_{n},
\underbrace{{\mathbf y}_{1},\ldots}_{m},\underbrace{{\mathbf
z}_{1},\ldots}_{l}) \vert\underbrace{{\mathbf x}_{1},\ldots}_{n},
\underbrace{{\mathbf y}_{1},\ldots}_{m},\underbrace{{\mathbf
z}_{1},\ldots}_{l}\rangle, \label{7}
\end{equation}
form a subspace $H_{a}$ of the space $H$. Let us show that the
state vectors (\ref{5}) [with conditions (\ref{6})] form an
orthonormalized basis in the subspace $H_{a}$. To this end it is
necessary to take into account that in calculating of the
following vacuum averages:
\begin{equation*}
\langle 0|{\hat\psi}_{1}({\mathbf x}_{1})\ldots
{\hat\psi}_{2}({\mathbf x}_{2})\ldots
{\hat\varphi}_{\alpha}({\mathbf
x}){\hat\varphi}_{\alpha^{\prime}}^{+}({\mathbf x}')\ldots
{\hat\psi}_{2}^{+}({\mathbf x}_{2}^{\prime})\ldots
{\hat\psi}_{1}^{+}({\mathbf x}_{1}^{\prime})\ldots|0 \rangle
\label{8}
\end{equation*}
we can use the Wick theorem with the following contractions:
\[
\underset{a}{{\hat\psi}_{i}}({\mathbf
x})\underset{a}{{\hat\psi}^{+}_{i'}}({\mathbf x}')=\langle 0|
{\hat\psi}_{i}({\mathbf x}){\hat\psi}^{+}_{i'}({\mathbf
x}')|0\rangle =\delta_{ii'}\delta({\mathbf x}-{\mathbf x}'),\qquad
\underset{a}{{\hat\psi}_{i}}({\mathbf
x})\underset{a}{{\hat\psi}_{i'}}({\mathbf x}')=0, \label{9}
\]
if to consider the operators ${\hat\psi}_{1}$, ${\hat\psi}_{2}$,
${\hat\varphi}$ referred to the moment of time $+0$, and operators
${\hat\varphi}^{+}$, ${\hat\psi}_{2}^{+}$, ${\hat\psi}_{1}^{+}$ to
the moment of time $-0$. In addition, we should remember that the
creation and annihilation operators ${\hat\varphi}_{\alpha}$,
${\hat\varphi}^{+}_{\alpha^{\prime}}$ depend  on ${\hat\psi}_{i},
{\hat\psi}_{i'}$ (see (\ref{2})). We also assume that the wave
function (\ref{1'}) of the atom differs from zero for $|{\mathbf
x}_{1}-{\mathbf x}_{2}|<r_{0}$. Taking into account (\ref{2}) and
noting that for $|{\mathbf x}_{1}^{\prime}-{\mathbf
x}_{2}^{\prime}|>a$
\[
\underset{ab}{{\hat\varphi}_{\alpha}}({\mathbf
z})\underset{b}{{\hat\psi}_{2}^{+}}({\mathbf
x}_{2}^{\prime})\underset{a}{{\hat\psi}_{1}^{+}}({\mathbf
x}_{1}^{\prime})=\int d{\mathbf z}_{1}d{\mathbf
z}_{2}\varphi_{\alpha}^{*}({\mathbf z}_{1}-{\mathbf
z_{2}})\delta({\mathbf z}-{\mathbf
Z})\underset{a}{{\hat\psi}_{1}}({\mathbf z}_{1})
\underset{b}{{\hat\psi}_{2}}({\mathbf
z}_{2})\underset{b}{{\hat\psi}_{2}^{+}}({\mathbf
x}_{2}^{\prime})\underset{a}{{\hat\psi}_{1}^{+}}({\mathbf
x}_{1}^{\prime})= \varphi_{\alpha}^{*}({\mathbf
x}_{1}^{\prime}-{\mathbf x_{2}}^{\prime})\delta({\mathbf z}-{\mathbf
X}')=0,
\]
\[
{\mathbf X}'={m_{1}{\mathbf x}_{1}^{\prime}+m_{2}{\mathbf
x}_{2}^{\prime}\over m_{1}+m_{2}}
\]
and also
\begin{multline*}
\underset{ab}{{\hat\varphi}_{\alpha}}({\mathbf
z})\underset{ba}{{\hat\varphi}_{\alpha^{\prime}}^{+}}({\mathbf
z}')=\int d{\mathbf z}_{1}d{\mathbf z}_{2}d{\mathbf
z}_{1}^{\prime}d{\mathbf
z}_{2}^{\prime}\varphi_{\alpha}^{*}({\mathbf z}_{1}-{\mathbf
z_{2}})\delta({\mathbf z}-{\mathbf Z})
\varphi_{\alpha^{\prime}}({\mathbf z}_{1}^{\prime}-{\mathbf
z_{2}}^{\prime})\delta({\mathbf z}'-{\mathbf Z}')
\underset{a}{{\hat\psi}_{1}}({\mathbf z}_{1})
\underset{b}{{\hat\psi}_{2}}({\mathbf
z}_{2})\underset{b}{{\hat\psi}_{2}^{+}}({\mathbf
z}_{2}^{\prime})\underset{a}{{\hat\psi}_{1}^{+}}({\mathbf
z}_{1}^{\prime})\\=\int d{\mathbf z}_{1}d{\mathbf
z}_{2}\varphi_{\alpha}^{*}({\mathbf z}_{1}-{\mathbf
z_{2}})\delta({\mathbf z}-{\mathbf
Z})\varphi_{\alpha^{\prime}}({\mathbf z}_{1}-{\mathbf
z}_{2})\delta({\mathbf z}'-{\mathbf
Z})=\delta_{\alpha\alpha'}\delta({\mathbf z}-{\mathbf z}')
\end{multline*}
(the double contractions correspond to the operators
$\hat{\varphi}_{\alpha}$, $\hat{\varphi}^{+}_{\alpha}$) we get
\begin{multline}
\langle 0|\underbrace{{\hat\psi}_{1}({\mathbf x}_{1})\ldots }_{n}
\underbrace{{\hat\psi}_{2}({\mathbf y}_{1})\ldots }_{m}
\underbrace{{\hat\varphi}_{\alpha_{1}}({\mathbf z}_{1})\ldots }_{l}
\underbrace{{\hat\varphi}^{+}_{\alpha_{1}^{\prime}}({\mathbf
z}_{1}^{\prime})\ldots }_{l'}\underbrace{{\hat\psi}_{2}^{+}({\mathbf
y}_{1}^{\prime})\ldots }_{m'}\underbrace{{\hat\psi}_{1}({\mathbf
x}_{1}^{\prime})\ldots }_{n'}|0\rangle=\\
=\delta_{nn'}\delta_{mm'}\delta_{ll'}\sum {\mathcal
P}_{x^{\prime}}{\mathcal P}_{y^{\prime}} \underbrace{\delta({\mathbf
x}_{1}-{\mathbf
x}_{1}^{\prime})\ldots}_{n}\underbrace{\delta({\mathbf y}-{\mathbf
y}_{1}^{\prime})\ldots}_{m}\underbrace{\delta({\mathbf
z}_{1}-{\mathbf
z}_{1}^{\prime})\delta_{\alpha_{1}\alpha_{1}^{\prime}}\ldots}_{l}.
\label{10}
\end{multline}
This relation shows that the vectors (\ref{7}) form the
orthonormalized basis in the subspace $H_{a}$ if to consider the
creation and annihilation operators
$\hat\varphi_{\alpha}^{+}({\mathbf z})$,
$\hat\varphi_{\alpha}({\mathbf z})$ as the Bose operators, which
commute with ${\hat{\psi}}_{i}({\mathbf x})$,
${\hat{\psi}}_{i}^{+}({\mathbf x})$. The quantity ${\mathcal
P}_{x^{\prime}}$ in (\ref{10}) is equal to $+1$ if the number of
permutations of the arguments ${\mathbf x}^{\prime}_{1}\ldots
{\mathbf x}^{\prime}_{n}$ is even and it is equal to $-1$ if the
number of these permutations is odd. The quantity ${\mathcal
P}_{y^{\prime}}$ is defined similarly. However, the case when it is
necessary to take into account a more complicated arrangement of the
contractions between $\varphi_{\alpha} $, $\varphi^{+}_{\alpha}$:
\[
\underset{a}{{\mathbf x}_{4}}\underset{b}{{\mathbf
x}_{4}^{\prime}}\quad\ldots\quad \underset{c}{{\mathbf
x}_{2}}\underset{d}{{\mathbf
x}_{2}^{\prime}}\quad\underset{e}{{\ldots}_{\,}}\quad
\underset{e}{{\mathbf x}_{1}}\underset{d}{{\mathbf
x}_{1}^{\prime}}\quad\ldots\quad \underset{c}{{\mathbf
x}_{3}}\underset{b}{{\mathbf
x}_{3}^{\prime}}\quad\underset{a}{{\ldots}_{\,}}
\]
can take place (the odd indices correspond to the creation operators
and even to annihilation operators). Since ${\mathbf
x}_{1}^{\prime}={\mathbf x}_{2}^{\prime}$, ${\mathbf x}_{2}={\mathbf
x}_{3}$, ${\mathbf x}_{4}^{\prime}={\mathbf x}_{3}^{\prime}$,
\[
|{\mathbf x}_{1}-{\mathbf x}_{1}^{\prime}| \lesssim r_{0}, \qquad
|{\mathbf x}_{1}^{\prime}-{\mathbf x}_{2}| \lesssim r_{0}, \qquad
 |{\mathbf x}_{3}^{\prime}-{\mathbf x}_{2}| \lesssim r_{0}, \qquad |{\mathbf
x}_{4}-{\mathbf x}_{3}^{\prime}|\lesssim r_{0},
\]
whence $|{\mathbf x}_{4}-{\mathbf x}_{1}|\lesssim 4r_{0}$.
Therefore, if $a\leq 4r_{0}$, then the above contractions are to be
taken into account. If however, we consider the $n$-particle states,
then the discussed arrangements of the contractions are not to be
taken into account for $n\lesssim a/r_{0}$ (we emphasize that
$r_{0}/a$ is a small parameter in our problem, see (\ref{18})). The
said above concerns to the majority of problems in non-relativistic
quantum theory, where a finite number of particles is usually
studied (see sections V, VI).

With the use of (\ref{10}) it is easy to find the projection
operator ${\mathcal P}_{H_{a}}$ onto the subspace $H_{a}$:
\[
{\mathcal P}_{H_{a}}=\sum_{k+m+l\leq n} {1\over k!}{1\over
m!}{1\over l!} \underbrace{\int d{\mathbf x}_{1}\ldots\int d{\mathbf
y}_{1}\ldots\int d{\mathbf z}_{1}\ldots}_{R_{a}}
|\underbrace{{\mathbf x}_{1}\ldots}_{k}\underbrace{{\mathbf
y}_{1}\ldots}_{m}\underbrace{{\mathbf z}_{1}\ldots}_{l}
\rangle\langle\underbrace{{\mathbf
x}_{1}\ldots}_{k}\underbrace{{\mathbf
y}_{1}\ldots}_{m}\underbrace{{\mathbf z}_{1}\ldots}_{l}|,
\]
such that the operators $\underline{A}$ acting in the subspace
$H_{a}\in H$ correspond to the operators of physical quantities
$A$ that act in $H$ (hereinafter the sums of $1,2,\ldots
n$-particle subspaces over the bound states are considered;
$n\thickapprox a/r_{0}>>1$),
\begin{eqnarray}
{\underline{A}}={\mathcal P}_{H_{a}}A{\mathcal
P}_{H_{a}}. \label{11}
\end{eqnarray}

Let us introduce now an auxiliary space $\tilde{H}$ with the Fermi
creation and annihilation operators ${\hat\chi}_{1}^{+}({\mathbf
x}), {\hat\chi}_{2}^{+}({\mathbf x})$, ${\hat\chi}_{1}({\mathbf x}),
{\hat\chi}_{2}({\mathbf x})$ and Bose creation and annihilation
operators ${\hat\eta}_{\alpha}^{+}({\mathbf x})$,
${\hat\eta}_{\alpha}({\mathbf x})$ and take the vectors
\[
|{\mathbf x}_{1},\ldots,{\mathbf y}_{1},\ldots,{\mathbf
z}_{1},\ldots)={\hat\chi}_{1}^{+}({\mathbf x}_{1})\ldots
{\hat\chi}_{2}^{+}({\mathbf
y}_{1})\ldots{\hat\eta}^{+}_{\alpha}({\mathbf z}_{1})\ldots |0),
\]
as a basis of this space, where $|0)$ is a vacuum vector in
$\tilde{H}$. Then the linear span of the vectors
\[
|{\mathbf x}_{1},\ldots,{\mathbf y}_{1},\ldots,{\mathbf
z}_{1},\ldots)\in {\tilde H}_{a}, \qquad {\mathbf x}, {\mathbf y},
{\mathbf z}\in R_{a} \label{12}
\]
determines the subspace $\tilde{H}_{a}$ of the space $\tilde{H}$.

Now we can easily establish the isomorphic correspondence between
$H_{a}$ and $\tilde{H}_{a}$:
\[
|{\mathbf x}_{1},\ldots,{\mathbf y}_{1},\ldots,{\mathbf
z}_{1},\ldots\rangle\iff |{\mathbf x}_{1},\ldots,{\mathbf
y}_{1},\ldots,{\mathbf z}_{1},\ldots),
\]
which preserves the scalar product
\[
\langle{\mathbf x}',\ldots,{\mathbf y}',\ldots,{\mathbf z}',\ldots |
{\mathbf x},\ldots,{\mathbf y},\ldots,{\mathbf z}\ldots
\rangle=({\mathbf x}',\ldots,{\mathbf y}',\ldots,{\mathbf z}',\ldots
| {\mathbf x},\ldots,{\mathbf y},\ldots,{\mathbf z}\ldots), \quad
{\mathbf x},{\mathbf y},{\mathbf z};\,{\mathbf x}',{\mathbf
y}',{\mathbf z}'\in R_{a}.
\]
We can also
establish the isomorphism between the operators ${\underline
A}\iff{\tilde A}$, acting in the spaces $H_{a}$ and ${\tilde
H}_{a}$ according to the formula:
\begin{equation}
\langle{\mathbf x}',\ldots,{\mathbf y}',\ldots,{\mathbf z}',\ldots
|{\underline A}| {\mathbf x},\ldots,{\mathbf y},\ldots,{\mathbf
z}\ldots \rangle=({\mathbf x}',\ldots,{\mathbf y}',\ldots,{\mathbf
z}',\ldots |{\tilde A} | {\mathbf x},\ldots,{\mathbf
y},\ldots,{\mathbf z}\ldots). \label{15}
\end{equation}
This isomorphic correspondence is remained after the
multiplication of the operator by a number, after addition of
operators, and after multiplication of operators:
\[
\lambda {\underline A}\iff\lambda{\tilde A}, \qquad {\underline
A}+{\underline B}\iff {\tilde A}+{\tilde B}, \qquad {\underline
A}\,\,{\underline B}\iff {\tilde A}{\tilde B}. \label{16}
\]
The formulas (\ref{11}),(\ref{15}) lead to
\begin{equation}
({\mathbf x}',\ldots,{\mathbf y}',\ldots,{\mathbf z}',\ldots
|{\tilde A} | {\mathbf x},\ldots,{\mathbf y},\ldots,{\mathbf
z}\ldots)=\langle{\mathbf x}',\ldots,{\mathbf y}',\ldots,{\mathbf
z}',\ldots |A| {\mathbf x},\ldots,{\mathbf y},\ldots,{\mathbf
z}\ldots \rangle. \label{17}
\end{equation}
This relation determines the operators of various physical
quantities ${\tilde A}$ acting in ${\tilde H}_{a}$ and, hence,
transfers the quantum theory in which the compound (the bound
states) and elementary particles exist on an equal basis from the
space of states $H$ onto the space of states ${\tilde H}_{a}$. We
would like to remind here that $\hat{\varphi}_{\alpha}^{+}$
entering (\ref{5}) is determined by (\ref{2}).

The relation (\ref{17}) defines the operator ${\tilde A}$ in
${\tilde H}_{a}$ uniquely, but it does not define it uniquely in
${\tilde H}$. It is evident that the operator ${\tilde A}$ acting in
${\tilde H}$ (continued from $\tilde{H}_{a}$ to the whole space
${\tilde H}$) is determined up to the term ${\tilde A}'$, the matrix
elements of which are zero in the space $\tilde{H}_{a}$
($\tilde{A}=\tilde{A}'+\tilde{A}''$). If to introduce the projection
operator ${\mathcal P}_{\tilde H_{a}}$ onto the subspace
$\tilde{H}_{a}$ and to require that the operator $\tilde{A}$ has no
nonzero matrix elements in the orthogonal subspace, then the
operator $\tilde{A}$ will be defined in $\tilde{H}$ uniquely,
$\tilde{A}={\mathcal P}_{\tilde H_{a}}\tilde {A}''{\mathcal
P}_{\tilde H_{a}}$. We will omit the projection operator ${\mathcal
{P}}_{\tilde H_{a}}$ when constructing the operator $\tilde{A}$
acting in ${\tilde H}$. The reason for this is the assumption that
the matrix elements of the operators (in the position space),
corresponding to a quite large external parameter $R\thicksim
|{\mathbf x}_{i}-{\mathbf x}_{j}|$ give a dominant contribution to
the quantum-mechanical processes. Further we will assume that
$R>>r_{0}$ (usually in case of particle collisions
$R^{-1}\sim\sqrt{m{\mathcal E}}$, where ${\mathcal E}$ is the
particle kinetic energy). The latter inequality makes it possible to
choose parameter $a$ (see (\ref{6})) as follows${}^{10}$:
\begin{equation}
R>>a>>r_{0}.  \label{18}
\end{equation}
The matrix elements of the operators should not depend on
parameter $a$ chosen by this way. Let us mentally decrease the
radius of a bound state, $r_{0}\rightarrow 0$. Then the whole
written scheme, as we have already noted, does not depend on $a$
up to the values $a=r_{0}$. Therefore, the subspace ${\tilde
H}_{a}$ can be identified with $\tilde{H}$ due to the inequality
$R>>a$. In other words one can considers that ${\mathcal P}_{\tilde
H_{a}}\rightarrow 1$ for $a\rightarrow 0$. At the same time we do
not break the quantum-mechanical description of bound states in
virtue of the inequality $a>>r_{0}$. From the physical point of
view the inequality $r_{0}<< R$ gives the stability domain for the
bound states considered as elementary particles. The calculation
of the following approximation is to be associated with accounting
of the difference of the subspace $H_{a}$ from the space $H$.

Finally we note that for an arbitrary vector $|\ )\in {\tilde
H}_{a}$ the following evident relations are valid:
\begin{equation}
\hat{\zeta}({\mathbf x})\hat{\xi}({\mathbf x}')|\ )=0, \quad (\ |
{\hat{\zeta}}^{+}({\mathbf x}){\hat{\xi}}^{+}({\mathbf x}')=0, \quad
|{\mathbf x}-{\mathbf x}'|<a, \label{19}
\end{equation}
or
\[
\hat{\zeta}({\mathbf x})\hat{\xi}({\mathbf x}'){\tilde H}_{a}=0,
\quad {\tilde H}_{a}{\hat{\zeta}}^{+}({\mathbf
x}){\hat{\xi}}^{+}({\mathbf x}')=0, \quad |{\mathbf x}-{\mathbf
x}'|<a,
\]
where $\hat{\zeta}$ and $\hat{\xi}$ are any of the annihilation
operators ${\hat\chi}_{1}, {\hat\chi}_{2}, {\hat\eta}$.

\section{The Structure of $\tilde{\hat{A}}({\mathbf
u},{\mathbf v})$ operators }

Here we consider the method for obtaining the operators
$\tilde{\hat{ A}}$. Let the operator $\hat{A}$ represents a
normal-ordered production of ${\hat\psi}_{i}({\mathbf v})$,
${\hat\psi}_{i}^{+}({\mathbf u})$, $(i=1,2)$:
\begin{equation}
{\hat A}({\mathbf u},{\mathbf v}) = {\hat\psi}_{1}^{+}({\mathbf
u}_{1})\ldots {\hat\psi}_{2}^{+}({\mathbf u}_{2})
{\hat\psi}_{1}({\mathbf v}_{1}) \ldots {\hat\psi}_{2}({\mathbf
v}_{2})\ldots \label{20}
\end{equation}
The operators of such type are the particle density operator
$\hat{\rho}_{i}({\mathbf x})$,
\[
{\hat{\rho}}_{i}({\mathbf x}) = {\hat\psi}_{i}^{+}({\mathbf
x}){\hat\psi}_{i}({\mathbf x}),
\]
the momentum density operator $\hat{\pi}_{i}({\mathbf x})$,
\[
\hat{\mbox{\boldmath$\pi$}}_{i}({\mathbf x})=-{i\over
2}\biggl({\hat\psi}_{i}^{+}({\mathbf
x}){\partial{\hat\psi}_{i}({\mathbf x})\over\partial{\mathbf
x}}-{\partial{\hat\psi}_{i}^{+}({\mathbf x})\over\partial{\mathbf
x}}{\hat\psi}_{i}({\mathbf x})\biggr),
\]
a Hamiltonian of the system, etc.

The matrix element in the right-hand side of (\ref{17}) may be
written as the following vacuum average:
\[
\langle 0|{\hat\psi}_{1}({\mathbf x}_{1})\ldots
{\hat\psi}_{2}({\mathbf x}_{2})\ldots
{\hat{\varphi}}_{\alpha}({\mathbf x})\ldots{\hat A}({\mathbf
u},{\mathbf v}) {\hat{\varphi}}_{\alpha^{\prime}}^{+}({\mathbf
x})\ldots {\hat\psi}_{2}({\mathbf x}_{2}^{\prime})\ldots
{\hat\psi}_{1}({\mathbf x}_{1}^{\prime})\ldots|0\rangle.
\]
Let us note that when calculating this average by using Wick's
theorem, the quantity, which is averaged over the vacuum state has
the meaning of a mixed $T$-product if to consider the operators
${\hat\psi}_{1},\ldots, {\hat\psi}_{2},\ldots, {\hat\varphi}\ldots$
referring to the moment of time $+0$, the operators
${\hat\psi}_{1}^{+},\ldots, {\hat\psi}_{2}^{+},$ $\ldots,
{\hat\varphi}^{+}\ldots$ to $-0$, and the normal-ordered operator
${\hat A}({\mathbf u},{\mathbf v})$ refereing to the moment of time
$0$. Thus, there is no need to place the contractions inside the
expression for ${\hat A}({\mathbf u},{\mathbf v})$. Let
\begin{equation}
A_{b}({\mathbf y};{\mathbf y}';{\mathbf u},{\mathbf v})\equiv
A_{b}({\mathbf y}_{1},\ldots,{\mathbf y}_{2},\ldots,{\mathbf
y},\ldots;{\mathbf y}_{1}^{\prime},\ldots,{\mathbf
y}_{2}^{\prime},\ldots,{\mathbf y}',\ldots;{\mathbf u},{\mathbf v})
\label{22}
\end{equation}
be the analytic expression that corresponds to the diagram "$b$".
For this diagram the operators with the arguments ${\mathbf u}$ are
related to the operators with the arguments ${\mathbf
y}_{1},\ldots,{\mathbf y}_{2}$, $\ldots,{\mathbf y},\ldots$. In
addition, the latter arguments are spaced apart by the distances
greater than $a$ and coincide with some of the arguments ${\mathbf
x}_{1},\ldots,{\mathbf x}_{2},\ldots,{\mathbf x},\ldots,$. The
similar statement should be also made concerning the arguments
${\mathbf y}_{1}^{\prime},\ldots, {\mathbf
y}_{2}^{\prime},\ldots,{\mathbf y}',\ldots$. Then, the operator
$\tilde{\hat A}({\mathbf u},{\mathbf v})$ acting in ${\tilde H}$ is
given, according to (\ref{6}), by
\begin{equation}
{\tilde{\hat A}}({\mathbf  u},{\mathbf  v})=\sum_{b}\int   {\hat
R}_{1}{\hat R}_{2}{\hat R}A_{b}({\mathbf  y};{\mathbf   y}';{\mathbf
u},{\mathbf v}){\hat R}_{1}^{\prime}{\hat R}_{2}^{\prime}{\hat
R}^{\prime}, \label{23}
\end{equation}
where
\begin{gather*}
{\hat R}_{1}=\prod{\hat\chi}^{+}({\mathbf y}_{1})d{\mathbf
y}_{1},\qquad {\hat R}_{2}=\prod{\hat\chi}^{+}({\mathbf
y}_{2})d{\mathbf y}_{2}, \qquad
{\hat R}=\prod{\hat\eta}^{+}({\mathbf y})d{\mathbf y}, \\
{\hat R}_{1}^{\prime}=\prod{\hat\chi}({\mathbf
y}_{1}^{\prime})d{\mathbf y}_{1}^{\prime}, \qquad {\hat
R}_{2}^{\prime}=\prod{\hat\chi}({\mathbf y}_{2}^{\prime})d{\mathbf
y}_{2}^{\prime}, \qquad {\hat R}^{\prime}=\prod{\hat\eta}({\mathbf
y}^{\prime})d{\mathbf y}^{\prime}
\end{gather*}
Here the summation is taken over all diagrams of the described
type.

If $\hat{A}({\mathbf v})=1$ (see the proof of (\ref{10})), then
${\tilde{\hat A}}({\mathbf v})=1$ on the subspace ${\tilde{H}}_{a}$.
Let now $\hat{A}({\mathbf u},{\mathbf v})={\hat\psi}_{1}({\mathbf
v})$. Then the only diagrams of the described type for the vacuum
average $\langle 0|\ldots {\hat\psi}_{1}({\mathbf v})\ldots|0
\rangle$ are the following diagrams:
\[
A_{b_{1}}=\underset{a}{{\hat\psi}_{1}}({\mathbf
v})\ldots\underset{a}{{\hat\psi}^{+}_{1}}({\mathbf y}^{\prime}),
\qquad A_{b_{2}}=\underset{a}{{\hat\psi}_{2}}({\mathbf
y}_{2})\ldots\underset{b}{{\hat\psi}_{1}}({\mathbf v})\ldots
\underset{ba}{{\hat\varphi}^{+}_{\alpha}}({\mathbf y}^{\prime}).
\]
The expressions
\[
A_{b_{1}}({\mathbf y}_{1}^{\prime};{\mathbf v})=\delta({\mathbf
v}-{\mathbf y}_{1}^{\prime}),
\]
\[
A_{b_{2}}({\mathbf y}_{2};{\mathbf y}^{\prime},{\mathbf v})=\int
d{\mathbf x}_{1} d{\mathbf x}_{2}\varphi_{\alpha}({\mathbf
x}_{1}-{\mathbf x}_{2})\delta({\mathbf y}'-{\mathbf
X})\delta({\mathbf v}-{\mathbf x}_{1})\delta({\mathbf
x}_{2}-{\mathbf y}_{2})=\varphi_{\alpha}({\mathbf v},{\mathbf
y}_{2},{\mathbf y}'),
\]
correspond to the above mentioned diagrams. Here
$\varphi_{\alpha}({\mathbf v},{\mathbf y}_{2},{\mathbf y}')$ is
defined in accordance with (\ref{1'}). Therefore, according to
(\ref{23}) we have
\begin{equation}
\tilde{\hat{\psi}}_{1}({\mathbf v})=\int d{\mathbf
y}_{1}^{\prime}A_{b_{1}}({\mathbf y}_{1}^{\prime},{\mathbf v}){\hat
\chi}_{1}({\mathbf y}_{1}^{\prime})+\int d{\mathbf y}_{2}d{\mathbf
y}^{\prime} A_{b_{2}}({\mathbf y}_{2};{\mathbf y}^{\prime},{\mathbf
v}){\hat \chi}_{2}^{+}({\mathbf y}_{2}){\hat\eta}({\mathbf
y}^{\prime})={\hat \chi}_{1}({\mathbf v})+\hat{O}_{1}({\mathbf
v}),\label{26}
\end{equation}
where
\begin{equation}
\hat{O}_{1}({\mathbf v})=\int d{\mathbf y}{\hat \varphi}({\mathbf
v},{\mathbf y}){\hat \chi}_{2}^{+}({\mathbf y}), \quad
{\hat\varphi}({\mathbf x}_{1},{\mathbf
x}_{2})=\varphi_{\alpha}({\mathbf x}){\hat\eta}_{\alpha}({\mathbf
X}) \label{26'}
\end{equation}
and ${\mathbf X}$ is given by (\ref{1'}), ${\mathbf x}={\mathbf
x}_{1}-{\mathbf x}_{2}$. Similarly, one finds
\begin{equation}
\tilde{\hat{\psi}}_{2}({\mathbf v})={\hat \chi}_{2}({\mathbf
v})+\hat{O}_{2}({\mathbf v}), \label{27}
\end{equation}
where
\[
\hat{O}_{2}({\mathbf v})=\int d{\mathbf y}{\hat
\chi}_{1}^{+}({\mathbf y}){\hat \varphi}({\mathbf y},{\mathbf v}).
\]
When deriving (\ref{26}), (\ref{27}), we have essentially used the
inequalities (\ref{6}).

Now let us consider ${\hat A}({\mathbf u}, {\mathbf
v})={\hat\psi}_{1}^{+}({\mathbf u}){\hat\psi}_{1}({\mathbf v})$. In
this case the following five diagrams
\[
A_{1}=\underset{a}{\hat{\psi}}\ldots\underset{a}{{\hat{\psi}}^{+}_{1}}({\mathbf
u})\underset{b}{\hat{\psi}_{1}}({\mathbf
v})\ldots\underset{b}{\hat{\psi}^{+}}, \qquad
A_{2}=\underset{a}{\hat{\psi}}\ldots \underset{b}{\hat{\psi}}\ldots
\underset{b}{\hat{\psi}^{+}_{1}}({\mathbf
u})\underset{c}{\hat{\psi}_{1}}({\mathbf
v})\ldots\underset{ca}{\hat{\varphi}^{+}},\qquad
A_{3}=\underset{ab}{\hat{\varphi}}\ldots\underset{b}{\hat{\psi}^{+}_{1}}({\mathbf
u})\underset{c}{\hat{\psi}_{1}}({\mathbf
v})\ldots\underset{c}{\hat{\psi}^{+}}
\ldots\underset{a}{\hat{\psi}^{+}},
\]
\[
A_{4}=\underset{ab}{\hat{\varphi}}\ldots\underset{b}{\hat{\psi}_{1}^{+}}({\mathbf
u})\underset{c}{\hat{\psi}_{1}}({\mathbf
v})\ldots\underset{ca}{\hat{\varphi}^{+}}, \qquad
A_{5}=\underset{a}{\hat{\psi}}\ldots
\underset{bd}{\hat{\varphi}}\ldots\underset{d}{\hat{\psi}^{+}_{1}}({\mathbf
u})\underset{c}{\hat{\psi}_{1}}({\mathbf v})
\ldots\underset{ca}{\hat{\varphi}^{+}}
\ldots\underset{b}{\hat{\psi}^{+}}
\]
correspond to this operator (the indices 1 and 2 for $\hat{\psi}$
and $\hat{\psi}^{+}$ can be easy restored if to take into account
(\ref{2}) and the definition of contractions). The analytic
expressions for these diagrams have the form
\begin{gather*}
\tilde{\hat A}_{1}={\hat\chi}_{1}^{+}({\mathbf u}_{1}){\hat
\chi}_{1}({\mathbf v}_{1}), \quad \tilde{\hat A}_{2}=\int d{\mathbf
z}_{1} d{\mathbf z}_{2}\,{\hat\varphi}({\mathbf z}_{1},{\mathbf
z}_{2})\delta({\mathbf z}_{1}-{\mathbf v}_{1}){\hat
\chi}_{1}^{+}({\mathbf u}_{1}){\hat\chi}_{2}^{+}({\mathbf
z}_{2})=\int d{\mathbf z}_{2}\,{\hat\varphi}({\mathbf
v}_{1},{\mathbf z}_{2}){\hat
\chi}_{1}^{+}({\mathbf u}_{1}){\hat\chi}_{2}^{+}({\mathbf z}_{2}),\\
\tilde{\hat A}_{3}=\int d{\mathbf z}_{1}d{\mathbf
z}_{2}\,{\hat\varphi}^{+}({\mathbf z}_{1},{\mathbf
z}_{2})\delta({\mathbf z}_{1}-{\mathbf u}_{1}){\hat
\chi}_{2}({\mathbf z}_{2}){\hat\chi}_{1}({\mathbf v}_{1})=\int
d{\mathbf z}_{2}\,{\hat\varphi}^{+}({\mathbf u}_{1},{\mathbf
z}_{2}){\hat \chi}_{2}({\mathbf z}_{2}){\hat\chi}_{1}({\mathbf v}_{1}),\\
\tilde{\hat A}_{4}=\int d{\mathbf z}_{1}d{\mathbf z}_{2}d{\mathbf
z}_{1}^{\prime}d{\mathbf z}_{2}^{\prime}\,{\hat\varphi}^{+}({\mathbf
z}_{1},{\mathbf z}_{2}){\hat\varphi}({\mathbf
z}_{1}^{\prime},{\mathbf z}_{2}^{\prime})\delta({\mathbf
z}_{1}-{\mathbf u}_{1})\delta({\mathbf v}_{1}-{\mathbf
z}_{1}^{\prime})\delta({\mathbf z}_{2}-{\mathbf
z}_{2}^{\prime})=\int d{\mathbf z}_{2}\,{\hat\varphi}^{+}({\mathbf
u}_{1},{\mathbf z}_{2}){\hat\varphi}({\mathbf
v}_{1},{\mathbf z}_{2}),\\
\tilde{\hat A}_{5}=-\int d{\mathbf z}_{1}d{\mathbf z}_{2}d{\mathbf
z}_{1}^{\prime}d{\mathbf z}_{2}^{\prime}\,{\hat\varphi}^{+}({\mathbf
z}_{1},{\mathbf z}_{2}){\hat\varphi}({\mathbf
z}_{1}^{\prime},{\mathbf z}_{2}^{\prime})\delta({\mathbf
z}_{1}-{\mathbf u}_{1})\delta({\mathbf v}_{1}-{\mathbf
z}_{1}^{\prime}){\hat\chi}_{2}^{+}({\mathbf
z}_{2}^{\prime}){\hat\chi}_{2}({\mathbf z}_{2}) \\ =-\int d{\mathbf
z}_{2}d{\mathbf z}_{2}^{\prime}\,{\hat\varphi}^{+}({\mathbf
u}_{1},{\mathbf z}_{2}){\hat\varphi}({\mathbf v}_{1},{\mathbf
z}_{2}^{\prime}){\hat\chi}_{2}^{+}({\mathbf
z}_{2}^{\prime}){\hat\chi}_{2}({\mathbf z}_{2}),
\end{gather*}
whence, we find the operators (\ref{23}) corresponding to the
diagrams (\ref{22}),
\[
\tilde{\hat A}_{1}={\hat\chi}_{1}^{+}({\mathbf u}_{1}){\hat
\chi}_{1}({\mathbf v}_{1}),\qquad \tilde{\hat A}_{2}={\hat
\chi}_{1}^{+}({\mathbf u}_{1}){\hat O}_{1}({\mathbf v}_{1}), \qquad
\tilde{\hat A}_{3}={\hat O}_{1}^{+}({\mathbf u}_{1}){\hat
\chi}_{1}({\mathbf v}_{1}),\qquad {\tilde{\hat A}_{4}+\tilde{\hat
A}_{5}}={\hat O}_{1}^{+}({\mathbf u}_{1}){\hat O}_{1}({\mathbf
v}_{1}).
\]
In obtaining the latter expression, we have taken into account the
anticommutative relations for ${\hat\chi}$, ${\hat\chi}^{+}$. Next
bearing in mind (\ref{26}), (\ref{27}) we obtain the final
expression for $\tilde{\hat A}({\mathbf u}, {\mathbf
v})=\widetilde{{\hat\psi}^{+}_{1}({\mathbf u}_{1}){\hat
\psi}_{1}({\mathbf v}_{1})}$ that corresponds to ${\hat A}({\mathbf
u},{\mathbf v})={\hat \psi}^{+}_{1}({\mathbf u}_{1}){\hat
\psi}_{1}({\mathbf v}_{1})$:
\begin{equation}
{\hat \psi}^{+}_{1}({\mathbf u}_{1}){\hat \psi}_{1}({\mathbf
v}_{1})\rightarrow \widetilde{{\hat \psi}^{+}_{1}({\mathbf
u}_{1}){\hat \psi}_{1}({\mathbf v}_{1})}=\tilde{{\hat
\psi}}^{+}_{1}({\mathbf u}_{1})\tilde{{\hat \psi}}_{1}({\mathbf
v}_{1}). \label{28}
\end{equation}
Similarly, one gets
\begin{alignat}{2}
{\hat \psi}_{1}({\mathbf u}_{1}){\hat \psi}_{1}({\mathbf
v}_{1})&\rightarrow \widetilde{{\hat \psi}_{1}({\mathbf u}_{1}){\hat
\psi}_{1}({\mathbf v}_{1})}&=\tilde{{\hat \psi}}_{1}({\mathbf
u}_{1})\tilde{{\hat \psi}}_{1}({\mathbf v}_{1}),\nonumber\\
{\hat \psi}^{+}_{1}({\mathbf u}_{1}){\hat \psi}_{2}({\mathbf
v}_{1})&\rightarrow \widetilde{{\hat \psi}^{+}_{1}({\mathbf
u}_{1}){\hat \psi}_{2}({\mathbf v}_{1})}&=\tilde{{\hat
\psi}}^{+}_{1}({\mathbf u}_{1})\tilde{{\hat \psi}}_{2}({\mathbf
v}_{1}),\label{29}\\
&\ldots \nonumber
\end{alignat}
In general case the following formula is valid:
\begin{equation}
{\hat \psi}^{+}_{i_{1}}({\mathbf u}_{i_{1}})\ldots {\hat
\psi}^{+}_{i_{n}}({\mathbf u}_{i_{n}}) {\hat \psi}_{j_{1}}({\mathbf
v}_{j_{1}})\ldots {\hat \psi}_{j_{m}}({\mathbf
v}_{j_{m}})\rightarrow \widetilde{{\hat \psi}^{+}_{i_{1}}({\mathbf
u}_{i_{1}})\ldots {\hat \psi}_{j_{m}}({\mathbf
v}_{j_{m}})}=\tilde{{\hat \psi}}^{+}_{i_{1}}({\mathbf u}_{1})\ldots
\tilde{{\hat \psi}}_{j_{m}}({\mathbf v}_{j_{m}}). \label{30}
\end{equation}
To explain this formula we note that each of ${\hat
\psi}_{j}({\mathbf v}_{j})$ (or ${\hat \psi}^{+}_{i}({\mathbf
u}_{i})$) entering ${\hat A}({\mathbf u},{\mathbf v})$ (see
(\ref{20})) is related to other operators ${\hat \psi}^{+}$ (or
${\hat \psi}$) not entering ${\hat A}({\mathbf u},{\mathbf v})$ only
by unique way, which leads to the binary relations (\ref{28}),
(\ref{29}). Therefore, we come to (\ref{30}) by sorting out all the
operators ${\hat \psi}_{i}$, ${\hat \psi}_{i'}$ containing in ${\hat
A}({\mathbf u},{\mathbf v})$.

The operators ${\hat\psi}_{i}({\mathbf x})$, ${\hat
\psi}_{j}({\mathbf x}')$ are anticommutative. For this reason there
is a question concerning the consistency of (\ref{29}), (\ref{30}).
The anticommutativity of $\tilde{{\hat \psi}}_{i}({\mathbf x})$,
$\tilde{{\hat \psi}}_{j}({\mathbf x}')$ (and also $\tilde{{\hat
\psi}}^{+}_{i}({\mathbf x})$, $\tilde{{\hat \psi}}^{+}_{j}({\mathbf
x}')$ ) represents the consistency condition of these formulas,
\[
\{ {\tilde{{\hat \psi}}}_{i}({\mathbf x}), \tilde{{\hat
\psi}}_{j}({\mathbf x}')\}=\{\tilde{{\hat \psi}}^{+}_{i}({\mathbf
x}), \tilde{{\hat \psi}}^{+}_{j}({\mathbf x}')\}=0.
\]
The validity of these anticommutative relations can be easily
proved if to use the definitions (\ref{26})-(\ref{27}) for
$\tilde{{\hat \psi}}$, $\tilde{{\hat \psi}}^{+}$ and the
commutative relations for ${\hat \chi}$, ${\hat \chi}^{+}$ and
${\hat \eta}_{\alpha}$, ${\hat \eta}_{\alpha}^{+}$.

\section{The Operators of Basic Physical Quantities}

In this section we consider the operators of basic physical
quantities, which act in the Hilbert space ${\tilde H}$. Let us
start from the density operator for particles of the first kind.
The corresponding operator acting in the original Hilbert space
$H$ is of the form
\[
{\hat\rho}_{1}({\mathbf x})={\hat\psi}_{1}^{+}({\mathbf
x}){\hat\psi}_{1}({\mathbf x}). \label{31}
\]
Hence, in accordance with (\ref{29}), one finds
\[
\tilde{\hat\rho}_{1}({\mathbf x})=\tilde{\hat\psi}^{+}_{1}({\mathbf
x})\tilde{\hat\psi}_{1}({\mathbf x})={\hat\chi}_{1}^{+}({\mathbf
x}){\hat\chi}_{1}({\mathbf x})+{\hat O}_{1}^{+}({\mathbf
x}){\hat\chi}_{1}({\mathbf x})+{\hat\chi}_{1}^{+}({\mathbf x}){\hat
O}_{1}({\mathbf x})+{\hat O}_{1}^{+}({\mathbf x}){\hat
O}_{1}({\mathbf x}). \label{32}
\]
Note that the operators with zero matrix elements in the subspace
${\tilde H}_{a}$ appear in the right-hand side of (\ref{28}) because
of points ${\mathbf u}_{1}$ and ${\mathbf v}_{1}$ are close to each
other. Since
\[
{\hat O}_{1}^{+}({\mathbf x}){\hat\chi}_{1}({\mathbf x})=\int
d{\mathbf y}{\hat\varphi}^{+}({\mathbf x},{\mathbf y}){\hat
\chi}_{2}({\mathbf y}){\hat \chi}_{1}({\mathbf x}), \qquad {\hat
\chi}_{1}^{+}({\mathbf x}){\hat O}_{1}({\mathbf x})= \int d{\mathbf
y} {\hat \chi}_{1}^{+}({\mathbf x}){\hat \chi}_{2}^{+}({\mathbf
y}){\hat \varphi}({\mathbf x},{\mathbf y})
\]
and ${\hat\varphi}({\mathbf x},{\mathbf y})$ differs from zero only
for $|{\mathbf x}-{\mathbf y}|\lesssim a$, these operators according
to (\ref{19}) do not have the matrix elements in the subspace
${\tilde H}_{a}$ and therefore can be omitted. Using the permutation
relation $\{{\hat \chi}_{2}^{+}({\mathbf z}_{1}),{\hat
\chi}_{2}({\mathbf z}_{2})\}=\delta ({\mathbf z}_{1}-{\mathbf
z}_{2})$ the operator ${\hat O}_{1}^{+}({\mathbf x}){\hat
O}_{1}({\mathbf x})$ is written as
\[
{\hat O}_{1}^{+}({\mathbf x}){\hat O}_{1}({\mathbf x})=\int
d{\mathbf z}{\hat\varphi}^{+}({\mathbf x},{\mathbf
z}){\hat\varphi}({\mathbf x},{\mathbf z})+\int d{\mathbf
z}_{1}d{\mathbf z}_{2}{\hat\varphi}^{+}({\mathbf x},{\mathbf
z}_{2}){\hat\varphi}({\mathbf x},{\mathbf
z}_{1}){\hat\chi}_{2}^{+}({\mathbf z}_{1}) {\hat\chi}_{2}({\mathbf
z}_{2}).
\]
The matrix element of the second term is zero in the subspace
${\tilde H}_{a}$ because of ${\mathbf z}_{1}\approx{\mathbf
z}_{2}\approx{\mathbf x}$ (see (\ref{19})). For this reason this
term can be omitted. Thus, with the use of the method that was
described in previous section we have
\begin{equation}
\tilde{\hat \rho}_{1}({\mathbf x})={\hat\chi}_{1}^{+}({\mathbf
x}){\hat\chi}_{1}({\mathbf x})+\int d{\mathbf
z}{\hat\varphi}^{+}({\mathbf x},{\mathbf z}){\hat\varphi}({\mathbf
x},{\mathbf z}). \label{33}
\end{equation}
Similarly, if ${\hat\rho}_{2}({\mathbf
x})={\hat\psi}_{2}^{+}({\mathbf x}){\hat\psi}_{2}({\mathbf x})$
represents the density operator for particles of the second kind,
then ${\hat\rho}_{2}({\mathbf x})\rightarrow
\tilde{{\hat\rho}}_{2}({\mathbf x})$, where
\begin{equation}
\tilde{\hat \rho}_{2}({\mathbf x})={\hat\chi}_{2}^{+}({\mathbf
x}){\hat\chi}_{2}({\mathbf x})+\int d{\mathbf
z}{\hat\varphi}^{+}({\mathbf z},{\mathbf x}){\hat\varphi}({\mathbf
z},{\mathbf x}). \label{34}
\end{equation}
Bearing in mind (\ref{26}) and the assumption concerning the
"small radius" of the bound state, one gets the following
formulas:
\[
\int d{\mathbf z}{\hat\varphi}^{+}({\mathbf x},{\mathbf
z}){\hat\varphi}({\mathbf x},{\mathbf z})\approx
{\hat\eta}^{+}_{\alpha}({\mathbf x}){\hat\eta}_{\alpha}({\mathbf
x}), \qquad \int d{\mathbf z}{\hat\varphi}^{+}({\mathbf z},{\mathbf
x}){\hat\varphi}({\mathbf z},{\mathbf x})\approx
{\hat\eta}^{+}_{\alpha}({\mathbf x}){\hat\eta}_{\alpha}({\mathbf
x}),
\]
which allow us to obtain the densities operators for particles of
the first and second kinds,
\begin{equation}
\tilde{\hat \rho}_{1}({\mathbf x})={\hat\chi}_{1}^{+}({\mathbf
x}){\hat\chi}_{1}({\mathbf x})+{\hat\eta}^{+}_{\alpha}({\mathbf
x}){\hat\eta}_{\alpha}({\mathbf x}), \qquad \tilde{\hat
\rho}_{2}({\mathbf x})={\hat\chi}_{2}^{+}({\mathbf
x}){\hat\chi}_{2}({\mathbf x})+{\hat\eta}^{+}_{\alpha}({\mathbf
x}){\hat\eta}_{\alpha}({\mathbf x}). \label{35}
\end{equation}
Thus, the operators ${\hat\eta}^{+}_{\alpha}({\mathbf x})$,
${\hat\eta}_{\alpha}({\mathbf x})$ can be interpreted as the
creation and annihilation operators of the bound states with quantum
numbers $\alpha$ at the point ${\mathbf x}$, and
${\hat\eta}^{+}_{\alpha}({\mathbf x}){\hat\eta}_{\alpha}({\mathbf
x})$ as the density operator of the bound states. For example, the
first formula from (\ref{35}) have a simple physical meaning: the
density of particles of the first kind is equal to the sum of
densities for free particles of the same kind and bound states (each
bound state contains one particle of the first kind).

Consider a state vector $\Phi({\mathbf X}) =
\hat{\eta}_{\alpha}^{+}({\mathbf X})|0\rangle$, which specifies a
compound particle at the point ${\mathbf X}$ (this state vector
corresponds to continuous spectrum). Then, in accordance with
(\ref{35}), we have
\[
(\Phi({\mathbf X}),\hat{\rho}_{1}({\mathbf x}_{1})\Phi({\mathbf
X}'))= \delta({\mathbf X}-{\mathbf X}')\biggl({M\over
m_{2}}\biggr)^{3}\bigg\vert\varphi_{\alpha}\biggl({M\over
m_{2}}({\mathbf x}_{1}-{\mathbf X})\biggr)\bigg\vert^{2}.
\]
For a wave packet
\[
\Psi_{{\mathbf X}_{0}}=\int d{\mathbf X}f_{{\mathbf X}_{0}}({\mathbf
X})\Phi({\mathbf X}), \qquad \int d{\mathbf X}\bigg\vert f_{{\mathbf
X}_{0}}({\mathbf X})\bigg\vert^{2}=1
\]
the quantity
\[
(\Psi_{{\mathbf X}_{0}},\hat{\rho}_{1}({\mathbf
x}_{1})\Psi_{{\mathbf X}_{0}})=\biggl({M\over m_{2}}\biggr)^{3}\int
d{\mathbf X}\bigg\vert f_{{\mathbf X}_{0}}({\mathbf
X})\bigg\vert^{2}\bigg\vert\varphi_{\alpha}\biggl({M\over
m_{2}}({\mathbf x}_{1}-{\mathbf X})\biggr)\bigg\vert^{2}
\]
should be treated as the probability density to find the first
particle at the point ${\mathbf x}_{1}$ if the atom is in a state
$\Psi_{{\mathbf X}_{0}}$. If the bound state is localized near a
point ${\mathbf X}_{0}$ (i.e., $\vert f_{{\mathbf X}_{0}}({\mathbf
X})\vert^{2}\rightarrow\delta({\mathbf X}-{\mathbf X}_{0})$), then
\[
(\Psi_{{\mathbf X}_{0}},\hat{\rho}_{1}({\mathbf
x}_{1})\Psi_{{\mathbf X}_{0}})\rightarrow \biggl({M\over
m_{2}}\biggr)^{3}\bigg\vert\varphi_{\alpha}\biggl({M\over
m_{2}}({\mathbf x}_{1}-{\mathbf X}_{0})\biggr)\bigg\vert^{2}.
\]
Since $(M/ m_{2})({\mathbf x}_{1}-{\mathbf X}_{0})={\mathbf
x}={\mathbf x}_{1}-{\mathbf x}_{2}$, we come (as it should be) to
the probability distribution for the space coordinate of the first
particle in atom (the atom is at the point ${\mathbf X}_{0}$).

Let us find now the momentum density operator in the space ${\tilde
H}$. In original Hilbert space, the momentum density operator
$\hat{\mbox{\boldmath$\pi$}}_{1}({\mathbf x})$ for particles of the
first kind is defined as
\[
{\hat{\mbox{\boldmath$\pi$}}}_{1}({\mathbf x})=-{i \over
2}\biggl({\hat \psi}_{1}^{+}({\mathbf x}){\partial{\hat
\psi}_{1}({\mathbf x})\over\partial {\mathbf x}} -{\partial {\hat
\psi}_{1}^{+}({\mathbf x})\over\partial {\mathbf x}} {\hat
\psi}_{1}({\mathbf x})\biggr).\label{36}
\]
Then according to (\ref{29})
\[
{\hat{\mbox{\boldmath$\pi$}}}_{1}({\mathbf x})\rightarrow
\tilde{\hat{\mbox{\boldmath$\pi$}}}_{1}({\mathbf x})=-{i \over
2}\biggl(\tilde{{\hat \psi}}_{1}^{+}({\mathbf x}){\partial
\tilde{{\hat \psi}}_{1}({\mathbf x})\over\partial {\mathbf x}} -
{\partial\tilde{{\hat \psi}}_{1}^{+}({\mathbf x})
\over\partial{\mathbf x}}\tilde{{\hat \psi}}_{1}({\mathbf
x})\biggr).
\]
Following the derivation of (\ref{33}), (\ref{34}) for
$\tilde{\hat\rho}_{1}({\mathbf x})$ and
$\tilde{\hat\rho}_{2}({\mathbf x})$ we obtain
\begin{equation}
\tilde{\hat{\mbox{\boldmath$\pi$}}}_{1}({\mathbf x})=-{i\over
2}\biggl(\hat{\chi}_{1}^{+}({\mathbf
x}){\partial\hat{\chi}_{1}({\mathbf x})\over\partial{\mathbf
x}}-{\partial\hat{\chi}_{1}^{+}({\mathbf x}) \over\partial{\mathbf
x}} \hat{\chi}_{1}({\mathbf x})\biggr) -{i\over 2}\int d{\mathbf
y}\biggl( {\hat\varphi}^{+}({\mathbf x},{\mathbf
y}){\partial{\hat\varphi}({\mathbf x},{\mathbf y})\over \partial
{\mathbf x}}- {\partial{\hat\varphi}^{+}({\mathbf x},{\mathbf
y})\over\partial{\mathbf x}}{\hat\varphi}({\mathbf x},{\mathbf
y})\biggr), \label{37}
\end{equation}
\begin{equation}
\tilde{\hat{\mbox{\boldmath$\pi$}}}_{2}({\mathbf x})=-{i\over
2}\biggl(\hat{\chi}_{2}^{+}({\mathbf
x}){\partial\hat{\chi}_{2}({\mathbf x}) \over\partial{\mathbf
x}}-{\partial\hat{\chi}_{2}^{+}({\mathbf x}) \over\partial{\mathbf
x}} \hat{\chi}_{2}({\mathbf x})\biggr) -{i\over 2}\int d{\mathbf
y}\biggl( {\hat\varphi}^{+}({\mathbf y},{\mathbf
x}){\partial{\hat\varphi}({\mathbf y},{\mathbf x})\over \partial
{\mathbf x}}- {\partial{\hat\varphi}^{+}({\mathbf y},{\mathbf
x})\over\partial{\mathbf x}}{\hat\varphi}({\mathbf y},{\mathbf
x})\biggr). \label{38}
\end{equation}
It is convenient for our further consideration to rewrite
(\ref{37}), (\ref{38}) in terms of the center of mass variables
${\mathbf y}={\mathbf y}_{1}-{\mathbf y}_{2}$ and ${\mathbf
Y}=(m_{1}{\mathbf y}_{1}+m_{2}{\mathbf y}_{2})/( m_{1}+m_{2})$:
\begin{equation}
\begin{split}
\tilde{\hat{\mbox{\boldmath$\pi$}}}_{1}({\mathbf x})&=-{i\over
2}\biggl({\hat\chi}_{1}^{+}({\mathbf x}){\partial
{\hat\chi}_{1}({\mathbf x})\over
\partial{\mathbf x}}-{\partial {\hat\chi}_{1}^{+}({\mathbf x})\over
\partial{\mathbf x}}{\hat\chi}_{1}({\mathbf x})\biggr)
-{i\over 2}\int d{\mathbf y}d{\mathbf Y}\delta\left({\mathbf
x}-{\mathbf Y}-{m_{2}\over M}{\mathbf y}\right) \biggl\{
\hat{\varphi}^{+}({\mathbf y},{\mathbf
Y}){\partial\hat{\varphi}({\mathbf y},{\mathbf Y})\over
\partial{\mathbf y}}
\\& -{\partial\hat{\varphi}^{+}({\mathbf y},{\mathbf Y})\over \partial{\mathbf y}}
\hat{\varphi}({\mathbf y},{\mathbf Y})+ {m_{1}\over
M}\biggl(\hat{\varphi}^{+}({\mathbf y},{\mathbf
Y}){\partial\hat{\varphi}({\mathbf y},{\mathbf Y})\over
\partial{\mathbf Y}}-{\partial\hat{\varphi}^{+}({\mathbf y},{\mathbf
Y})\over \partial{\mathbf
Y}}\hat{\varphi}({\mathbf y},{\mathbf Y})\biggr)\biggr\},\\
\tilde{\hat{\mbox{\boldmath$\pi$}}}_{2}({\mathbf x})&=-{i\over
2}\biggl( {\hat\chi}_{2}^{+}({\mathbf x}){\partial
{\hat\chi}_{2}({\mathbf x})\over
\partial{\mathbf x}}-{\partial {\hat\chi}_{2}^{+}({\mathbf x})\over
\partial{\mathbf x}}{\hat\chi}_{2}({\mathbf x})\biggr)
-{i\over 2}\int d{\mathbf y}d{\mathbf Y}\delta\left({\mathbf
x}-{\mathbf Y}+{m_{1}\over M}{\mathbf y}\right) \biggl\{
-\hat{\varphi}^{+}({\mathbf y},{\mathbf
Y}){\partial\hat{\varphi}({\mathbf y},{\mathbf Y})\over \partial{\mathbf y}}\\
&+ {\partial\hat{\varphi}^{+}({\mathbf y},{\mathbf Y})\over
\partial{\mathbf y}} \hat{\varphi}({\mathbf y},{\mathbf Y})+{m_{2}\over
M}\biggl(\hat{\varphi}^{+}({\mathbf y},{\mathbf
Y}){\partial\hat{\varphi}({\mathbf y},{\mathbf Y})\over
\partial{\mathbf Y}}-{\partial\hat{\varphi}^{+}({\mathbf y},{\mathbf
Y})\over \partial{\mathbf Y}}\hat{\varphi}({\mathbf y},{\mathbf
Y})\biggr)\biggr\}, \label{39}
\end{split}
\end{equation}
where $\hat{\varphi}({\mathbf y}_{1},{\mathbf
y}_{2})\equiv\hat{\varphi}({\mathbf y},{\mathbf Y})$. Note that in
terms of these variables the operators
$\tilde{\hat\rho}_{1}({\mathbf x})$ and
$\tilde{\hat\rho}_{2}({\mathbf x})$ have the form
\begin{eqnarray}
\tilde{\hat\rho}_{1}({\mathbf x})=\hat{\chi}^{+}_{1}({\mathbf
x})\hat{\chi}_{1}({\mathbf x})+\int d{\mathbf y}d{\mathbf
Y}\delta\left({\mathbf x}-{\mathbf Y}-{m_{2}\over M}{\mathbf
y}\right)\hat{\varphi}^{+}({\mathbf y},{\mathbf
Y})\hat{\varphi}({\mathbf y},{\mathbf Y}), \nonumber
\\
\tilde{\hat\rho}_{2}({\mathbf x})=\hat{\chi}^{+}_{2}({\mathbf
x})\hat{\chi}_{2}({\mathbf x})+\int d{\mathbf y}d{\mathbf
Y}\delta\left({\mathbf x}-{\mathbf Y}+{m_{1}\over M}{\mathbf
y}\right)\hat{\varphi}^{+}({\mathbf y},{\mathbf
Y})\hat{\varphi}({\mathbf y},{\mathbf Y}).\label{40}
\end{eqnarray}
It is clear that (\ref{39}), (\ref{40}) can be expressed through the
creation and annihilation operators
$\hat{\eta}^{+}_{\alpha}({\mathbf x})$,
$\hat{\eta}_{\alpha}({\mathbf x})$ of atoms if to employ
(\ref{26'}). Taking into account (\ref{39}) it is easy to find the
operator $\tilde{\hat{\mbox{\boldmath$\pi$}}}=
\tilde{\hat{\mbox{\boldmath$\pi$}}}_{1}({\mathbf
x})+\tilde{\hat{\mbox{\boldmath$\pi$}}}_{2}({\mathbf x})$ of the
total momentum density of the system in the approximation, where the
radius of bound state is small,
\begin{gather*}
\tilde{\hat{\mbox{\boldmath$\pi$}}}=-{i\over 2}\biggl(
{\hat\chi}_{1}^{+}({\mathbf x}){\partial {\hat\chi}_{1}({\mathbf
x})\over
\partial{\mathbf x}}-{\partial {\hat\chi}_{1}^{+}({\mathbf x})\over
\partial{\mathbf x}}{\hat \chi}_{1}({\mathbf x})\biggr)
-{i\over 2}\biggl( {\hat\chi}_{2}^{+}({\mathbf x}){\partial
{\hat\chi}_{2}({\mathbf x})\over
\partial{\mathbf x}}-{\partial {\hat\chi}_{2}^{+}({\mathbf x})\over
\partial{\mathbf x}}{\hat\chi}_{2}({\mathbf x})\biggr)-{i\over 2}\biggl(
{\hat\eta}_{\alpha}^{+}({\mathbf x}){\partial
{\hat\eta}_{\alpha}({\mathbf x})\over
\partial{\mathbf x}}-{\partial {\hat\eta}_{\alpha}^{+}({\mathbf x})\over
\partial{\mathbf x}}{\hat\eta}_{\alpha}({\mathbf x})\biggr). \label{41}
\end{gather*}
The third term in this formula is in accordance with interpretation
of ${\hat\eta}_{\alpha}^{+}({\mathbf x}),
{\hat\eta}_{\alpha}({\mathbf x})$ as the creation and annihilation
operators of the bound state with quantum numbers $\alpha$ at the
point ${\mathbf x}$.

Finally let us consider a Hamiltonian in the space ${\tilde H}$.
We suppose that this Hamiltonian has a standard form in the
Hilbert space $H$ and can be written as
\[
\hat{\mathcal H}=\hat{\mathcal H}_{0}+{\hat V}, \label{42}
\]
where $\hat{\mathcal H}_{0}$ and ${\hat V}$ are the operators of
kinetic energy and potential energy given by
\begin{gather}
\hat{\mathcal H}_{0}=\sum_{i=1}^{2}{1\over 2m_{i}}\int d{\mathbf
x}{\partial \hat{\psi}_{i}^{+}({\mathbf x})\over\partial{\mathbf
x}}{\partial\hat{\psi}_{i}({\mathbf x})\over\partial{\mathbf x}},
\\
\hat{V}={1\over 2}\sum_{i,j=1}^{2}\int d{\mathbf x} d{\mathbf
x}'\hat{\psi}_{i}^{+}({\mathbf x})\hat{\psi}_{j}^{+}({\mathbf
x}')\nu_{ij}({\mathbf x}-{\mathbf x}')\hat{\psi}_{j}({\mathbf
x}')\hat{\psi}_{i}({\mathbf x}), \nonumber \label{43}
\end{gather}
and $\nu_{ij}({\mathbf x}-{\mathbf x}')$ is a potential energy of
interaction for particles of the kinds $i$ and $j$. After the
similar calculations that led us to the expressions for
$\tilde{\hat\rho}$, $\tilde{\hat{\mbox{\boldmath$\pi$}}}$, we obtain
\begin{equation}
\hat{\mathcal H}_{0}\rightarrow\tilde{\hat{\mathcal
H}}_{0}=\sum_{i=1}^{2}{1\over 2m_{i}}\biggr(\int d{\mathbf
x}{\partial\hat{\chi}_{i}^{+}({\mathbf x})\over\partial{\mathbf
x}}{\partial\hat{\chi}_{i}({\mathbf x})\over\partial{\mathbf
x}}+\int d{\mathbf x}_{1}d{\mathbf
x}_{2}{\partial\hat{\varphi}^{+}({\mathbf x}_{1},{\mathbf
x}_{2})\over\partial{\mathbf x}_{i}}{\partial\hat{\varphi}({\mathbf
x}_{1},{\mathbf x}_{2}) \over\partial{\mathbf x}_{i}}\biggr).
\label{44}
\end{equation}
Next changing to the center of mass variables (see (\ref{26'}))
and noting that
\[
{\partial\over\partial{\mathbf x}_{1}}={\partial\over\partial
{\mathbf x}}+{m_{1}\over M} {\partial\over\partial {\mathbf X}},
\qquad {\partial\over\partial{\mathbf
x}_{2}}=-{\partial\over\partial {\mathbf x}}+{m_{2}\over M}
{\partial\over\partial {\mathbf X}},
\]
one gets
\begin{equation}
\tilde{\hat{\mathcal H}}_{0}=\sum_{i=1}^{2}{1\over 2m_{i}}\int
d{\mathbf x}{\partial\hat{\chi}_{i}^{+}({\mathbf
x})\over\partial{\mathbf x}}{\partial\hat{\chi}_{i}({\mathbf x})
\over\partial{\mathbf x}} -{1\over 2\mu}\int d{\mathbf x}d{\mathbf
X}\hat{\eta}_{\alpha}^{+}({\mathbf X})\hat{\eta}_{\beta}({\mathbf
X}){\partial{\varphi}_{\alpha}^{*}({\mathbf x})\over\partial{\mathbf
x}}{\partial{\varphi}_{\beta}({\mathbf x})\over\partial{\mathbf
x}}+{1\over 2M}\int d{\mathbf
X}{\partial\hat{\eta}_{\alpha}^{+}({\mathbf X})
\over\partial{\mathbf X}} {\partial\hat{\eta}_{\alpha}({\mathbf X})
\over\partial{\mathbf X}} \label{45}
\end{equation}
where $\mu=m_{1}m_{2}/(m_{1}+m_{2})$ is a reduced mass.

Let us find now $\tilde{\hat V}$ (${\hat V}\rightarrow \tilde{\hat
V}$). According to (\ref{29}) we have
\[
\tilde{\hat V}={1\over 2}\sum_{i,j=1}^{2}\int d{\mathbf x}d{\mathbf
x}'\tilde{\hat{\psi}}_{i}^{+}({\mathbf
x})\tilde{\hat{\psi}}_{j}^{+}({\mathbf x}')\nu_{ij}({\mathbf
x}-{\mathbf x}')\tilde{\hat{\psi}}_{j}({\mathbf
x}')\tilde{\hat{\psi}}_{i}({\mathbf x}),
\]
where $\tilde{\hat\psi}_{i}({\mathbf x})=\hat{\chi}_{i}({\mathbf
x})+{\hat O}_{i}({\mathbf x})$ (see (\ref{29})). Thus, $\tilde{\hat
V}$ can be represented in the form
\[
\tilde{\hat V}=\tilde{\hat V}_{0}+\tilde{\hat V}_{1}+\tilde{\hat
V}_{2}+\tilde{\hat V}_{3}+\tilde{\hat V}_{4},
\]
where $\tilde{\hat V}_{k}$ ($k=0,\ldots 4$) contains $k$ multipliers
of type $\hat{\chi}$ and $4-k$ multipliers of type $\hat O$. The
operators ${\hat O}_{i}({\mathbf x})$ have, according to (\ref{26}),
(\ref{27}), the form
\begin{gather}
{\hat O}_{i}({\mathbf x})=\int d{\mathbf
y}\hat{\varphi}_{i}({\mathbf x},{\mathbf
y})\hat{\chi}_{i'}^{+}({\mathbf y}), \nonumber \\
\hat{\varphi}_{1}({\mathbf x},{\mathbf y})=\hat{\varphi}({\mathbf
x},{\mathbf y}), \quad \hat{\varphi}_{2}({\mathbf x},{\mathbf
y})=\hat{\varphi}({\mathbf y},{\mathbf x}), \label{46}
\end{gather}
where index $i'$ is defined as $1'=2$, $2'=1$. Then $\tilde{\hat
V}_{0}$ is of the form
\begin{multline}
\tilde{\hat V}_{0}={1\over 2}\sum_{i,j=1}^{2}\int d{\mathbf
x}_{1}d{\mathbf x}_{2}\nu_{ij}({\mathbf x}_{1}-{\mathbf x}_{2}) \\
\times \int d{\mathbf y}_{1}d{\mathbf y}_{2}d{\mathbf
y}_{3}d{\mathbf y}_{4}\hat{\varphi}_{i}^{+}({\mathbf x}_{1},{\mathbf
y}_{1})\hat{\varphi}_{j}^{+}({\mathbf x}_{2},{\mathbf y}_{2})
\hat{\varphi}_{j}({\mathbf x}_{2},{\mathbf
y}_{3})\hat{\varphi}_{i}({\mathbf x}_{1},{\mathbf
y}_{4})\hat{\chi}_{i'}({\mathbf y}_{1})\hat{\chi}_{j'}({\mathbf
y}_{2})\hat{\chi}_{j'}^{+}({\mathbf
y}_{3})\hat{\chi}_{i'}^{+}({\mathbf y}_{4}). \label{47}
\end{multline}
Note that the operators $\hat\varphi$ and $\hat\varphi^{+}$ in
(\ref{47}) are normally ordered, whereas $\hat\chi$ and
$\hat\chi^{+}$ are not normally ordered. Therefore, we put them in
order using Wick's theorem:
\begin{gather}
\hat{\chi}_{i'}({\mathbf y}_{1})\hat{\chi}_{j'}({\mathbf
y}_{2})\hat{\chi}_{j'}^{+}({\mathbf
y}_{3})\hat{\chi}_{i'}^{+}({\mathbf
y}_{4})=:\hat{\chi}_{i'}({\mathbf y}_{1})\hat{\chi}_{j'}({\mathbf
y}_{2})\hat{\chi}_{j'}^{+}({\mathbf
y}_{3})\hat{\chi}_{i'}^{+}({\mathbf
y}_{4}):+:\underset{a}{\hat{\chi}_{i'}}({\mathbf
y}_{1})\hat{\chi}_{j'}({\mathbf
y}_{2})\underset{a}{\hat{\chi}_{j'}^{+}}({\mathbf
y}_{3})\hat{\chi}_{i'}^{+}({\mathbf y}_{4}): \nonumber\\
+:\underset{a}{\hat{\chi}_{i'}}({\mathbf
y}_{1})\hat{\chi}_{j'}({\mathbf y}_{2})\hat{\chi}_{j'}^{+}({\mathbf
y}_{3})\underset{a}{\hat{\chi}_{i'}^{+}}({\mathbf y}_{4}):+:
\hat{\chi}_{i'}({\mathbf
y}_{1})\underset{a}{\hat{\chi}_{j'}}({\mathbf
y}_{2})\underset{a}{\hat{\chi}_{j'}^{+}}({\mathbf
y}_{3})\hat{\chi}_{i'}^{+}({\mathbf y}_{4}):+:
\hat{\chi}_{i'}({\mathbf
y}_{1})\underset{a}{\hat{\chi}_{j'}}({\mathbf
y}_{2})\hat{\chi}_{j'}^{+}({\mathbf
y}_{3})\underset{a}{\hat{\chi}_{i'}^{+}}({\mathbf y}_{4}): \nonumber \\
+: \underset{a}{\hat{\chi}_{i'}}({\mathbf
y}_{1})\underset{b}{\hat{\chi}_{j'}}({\mathbf
y}_{2})\underset{a}{\hat{\chi}_{j'}^{+}}({\mathbf
y}_{3})\underset{b}{\hat{\chi}_{i'}^{+}}({\mathbf y}_{4}):+:
\underset{a}{\hat{\chi}_{i'}}({\mathbf
y}_{1})\underset{b}{\hat{\chi}_{j'}}({\mathbf
y}_{2})\underset{b}{\hat{\chi}_{j'}^{+}}({\mathbf
y}_{3})\underset{a}{\hat{\chi}_{i'}^{+}}({\mathbf y}_{4}):.
\label{48}
\end{gather}
The operator $\hat{\varphi}({\mathbf x},{\mathbf y})\equiv
\varphi_{\alpha}({\mathbf x}-{\mathbf
y})\hat{\eta}_{\alpha}({m_{1}{\mathbf x}+m_{2}{\mathbf y}\over
m_{1}+m_{2}})$ differs from zero only for ${\mathbf x}\approx
{\mathbf y}$, ($|{\mathbf x}-{\mathbf y}|<a$). Thus, only those of
${\varphi}_{i}$, for which $|{\mathbf y}_{1}-{\mathbf y}_{4}|<a$
contribute to the integral over ${\mathbf y}$ in (\ref{47}). This
means that  the first term in (\ref{48}) in virtue of (\ref{19})
does not contribute to the matrix element of $\tilde{\hat V}_{0}$
taken between the states belonging to $\tilde{H}_{a}$ because of
$\hat{\varphi}_{i}({\mathbf x}_{1},{\mathbf
y}_{4})\hat{\chi}_{i}({\mathbf y}_{1})\Phi=0$. Similarly, one can
prove that the terms, which contain the single contractions in
(\ref{48}) do not give a contribution to the matrix element of
$\tilde{\hat V}_{0}$ taken between the states in $\tilde{H}_{a}$.
The penultimate term in (\ref{48}) containing the double
contractions does not also contribute to the above mentioned matrix
element. Indeed, the penultimate term in (\ref{48}) equals to
$\delta({\mathbf y}_{1}-{\mathbf y}_{3})\delta({\mathbf
y}_{2}-{\mathbf y}_{4})$. In this case the nonzero matrix element
exists for ${\mathbf x}_{1}\approx{\mathbf x}_{2}$ and
$\hat{\varphi}_{j}({\mathbf x}_{2},{\mathbf
y}_{3})\hat{\varphi}_{i}({\mathbf x}_{1},{\mathbf y}_{4})\Phi=0$ in
virtue of (\ref{19}). Thus, only the latter term in (\ref{48}) equal
$\delta({\mathbf y}_{2}-{\mathbf y}_{3})\delta({\mathbf
y}_{1}-{\mathbf y}_{4})$ can give the contribution to the matrix
element of $\tilde{\hat V}_{0}$. Therefore, not changing the matrix
elements in $\tilde{H}_{a}$, the operator $\tilde{\hat V}_{0}$ can
be represented in the form
\begin{equation*}
\begin{split}
\tilde{\hat V}_{0}={1\over 2}\sum_{i,j=1}^{2}\int d{\mathbf x}_{1}
d{\mathbf x}_{2} d{\mathbf y}_{1} d{\mathbf
y}_{2}\hat{\varphi}^{+}_{i}({\mathbf x}_{1},{\mathbf
y}_{1})\hat{\varphi}^{+}_{j}({\mathbf x}_{2},{\mathbf
y}_{2})\nu_{ij}({\mathbf x}_{1}-{\mathbf
x}_{2})\hat{\varphi}_{j}({\mathbf x}_{2},{\mathbf
y}_{2})\hat{\varphi}_{i}({\mathbf x}_{1},{\mathbf y}_{1}),
\end{split}
\end{equation*}
or, according to (\ref{46}),
\begin{multline*}
\tilde{\hat V}_{0}={1 \over 2}\int d{\mathbf x}_{1} d{\mathbf x}_{2}
d{\mathbf y}_{1} d{\mathbf y}_{2}\hat{\varphi}^{+}({\mathbf
x}_{1},{\mathbf
y}_{1})\hat{\varphi}^{+}({\mathbf x}_{2},{\mathbf y}_{2})\\
\times\biggl\{\nu_{11}({\mathbf x}_{1}-{\mathbf
x}_{2})+\nu_{22}({\mathbf y}_{1}-{\mathbf y}_{2})+\nu_{12}({\mathbf
x}_{1}-{\mathbf y}_{2})+\nu_{21}({\mathbf y}_{1}-{\mathbf
x}_{2})\biggr\} \hat{\varphi}({\mathbf x}_{2},{\mathbf
y}_{2})\hat{\varphi}({\mathbf x}_{1},{\mathbf y}_{1}). \label{49}
\end{multline*}
Similarly, noting that
\begin{multline*}
\tilde{\hat V}_{1}={1 \over 2}\sum_{i,j=1}^{2}\int d{\mathbf
x}_{1}d{\mathbf x}_{2}d{\mathbf y}\nu_{ij}({\mathbf x}_{1}-{\mathbf x}_{2})\\
\times \biggl\{\hat{\varphi}_{i}^{+}({\mathbf x}_{1},{\mathbf
y}){\hat \chi}_{i'}({\mathbf y}){\hat \chi}_{j}^{+}({\mathbf
x}_{2}){\hat \chi}_{j}({\mathbf x}_{2}){\hat \chi}_{i}({\mathbf
x}_{1})+ \hat{\varphi}_{j}^{+}({\mathbf x}_{2},{\mathbf y}){\hat
\chi}_{i}({\mathbf x}_{1}){\hat \chi}_{j'}^{+}({\mathbf y}){\hat
\chi}_{j}({\mathbf x}_{2}){\hat \chi}_{i}({\mathbf
x}_{1})+h.c.\biggr\},
\end{multline*}
\begin{multline*}
\tilde{\hat V}_{3}={1\over 2}\sum_{i,j=1}^{2}\int d{\mathbf
x}_{1}d{\mathbf x}_{2}d{\mathbf y}_{1} d{\mathbf y}_{2} d{\mathbf
y}_{3}\nu_{ij}({\mathbf x}_{1}-{\mathbf
x}_{2})\biggl\{\hat{\varphi}_{i}^{+}({\mathbf x}_{1},{\mathbf
y}_{1})\hat{\varphi}_{j}^{+}({\mathbf x}_{2},{\mathbf
y}_{2})\hat{\varphi}_{j}({\mathbf x}_{2},{\mathbf y}_{3}){\hat
\chi}_{i'}({\mathbf y}_{1}){\hat \chi}_{j'}({\mathbf y}_{2}){\hat
\chi}_{j'}^{+}({\mathbf y}_{3}){\hat \chi}_{i}({\mathbf x}_{1})\\
+\hat{\varphi}_{i}^{+}({\mathbf x}_{1},{\mathbf
y}_{1})\hat{\varphi}_{j}^{+}({\mathbf x}_{2},{\mathbf
y}_{2})\hat{\varphi}_{i}({\mathbf x}_{1},{\mathbf y}_{3}){\hat
\chi}_{i'}({\mathbf y}_{1}){\hat \chi}_{j'}({\mathbf y}_{2}){\hat
\chi}_{j}({\mathbf x}_{2}){\hat \chi}_{i'}^{+}({\mathbf
y}_{3})+h.c.\biggr\}
\end{multline*}
and performing the same derivation as for $\tilde{\hat V}_{0}$, it
is easy to verify (using the anticommutative relations for
${\hat\chi}$, ${\hat \chi}^{+}$ and (\ref{48})) that we can
consider $\tilde{\hat V}_{1}=\tilde{\hat V}_{3}=0$ not changing
the matrix elements of $\tilde{\hat V}_{1}$ and $\tilde{\hat
V}_{3}$ in the subspace $\tilde{H}_{a}$ . Next it is evident that
the following formula is valid:
\begin{equation}
\tilde{\hat V}_{4}={1\over 2}\sum_{i,j=1}^{2}\int d{\mathbf
x}_{1}d{\mathbf x}_{2}\hat{\chi}_{i}^{+}({\mathbf x}_{1})
\hat{\chi}_{j}^{+}({\mathbf x}_{2})\nu_{ij}({\mathbf x}_{1}-{\mathbf
x}_{2})\hat{\chi}_{j}({\mathbf x}_{2})\hat{\chi}_{i}({\mathbf
x}_{1}).
\end{equation}
Finally let us find $\tilde{\hat V}_{2}$,
\begin{gather*}
\tilde{\hat V}_{2}={1\over 2}\sum_{i,j=1}^{2}\int d{\mathbf
x}_{1}d{\mathbf x}_{2}\nu_{ij}({\mathbf x}_{1}-{\mathbf
x}_{2})\biggl\{{\hat O}_{i}^{+}({\mathbf x}_{1}){\hat
O}_{j}^{+}({\mathbf x}_{2}){\hat \chi}_{j}({\mathbf x}_{2}){\hat
\chi}_{i}({\mathbf x}_{1})+{\hat O}_{i}^{+}({\mathbf x}_{1}){\hat
\chi}_{j}^{+}({\mathbf x}_{2}){\hat O}_{j}({\mathbf x}_{2}){\hat
\chi}_{i}({\mathbf
x}_{1})+h.c.\\
+{\hat O}_{i}^{+}({\mathbf x}_{1}){\hat \chi}_{j}^{+}({\mathbf
x}_{2}){\hat \chi}_{j}({\mathbf x}_{2}){\hat O}_{i}({\mathbf
x}_{1})+{\hat \chi}_{i}^{+}({\mathbf x}_{1}){\hat
O}_{j}^{+}({\mathbf x}_{2}){\hat O}_{j}({\mathbf x}_{2}){\hat
\chi}_{i}({\mathbf x}_{1})\biggr\}.
\end{gather*}
It can be easily seen that the first two terms and the
corresponding Hermitian conjugate terms do not give a contribution
to the matrix element of $\tilde{\hat V}_{2}$ in the subspace
$\tilde{H}_{a}$. Therefore, we can consider
\begin{multline*}
\tilde{\hat V}_{2}={1\over 2}\sum_{i,j=1}^{2}\int d{\mathbf x}_{1}
d{\mathbf x}_{2} d{\mathbf y}_{1} d{\mathbf y}_{2}\nu_{ij}({\mathbf
x}_{1}-{\mathbf x}_{2})\biggl\{\hat{\varphi}_{i}^{+}({\mathbf
x}_{1},{\mathbf y}_{1})\hat{\varphi}_{i}({\mathbf x}_{1},{\mathbf
y}_{2}){\hat \chi}_{i'}({\mathbf y}_{1}){\hat \chi}_{j}^{+}({\mathbf
x}_{2}){\hat
\chi}_{j}({\mathbf x}_{2}){\hat \chi}_{i'}^{+}({\mathbf y}_{2})\\
+\hat{\varphi}_{j}^{+}({\mathbf x}_{2},{\mathbf
y}_{1})\hat{\varphi}_{j}({\mathbf x}_{2},{\mathbf y}_{2}){\hat
\chi}_{i}^{+}({\mathbf x}_{1}){\hat \chi}_{j'}({\mathbf y}_{1}){\hat
\chi}_{j'}^{+}({\mathbf y}_{2}){\hat \chi}_{i}({\mathbf
x}_{1})\biggr\}.
\end{multline*}
The first and second terms in this expression contribute to the
matrix element of $\tilde{\hat V}_{2}$ in $\tilde{H}_{a}$ under
the following arrangements of contractions:
\[
:\underset{a}{{\hat \chi}_{i'}}({\mathbf y}_{1})\underset{a}{{\hat
\chi}_{j}^{+}}({\mathbf x}_{2})\underset{b}{{\hat
\chi}_{j}}({\mathbf x}_{2})\underset{b}{{\hat
\chi}_{i'}^{+}}({\mathbf y}_{2}):+: \underset{a}{{\hat
\chi}_{i'}}({\mathbf y}_{1}){\hat \chi}_{j}^{+}({\mathbf
x}_{2}){\hat \chi}_{j}({\mathbf x}_{2})\underset{a}{{\hat
\chi}_{i'}^{+}}({\mathbf y}_{2}) := \delta_{ji'}\delta({\mathbf
y}_{1}-{\mathbf x}_{2})\delta({\mathbf y}_{2}-{\mathbf
x}_{2})+\delta({\mathbf y}_{1}-{\mathbf
y}_{2})\hat{\chi}^{+}_{j}({\mathbf x}_{2})\hat{\chi}_{j}({\mathbf
x}_{2}),
\]
\[
: {\hat \chi}_{i}^{+}({\mathbf x}_{1})\underset{a}{{\hat
\chi}_{j'}}({\mathbf y}_{1})\underset{a}{{\hat
\chi}_{j'}^{+}}({\mathbf y}_{2}){\hat \chi}_{i}({\mathbf x}_{1})
:=\delta({\mathbf y}_{1}-{\mathbf y}_{2})\hat{\chi}^{+}_{i}({\mathbf
x}_{1})\hat{\chi}_{i}({\mathbf x}_{1}).
\]
Thus, we have
\begin{multline}
\tilde{\hat V}_{2}=\int d{\mathbf x}_{1} d{\mathbf
x}_{2}\nu_{12}({\mathbf x}_{1}-{\mathbf
x}_{2})\hat{\varphi}^{+}({\mathbf x}_{1},{\mathbf
x}_{2})\hat{\varphi}({\mathbf x}_{1},{\mathbf x}_{2})+\int d{\mathbf
x}_{1} d{\mathbf x}_{2}\hat{\varphi}^{+}({\mathbf x}_{2},{\mathbf
y}_{2})\hat{\varphi}({\mathbf x}_{2},{\mathbf y}_{2})
\\\times \biggl\{\nu_{11}({\mathbf x}_{1}-{\mathbf x}_{2})\hat{\chi}_{1}^{+}({\mathbf
x}_{1})\hat{\chi}_{1}({\mathbf x}_{1})+\nu_{21}({\mathbf
x}_{1}-{\mathbf y}_{2})\hat{\chi}_{1}^{+}({\mathbf
x}_{1})\hat{\chi}_{1}({\mathbf x}_{1}) +\nu_{22}({\mathbf
x}_{1}-{\mathbf y}_{2})\hat{\chi}_{2}^{+}({\mathbf
x}_{1})\hat{\chi}_{2}({\mathbf x}_{1})+\nu_{12}({\mathbf
x}_{1}-{\mathbf x}_{2})\hat{\chi}_{2}^{+}({\mathbf
x}_{1})\hat{\chi}_{2}({\mathbf x}_{1})\biggr\}.\label{52}
\end{multline}
The first term in this formula quadratic in field operators can be
combined with the latter term in (\ref{45}). As a result we obtain
\begin{equation*}
\begin{split}
\int d{\mathbf x}_{1}d{\mathbf x}_{2}\nu_{12}({\mathbf
x}_{1}-{\mathbf x}_{2})\hat{\varphi}^{+}({\mathbf x}_{1},{\mathbf
x}_{2})\hat{\varphi}({\mathbf x}_{1},{\mathbf x}_{2})&-{1\over
2\mu}\int d{\mathbf x}d{\mathbf X}\hat{\eta}_{\alpha}^{+}({\mathbf
X})\hat{\eta}_{\alpha}({\mathbf
X}){\partial\varphi_{\alpha}^{*}({\mathbf x})\over\partial {\mathbf
x}}{\partial\varphi_{\alpha}({\mathbf x})\over\partial {\mathbf x}}= \\
&\int d{\mathbf x}d{\mathbf X}\hat{\eta}_{\alpha}^{+}({\mathbf
X})\hat{\eta}_{\beta}({\mathbf X})\varphi_{\alpha}^{*}({\mathbf
x})\biggl\{ -{1\over 2\mu}\Delta_{\mathbf x}+\nu_{12}({\mathbf
x})\biggl\}\varphi_{\beta}({\mathbf x}).
\end{split}
\end{equation*}
Since $\varphi_{\beta}({\mathbf x})$ satisfies the Schroedinger
equation
\[
\biggl\{-{1\over 2\mu}\Delta_{\mathbf x}+\nu_{12}({\mathbf
x})\biggl\}\varphi_{\beta}({\mathbf
x})={\varepsilon}_{\beta}\varphi_{\beta}({\mathbf x}),
\]
where ${\varepsilon}_{\beta}$ are the atomic energy levels, the
latter formula takes the form
\[
\int d{\mathbf
X}\sum_{\alpha}{\varepsilon}_{\alpha}\hat{\eta}_{\alpha}^{+}({\mathbf
X})\hat{\eta}_{\alpha}({\mathbf X}).
\]
Hence, taking into account (\ref{44}), (\ref{45}), (\ref{52}), the
Hamiltonian of the system $\tilde{\hat{\mathcal{H}}}$ is of the form
\[
\tilde{\hat{\mathcal{H}}}=\tilde{\hat{\mathcal{H}}}_{0}+\tilde{\hat{\mathcal
H}}^{1}_{int}+\tilde{\hat{\mathcal H}}^{2}_{int}+\tilde{\hat{\mathcal
H}}^{3}_{int},
\]
where
\begin{equation}
\tilde{\hat{\mathcal{H}}}_{0}=\sum_{j=1}^{2}{1\over 2m_{j}} \int
d{\mathbf x}{\partial\hat{\chi}_{j}^{+}({\mathbf x})
\over\partial{\mathbf x}} {\partial\hat{\chi}_{j}({\mathbf x})
\over\partial{\mathbf x}}+\sum_{\alpha}\int d{\mathbf
X}\biggl\{{1\over 2M}{\partial\hat{\eta}_{\alpha}^{+}({\mathbf X})
\over\partial{\mathbf X}}{\partial\hat{\eta}_{\alpha}({\mathbf X})
\over\partial{\mathbf
X}}+\varepsilon_{\alpha}\hat{\eta}_{\alpha}^{+}({\mathbf
X})\hat{\eta}_{\alpha}({\mathbf X})\biggr\} \label{53}
\end{equation}
is the Hamiltonian for free particles and bound states, and
\begin{widetext}
\begin{multline}
\tilde{\hat{\mathcal H}}_{int}^{1}=\int d{\mathbf x}_{1} d{\mathbf
x}_{2} d{\mathbf y}_{2}\hat{\varphi}^{+}({\mathbf x}_{2},{\mathbf
y}_{2})\hat{\varphi}({\mathbf x}_{2}, {\mathbf
y}_{2})\biggl\{\biggl(\nu_{11}({\mathbf x}_{1}-{\mathbf x}_{2}) +
\nu_{21}({\mathbf x}_{1}-{\mathbf
y}_{2})\biggr)\hat{\chi}_{1}^{+}({\mathbf
x}_{1}) \hat{\chi}_{1}({\mathbf x}_{1}) \\
+\biggl(\nu_{22}({\mathbf x}_{1}-{\mathbf y}_{2})+\nu_{12}({\mathbf
x}_{1}-{\mathbf x}_{2})\biggr) \hat{\chi}_{2}^{+}({\mathbf x}_{1})
\hat{\chi}_{2}({\mathbf x}_{1})\biggr\}, \label{54}
\end{multline}
\begin{multline}
\tilde{\hat{\mathcal H}}_{int}^{2}={1\over 2}\int d{\mathbf
x}_{1}d{\mathbf x}_{2} d{\mathbf y}_{1}d{\mathbf y}_{2}
\hat{\varphi}^{+}({\mathbf x}_{1},{\mathbf
y}_{1})\hat{\varphi}^{+}({\mathbf x}_{2},{\mathbf y}_{2})
\hat{\varphi}({\mathbf x}_{2},{\mathbf y}_{2})\hat{\varphi}({\mathbf
x}_{1},{\mathbf y}_{1})\\ \times \biggl\{\nu_{11}({\mathbf
x}_{1}-{\mathbf x}_{2})+\nu_{22}({\mathbf y}_{1}-{\mathbf
y}_{2})+\nu_{12}({\mathbf x}_{1}-{\mathbf y}_{2})+ \nu_{21}({\mathbf
y}_{1}-{\mathbf x}_{2}) \biggr\}, \label{55}
\end{multline}
\begin{multline}
\tilde{\hat{\mathcal H}}_{int}^{3}={1\over 2}\int d{\mathbf
x}_{1}d{\mathbf x}_{2}\biggl\{\nu_{11}({\mathbf x}_{1}-{\mathbf
x}_{2})\hat{\chi}_{1}^{+}({\mathbf x}_{1})
\hat{\chi}_{1}^{+}({\mathbf
x}_{2})\hat{\chi}_{1}({\mathbf x}_{2})\hat{\chi}_{1}({\mathbf x}_{1})\\
+\nu_{22}({\mathbf x}_{1}-{\mathbf
x}_{2})\hat{\chi}_{2}^{+}({\mathbf x}_{1})
\hat{\chi}_{2}^{+}({\mathbf x}_{2})\hat{\chi}_{2}({\mathbf
x}_{2})\hat{\chi}_{2}({\mathbf x}_{1})+ 2\nu_{12}({\mathbf
x}_{1}-{\mathbf x}_{2})\hat{\chi}_{1}^{+}({\mathbf x}_{1})
\hat{\chi}_{1}({\mathbf x}_{1})\hat{\chi}_{2}^{+}({\mathbf x}_{2})
\hat{\chi}_{2}({\mathbf x}_{2})\biggr\} \label{56}
\end{multline}
\end{widetext}
are the Hamiltonians of interaction. The Hamiltonian
$\tilde{\hat{\mathcal H}}^{1}_{int}$ corresponds to scattering of
particles of the first and second kinds by bound states; the
Hamiltonian $\tilde{\hat{\mathcal H}}^{2}_{int}$ corresponds to
scattering of bound states by bound states; finally, the Hamiltonian
$\tilde{\hat{\mathcal H}}^{3}_{int}$ corresponds to scattering of
particles of the first and second kinds by particles of the same
kinds. The Hamiltonians of interaction (\ref{54}), (\ref{55}) may be
written through the creation $\hat{\eta}_{\alpha}^{+}({\mathbf x})$
and annihilation $\hat{\eta}_{\alpha}({\mathbf x})$ operators of
atoms by using of (\ref{26'}). We want to emphasize that the
obtained Hamiltonians of interaction do not lead to decay processes
and formation of compound particles as it should be in the
low-energy approximation. This fact reflects that atoms are
absolutely stable in the main approximation.

In conclusion of this section we address to the Galilean invariance
of the developed theory. We suppose that the initial theory based
only on the field operators ${\hat \psi}_{i}({\mathbf x}), {\hat
\psi}_{i}^{+}({\mathbf x}), \ (i=1,2)$ is Galilean invariant. This
means that under the following transformations:
\[
{\hat \psi}_{i}({\mathbf x})\rightarrow {\hat
\psi}_{i}^{\prime}({\mathbf x}) = e^{im_{i}{\mathbf v}{\mathbf
x}}{\hat \psi}_{i}({\mathbf x}) \label{57}
\]
the operators for particles number ${\hat N}_{i}$, momentum ${\hat
{\mathcal P}}_{k}$ and energy ${\hat{\mathcal H}}$ transform as follows:
\begin{gather*}
{\hat N}_{i}\rightarrow {\hat N}_{i}^{\prime}={\hat N}_{i},\quad
i=1,2,\\ {\hat {\mathcal P}}_{k}\rightarrow {\hat {\mathcal
P}}_{k}^{\prime}={\hat {\mathcal P}}_{k}+v_{k}({\hat N}_{1}+{\hat
N}_{2}), \qquad {\hat {\mathcal H}}\rightarrow {\hat {\mathcal
H}}^{\prime}={\hat {\mathcal H}}+v_{k}{\hat {\mathcal P}}_{k}+{1\over
2}m_{1}v^{2}{\hat N}_{1} + {1\over 2}m_{2}{\hat N}_{2}v^{2}.
\end{gather*}
Since ${\hat \psi}_{i}({\mathbf x})$ and ${\hat
\psi}_{i}^{\prime}({\mathbf x})$ meet the same commutation
relations, they are related to each other by the unitary
transformation
\[
{\hat \psi}_{i}^{\prime}({\mathbf x})=U_{v}{\hat \psi}_{i}({\mathbf
x})U_{v}^{+},
\]
where the unitary operator $U_{v}$, is defined by${}^{8}$:
\begin{equation}
U_{v} = \exp\biggl( -i{\mathbf v}m_{1}\int d{\mathbf x} {\mathbf
x}{\hat \rho}_{1}({\mathbf x}) - i{\mathbf v}m_{2}\int d{\mathbf x}
{\mathbf x}{\hat \rho}_{2}({\mathbf x})\biggr). \label{58}
\end{equation}
Let us show that
\[
{\tilde U}_{v} = \exp\biggl( -i{\mathbf v}m_{1}\int d{\mathbf x}
{\mathbf x} {\tilde{\hat \rho}}_{1}({\mathbf x}) - i{\mathbf
v}m_{2}\int d{\mathbf x} {\mathbf x}{\tilde {\hat
\rho}}_{2}({\mathbf x})\biggr)
\]
defines the Galilei transformations in the presence of bound
states of particles (compare to (\ref{58})). To this end we note
that this operator can be written, according to (\ref{40}), in the
form
\[
{\tilde U}_{v} = \exp\biggl( -i{\mathbf v}\sum_{i=1}^{2}m_{i}\int
d{\mathbf x}{\hat \chi}_{i}^{+}({\mathbf x}){\hat \chi}_{i}({\mathbf
x}) - i{\mathbf v} {\hat{\mathbf s}}\biggr),
\]
where
\begin{equation*}
\begin{split}
{\hat{\mathbf s}}=m_{1}\int d{\mathbf x}_{1}d{\mathbf x}_{2}
{\mathbf x}_{1} {\hat{\varphi}}^{+}({\mathbf x}_{1},{\mathbf x}_{2})
{\hat{\varphi}}({\mathbf x}_{1},{\mathbf x}_{2})& + m_{2}\int
d{\mathbf x}_{1}d{\mathbf x}_{2} {\mathbf x}_{1}
{\hat{\varphi}}^{+}({\mathbf
x}_{2},{\mathbf x}_{1}) {\hat{\varphi}}({\mathbf x}_{2},{\mathbf x}_{1})\\
&=\int d{\mathbf x}_{1} d{\mathbf x}_{2}(m_{1}{\mathbf
x}_{1}+m_{2}{\mathbf x}_{2}){\hat{\varphi}}^{+}({\mathbf
x}_{1},{\mathbf x}_{2}) {\hat{\varphi}}({\mathbf x}_{1},{\mathbf
x}_{2}).
\end{split}
\end{equation*}
Since
\[
{\hat{\varphi}}({\mathbf x}_{1},{\mathbf
x}_{2})={\varphi}_{\alpha}({\mathbf x}){\hat \eta}_{\alpha}({\mathbf
X}), \qquad \int\,d{\mathbf x}{\varphi}_{\alpha}^{*}({\mathbf
x}){\varphi}_{\beta}({\mathbf x})=\delta_{\alpha \beta},
\]
changing to the center of mass variables one finds,
\[
{\hat{\mathbf s}}=(m_{1}+m_{2})\int\,d{\mathbf X}{\mathbf
X}\sum_{\alpha}{\hat \eta}_{\alpha}^{+}({\mathbf X}){\hat
\eta}_{\alpha}({\mathbf X}).
\]
Therefore, just as we expected
\[
{\tilde U}_{v}=\exp\biggl( -i{\mathbf v}\sum_{i=1}^{2}m_{i}\int
d{\mathbf x}{\hat \chi}_{i}^{+}({\mathbf x}){\hat \chi}_{i}({\mathbf
x})-i{\mathbf v}(m_{1}+m_{2})\int\,d{\mathbf X}{\mathbf
X}\sum_{\alpha}{\hat \eta}_{\alpha}^{+}({\mathbf X}){\hat
\eta}_{\alpha}({\mathbf X})\biggr).
\]
The employment of this formula and the canonic permutation
relations result in
\begin{equation}
{\tilde U}_{v}{\hat \chi}_{i}({\mathbf x}){\tilde
U}_{v}^{+}=e^{im_{i}{\mathbf v}{\mathbf x}}{\hat \chi}_{i}({\mathbf
x})\equiv {\hat \chi}_{i}^{\prime}({\mathbf x}), \qquad {\tilde
U}_{v}{\hat \eta}_{\alpha}({\mathbf X}){\tilde
U}_{v}^{+}=e^{i(m_{1}+m_{2}){\mathbf v}{\mathbf X}}{\hat
\eta}_{\alpha}({\mathbf X})\equiv {\hat
\eta}_{\alpha}^{\prime}({\mathbf X}). \label{60}
\end{equation}
Since the quantity ${\hat \varphi}^{+}({\mathbf x},{\mathbf
X}){\hat\varphi}({\mathbf x},{\mathbf X})$ is not changed under such
transformation, the Hamiltonians of interaction
$\tilde{\hat{\mathcal H}}^{1}_{int}$, $\tilde{\hat{\mathcal
H}}^{2}_{int}$, $\tilde{\hat{\mathcal H}}^{3}_{int}$ (see (\ref{54})
- (\ref{56})) are invariant with respect to (\ref{60}). It is easy
to verify that the Hamiltonian $\tilde{\hat {\mathcal{H}}}_{0}$ for
free particles and bound states transforms according to the law
\[
{\tilde U}_{v}\tilde{\hat {\mathcal{H}}}_{0}{\tilde U}_{v}^{+} =
\tilde{\hat {\mathcal{H}}}_{0} + v_{k}{\tilde{\hat {\mathcal
P}}}_{k} + {m_{1}v^{2}\over 2}\tilde{\hat N}_{1} + {m_{2}v^{2}\over
2}\tilde{\hat N}_{2} + {(m_{1}+m_{2})v^{2}\over 2}\tilde{\hat
N}_{b}.
\]
Hence,
\[
{\tilde U}_{v}\tilde{\hat {\mathcal{H}}}{\tilde U}_{v}^{+} =
\tilde{\hat {\mathcal{H}}} + v_{k}{\tilde {\hat {\mathcal P}}}_{k} +
{m_{1}v^{2}\over 2}\tilde{\hat N}_{1} + {m_{2}v^{2}\over
2}\tilde{\hat N}_{2} + {(m_{1}+m_{2})v^{2}\over 2}\tilde{\hat
N}_{b},
\]
\[
{\tilde U}_{v}\tilde{\hat {{\mathcal{P}}}}_{k}{\tilde U}_{v}^{+} =
\tilde{\hat {{\mathcal{P}}}}_{k} + m_{1}v_{k}\tilde{\hat N}_{1} +
m_{2}v_{k}\tilde{\hat N}_{2} + (m_{1}+m_{2})v_{k}\tilde{\hat N}_{b},
\]
where $\tilde{\hat {{\mathcal{P}}}}_{k}$, $\tilde{\hat N}_{1}$,
$\tilde{\hat N}_{2}$, $\tilde{\hat N}_{b}$ are the momentum operator
and the particle number operators (for free particles of the first
and second kinds and their bound states). These formulas prove the
Galilean invariance of the developed theory. Finally we note the
validity of the following formula:
\[
{\tilde U}_{v}\tilde{\hat{\mbox{\boldmath$\pi$}}}_{i}({\mathbf x})
{\tilde U}_{v}^{+} =
\tilde{\hat{\mbox{\boldmath$\pi$}}}_{i}({\mathbf x})
 + m_{i}{\mathbf v}{\tilde {\hat \rho}}_{i}({\mathbf x}),
\]
which follows from (\ref{39}), (\ref{40}) by using (\ref{60}).

\section{Electromagnetic Interaction}

Here we consider the electromagnetic interaction assuming that the
formation of bound states of particles is caused by Coulomb's
(electromagnetic) forces. Therefore, the potential energy
$\nu_{ij}({\mathbf x}-{\mathbf x}')$ entering (\ref{54})--(\ref{56})
should be written as
\begin{equation}
\nu_{ij}({\mathbf x}-{\mathbf x}')={e_{i}e_{j}\over |{\mathbf
x}-{\mathbf x}'|},\quad i,j =1,2. \label{61}
\end{equation}
We also introduce the additional interactions of particles (see
below) with an external electromagnetic field ${\mathbf
A}^{(e)}({\mathbf x},t)$, $\varphi^{(e)}({\mathbf x},t)$ and
quantized electromagnetic field specified by a potential
$\hat{\mathbf a}({\mathbf x})$ (Coulomb's gauge)
\[
 \hat{\mathbf A}({\mathbf x},t)=\hat{\mathbf
a}({\mathbf x})+{\mathbf A}^{(e)}({\mathbf x},t),
\]
where${}^{8}$
\begin{equation*}
\hat{\mathbf a}({\mathbf x})=\sum_{\mathbf
k}\sum_{\lambda=1}^{2}\biggl({2\pi\over {\mathcal
V}\omega_{k}}\biggr)^{1/2}\biggl({\mathbf e}_{{\mathbf
k}\lambda}\hat{C}_{{\mathbf k}\lambda}e^{i{\mathbf k}{\mathbf
x}}+h.c. \biggr)
\end{equation*}
($\hat{C}_{{\mathbf k}\lambda}$ is the annihilation operator of a
photon with momentum $k$ and polarization ${\mathbf e}_{{\mathbf
k}\lambda}$). As in section II, we identify the subspace
$\tilde{H}_{a}$ to $\tilde H$ assuming that the matrix elements of
the operators in coordinate representation give the main
contribution to the quantum electrodynamics processes and correspond
to the space scale $\Delta x\gtrsim a$ ($a\rightarrow 0$,
$a>>r_{0}$).

The density $\varepsilon({\mathbf x})$ of a Hamiltonian in the
second quantization representation has the well-known form
\begin{multline*}
\hat{\varepsilon}({\mathbf x})=\hat{\varepsilon}_{f}({\mathbf x})+
\hat{\varepsilon}_{C}({\mathbf x})+
\sum_{i=1}^{2}e_{i}\hat{\psi}^{+}_{i}({\mathbf
x})\hat{\psi}_{i}({\mathbf x})\varphi^{(e)}({\mathbf
x},t)+\sum_{i=1}^{2}{1\over
2m_{i}}\biggl({\partial\over\partial{\mathbf x}} -
ie_{i}\hat{\mathbf A}({\mathbf
x},t)\biggr)\hat{\psi}^{+}_{i}({\mathbf
x})\biggl({\partial\over\partial{\mathbf x}} + ie_{i}\hat{\mathbf
A}({\mathbf x},t)\biggr)\hat{\psi}_{i}({\mathbf x}),
\end{multline*}
where $\hat{\varepsilon}_{f}({\mathbf x})$ is the energy density of
a free electromagnetic field, $\hat{\varepsilon}_{C}({\mathbf x})$
is the energy density of Coulomb's interaction, the third term
describes the interaction between the particles and external scalar
potential. This formula leads to the following Hamiltonian of the
system:
\begin{equation}
\hat{\mathcal H}(t)=\hat{\mathcal H}_{0}+\hat{\mathcal H}_{int}+\hat{V}(t),
\quad \hat{\mathcal H}_{0}=\hat{\mathcal H}_{f}+\hat{\mathcal H}_{p},
\label{62}
\end{equation}
where
\begin{equation}
\hat{\mathcal H}_{f}\equiv\int d{\mathbf
x}\hat{\varepsilon}_{f}({\mathbf x})=\sum_{{\mathbf
k},\lambda}\omega_{k}\hat{C}_{{\mathbf
k}\lambda}^{+}\hat{C}_{{\mathbf k}\lambda}, \label{63}
\end{equation}
\begin{equation}
\hat{\mathcal H}_{p}=\sum_{i=1}^{2}{1\over 2m_{i}}\int d{\mathbf
x}{\partial\hat{\psi}^{+}_{i}({\mathbf x}) \over\partial{\mathbf x}}
{\partial\hat{\psi}_{i}({\mathbf x}) \over\partial{\mathbf x}},
\label{64}
\end{equation}
\begin{equation}
\hat{\mathcal H}_{int}=\int d{\mathbf
x}\hat{\varepsilon}_{C}({\mathbf x}) \label{64'}
\end{equation}
are the Hamiltonians for free photons, free particles, and Coulomb's
interaction respectively. The operator $\hat{V}(t)$ represents a
Hamiltonian that describes the interaction of particles with the
electromagnetic fields $\hat{\mathbf A}({\mathbf x},t)$,
$\varphi^{(e)}({\mathbf x},t)$
\begin{equation}
\hat{V}(t)=-\int d{\mathbf x}\hat{\mathbf A}({\mathbf
x},t)\hat{\mathbf J}({\mathbf x},t)-{1\over 2}\int d{\mathbf
x}\hat{\mathbf A}^{2}({\mathbf x},t)\sum_{i=1}^{2}{e_{i}\over
m_{i}}\hat{\sigma}_{i}({\mathbf x})+\int d{\mathbf
x}\varphi^{(e)}({\mathbf x},t)\hat{\sigma}, \label{65}
\end{equation}
\begin{equation}
\hat{\mathbf J}({\mathbf x},t)=-\hat{\mathbf A}({\mathbf
x},t)\sum_{i=1}^{2}{e_{i}\over m_{i}}\hat{\sigma}_{i}({\mathbf
x})+\hat{\mathbf j}_{0}({\mathbf x}), \quad \hat{\mathbf
j}_{0}({\mathbf x})=\sum_{i=1}^{2}{e_{i}\over
m_{i}}\hat{\mbox{\boldmath$\pi$}_{i}}({\mathbf x}), \label{65'}
\end{equation}
where $\hat{\sigma}_{i}(x)=e_{i}\hat{\rho}_{i}(x)$ is the charge
density for particles of the kind $i$,
$\hat{\sigma}=\hat{\sigma}_{1}+\hat{\sigma}_{2}$. This Hamiltonian
is exact and describes (in this sense) the processes in which the
bound states of particles are involved. Since $\hat{\mathbf
a}({\mathbf x})$ commute among themselves and with
$\hat{\psi}_{i}({\mathbf x})$, $\hat{\psi}^{+}_{i}({\mathbf x})$, in
order to find the effective Hamiltonian $\tilde{\hat{\mathcal
H}}(t)$ that describes electrodynamic processes at low-energies in
the presence of bound states of particles we need, in accordance
with section IV, to replace the operators $\hat{\psi}_{i}$,
$\hat{\psi}^{+}_{i}$, $\hat{\mathbf a}$ by $\tilde{\hat{\psi}}_{i}$,
$\tilde{\hat{\psi}}^{+}_{i}$, $\tilde{\hat{\mathbf a}}$:
\[
\tilde{\hat{\mathcal H}}(t)=\hat{\mathcal
H}(t)\vert_{\hat{\psi}_{1}\rightarrow
\tilde{\hat{\psi}}_{1},\hat{\psi}_{2}\rightarrow
\tilde{\hat{\psi}}_{2}, \hat{\mathbf
a}\rightarrow\tilde{\hat{\mathbf a}}},
\]
where
\[
\tilde{\hat{\psi}}_{1}({\mathbf x})=\hat{\chi}_{1}({\mathbf x})+\int
d{\mathbf y}\hat{\varphi}({\mathbf x},{\mathbf
y})\hat{\chi}_{2}^{+}({\mathbf y}), \qquad
\tilde{\hat{\psi}}_{2}({\mathbf x})=\hat{\chi}_{2}({\mathbf x})+\int
d{\mathbf y}\hat{\varphi}({\mathbf y},{\mathbf
x})\hat{\chi}_{1}^{+}({\mathbf y}),\qquad \tilde{\hat{\mathbf
a}}=\hat{\mathbf a}
\]
($\tilde{\hat{\mathbf a}}({\mathbf x})$ coincides with $\hat{\mathbf
a}({\mathbf x})$ and acts in the space $\tilde{H}$). As a result the
effective Hamiltonian is defined by (\ref{62})-(\ref{65'}), in which
\begin{equation}
\hat{\sigma}_{i}({\mathbf
x})\rightarrow\tilde{\hat{\sigma}}_{i}({\mathbf
x})=e_{i}\tilde{\hat\rho}_{i}({\mathbf x}), \qquad
\hat{\mbox{\boldmath$\pi$}_{i}}({\mathbf x})\rightarrow
\tilde{\hat{\mbox{\boldmath$\pi$}}}_{i}({\mathbf x}), \qquad
\hat{\psi}_{i}({\mathbf x})\rightarrow \tilde{\hat\psi}_{i}({\mathbf
x}), \label{66}
\end{equation}
moreover, the operators $\tilde{\hat\rho}_{i}({\mathbf x})$,
$\tilde{\hat{\mbox{\boldmath$\pi$}}}_{i}({\mathbf x})$,
$\tilde{\hat\psi}_{i}({\mathbf x})$ containing the creation
$\hat{\eta}_{\alpha}^{+}$ and annihilation $\hat{\eta}_{\alpha}$
operators of bound states are determined by (\ref{39}), (\ref{40}),
(\ref{26})-(\ref{27}).

Let us define the operator of magnetic field
\begin{equation}
\hat{\mathbf H}({\mathbf x},t)=\hat{\mathbf h}({\mathbf
x})+\hat{\mathbf H}^{(e)}({\mathbf x},t), \label{67}
\end{equation}
where
\[
\hat{\mathbf h}({\mathbf x})={\rm rot}\,\hat{\mathbf a}({\mathbf
x}), \qquad \hat{\mathbf H}^{(e)}({\mathbf x},t)={\rm rot}\,{\mathbf
A}^{(e)}({\mathbf x},t),
\]
and the operator of electric field
\[
\hat{\mathbf E}({\mathbf x},t)=\hat{\mathbf e}({\mathbf
x})+\hat{\mathbf E}^{(e)}({\mathbf x},t), \label{68}
\]
where
\[
\hat{\mathbf e}({\mathbf x})=\hat{\mathbf e}^{t}({\mathbf
x})+\hat{\mathbf e}^{l}({\mathbf x}), \qquad \hat{\mathbf
E}^{(e)}({\mathbf x},t)=-{\partial\over\partial t}{\mathbf
A}^{(e)}({\mathbf x},t)-{\partial\over\partial {\mathbf
x}}\varphi^{(e)}({\mathbf x},t).
\]
The transverse $\hat{\mathbf e}^{t}({\mathbf x})$ and longitudinal
$\hat{\mathbf e}^{l}({\mathbf x})$ components of the electric field
created by particles are defined as follows:
\begin{equation}
\hat{\mathbf e}^{l}({\mathbf x})=-{\partial\over\partial {\mathbf
x}}\hat{\mathbf a}_{0}({\mathbf x}), \qquad \hat{\mathbf
e}^{t}({\mathbf x})=-\dot{\hat{\mathbf a}}({\mathbf x})\equiv
-i[\tilde{\hat{\mathcal H}}(t),\hat{\mathbf a}({\mathbf x})].
\label{69}
\end{equation}
Here $\hat{\mathbf a}_{0}({\mathbf x})$ represents the operator of
the scalar potential,
\begin{equation}
\hat{\mathbf a}_{0}({\mathbf x})=\int d{\mathbf x}'{\tilde{\hat
\sigma}({\mathbf x}')\over |{\mathbf x}-{\mathbf x}'|}. \label{70}
\end{equation}
Next it is easy to verify that
\[
[\hat{\eta}_{\alpha}({\mathbf X}),\hat{\eta}_{\beta}^{+}({\mathbf
X}')]=\delta_{\alpha\beta}\delta({\mathbf X}-{\mathbf X}'), \qquad
\sum_{\alpha}\varphi_{\alpha}^{*}({\mathbf
x})\varphi_{\alpha}({\mathbf x}')=\delta({\mathbf x}-{\mathbf x}')
\label{72}
\]
lead to the following commutation relation:
\begin{equation}
[\hat{\varphi}({\mathbf x},{\mathbf X}),\hat{\varphi}^{+}({\mathbf
x}',{\mathbf X}')]=\delta({\mathbf x}-{\mathbf x}')\delta({\mathbf
X}-{\mathbf X}'), \label{71}
\end{equation}
where $\hat{\varphi}({\mathbf x},{\mathbf
X})\equiv\hat{\varphi}({\mathbf x}_{1},{\mathbf x}_{2})$ is
determined by (\ref{26'}). Hereinafter we assume that a set of wave
functions $\varphi_{\alpha}({\mathbf x}')$ is complete. Therefore
this set accounts for the wave functions of continuous spectrum. In
addition, if we introduce the operator
\[
{\hat n}({\mathbf x},{\mathbf X})=\hat{\varphi}^{+}({\mathbf
x},{\mathbf X})\hat{\varphi}({\mathbf x},{\mathbf X}), \label{73}
\]
then
\[
[{\hat n}({\mathbf x},{\mathbf X}),{\hat n}({\mathbf x}',{\mathbf
X}')]=0. \label{74}
\]
Thus, it is clear that the Hamiltonian $\tilde{\hat {\mathcal
H}}_{int}$, which describes the Coulomb interaction between
particles and bound states meets the commutation relation
\[
[\tilde{\hat {\mathcal H}}_{int},\tilde{\hat\rho}_{i}({\mathbf
x})]=0. \label{75}
\]
Let us find now a commutator $[\tilde{\hat\rho}_{i}({\mathbf
x}),\tilde{\hat {\mathcal H}}_{0}]$ ($i=1,2$), where
$\tilde{\hat\rho}_{i}({\mathbf x})$ and $\tilde{\hat{\mathcal
H}}_{0}$ are defined by (\ref{40}), (\ref{63}), (\ref{64}). In order
to do this we consider $\tilde{\hat\rho}_{1}({\mathbf x})$ in the
form
\[
\tilde{\hat\rho}_{1}({\mathbf
x})=\tilde{\hat\rho}_{1}^{\prime}({\mathbf
x})+\tilde{\hat\rho}_{1}^{\prime\prime}({\mathbf x}),
\]
where
\[
\tilde{\hat\rho}_{1}^{\prime}({\mathbf
x})=\hat{\chi}^{+}_{1}({\mathbf x})\hat{\chi}_{1}({\mathbf x}),
\qquad \tilde{\hat\rho}_{1}^{\prime\prime}({\mathbf x})=\int
d{\mathbf y}d{\mathbf Y}\delta\left({\mathbf x}-{\mathbf
Y}-{m_{2}\over M}{\mathbf y}\right)\hat{\varphi}^{+}({\mathbf
y},{\mathbf Y})\hat{\varphi}({\mathbf y},{\mathbf Y}).
\]
Then in accordance with the commutation relations for the field
operators $\hat{\chi}_{i}({\mathbf x})$,
\[
[\hat{\chi}_{i}({\mathbf x}),\hat{\chi}_{j}^{+}({\mathbf
x}')]=\delta_{ij}\delta({\mathbf x}-{\mathbf x}'),
\]
one gets
\[
\left[\tilde{\hat\rho}_{1}^{\prime}({\mathbf x}),\int d{\mathbf
x}'\sum_{j=1}^{2}{1\over 2m_{j}}{\partial\hat{\chi}_{j}^{+}({\mathbf
x}') \over\partial{\mathbf x}'} {\partial\hat{\chi}_{j}({\mathbf
x}') \over\partial{\mathbf x}'} \right]=-{i\over
2m_{1}}{\partial\over\partial
x_{k}}\biggl(\hat{\chi}_{1}^{+}({\mathbf x}){\partial
\hat{\chi}_{1}({\mathbf x})\over\partial x_{k}}-{\partial
\hat{\chi}_{1}^{+}({\mathbf x})\over\partial
x_{k}}\hat{\chi}_{1}({\mathbf x}) \biggr).
\]
Next according to (\ref{44}) we have
\begin{multline*}
\begin{split}
i[\tilde{\hat\rho}_{1}^{\prime\prime}({\mathbf
x}),\tilde{\hat{\mathcal H}}_{0}]&={1\over 2\mu}\int d{\mathbf
y}d{\mathbf Y}\delta\left({\mathbf x}-{\mathbf Y}-{m_{2}\over
M}{\mathbf y}\right) {\partial\over
\partial y_{k}} \biggl(-\hat{\varphi}^{+}({\mathbf y},{\mathbf
Y}){\partial\hat{\varphi}({\mathbf y},{\mathbf Y}) \over\partial
y_{k}}+{\partial\hat{\varphi}^{+}({\mathbf y},{\mathbf Y})
\over\partial
y_{k}}\hat{\varphi}({\mathbf y},{\mathbf Y})\biggr)\\
&+{1\over 2M}\int d{\mathbf y}d{\mathbf Y}\delta\left({\mathbf
x}-{\mathbf Y}-{m_{2}\over M}{\mathbf y}\right){\partial\over
\partial Y_{k}}\biggl(-\hat{\varphi}^{+}({\mathbf y},{\mathbf
Y}){\partial\hat{\varphi}({\mathbf y},{\mathbf Y}) \over\partial
Y_{k}}+{\partial\hat{\varphi}^{+}({\mathbf y},{\mathbf Y})
\over\partial Y_{k}}\hat{\varphi}({\mathbf y},{\mathbf Y})\biggr).
\end{split}
\end{multline*}
Bearing in mind
\[
-{\partial\over \partial y_{k}}\delta\left({\mathbf x}-{\mathbf
Y}-{m_{2}\over M}{\mathbf y}\right)={m_{2}\over M}{\partial\over
\partial x_{k}}\delta\left({\mathbf x}-{\mathbf Y}-{m_{2}\over M}{\mathbf
y}\right), \qquad -{\partial\over \partial
Y_{k}}\delta\left({\mathbf x}-{\mathbf Y}-{m_{2}\over M}{\mathbf
y}\right)={\partial\over
\partial x_{k}}\delta\left({\mathbf x}-{\mathbf Y}-{m_{2}\over M}{\mathbf y}\right)
\]
and taking into account the definitions of $\tilde{\hat\rho}_{i},
\tilde{\hat{\mbox{\boldmath$\pi$}}}_{i}$ ($i=1,2$), one finds
\begin{equation}
i[\tilde{\hat\rho}_{1}({\mathbf x}),\tilde{\hat{\mathcal
H}}_{0}+\tilde{\hat{\mathcal H}}_{int}]={1\over m_{1}}{\rm div}\,
\tilde{\hat{\mbox{\boldmath$\pi$}}}_{1}({\mathbf x}). \label{76}
\end{equation}
Similarly,
\begin{equation}
i[\tilde{\hat\rho}_{2}({\mathbf x}),\tilde{\hat{\mathcal
H}}_{0}+\tilde{\hat{\mathcal H}}_{int}]={1\over m_{2}}{\rm div}\,
\tilde{\hat{\mbox{\boldmath$\pi$}}}_{2}({\mathbf x}). \label{77}
\end{equation}
Let us find now a commutator $[\tilde{\hat\rho}_{i}({\mathbf
x}'),\tilde{\hat{\mbox{\boldmath$\pi$}}}_{k}({\mathbf x})]$,
($i,k=1,2$). With the use of (\ref{39}) we have
\[
[\tilde{\hat\rho}_{i}^{\prime\prime}({\mathbf
x}'),\tilde{\hat{\mbox{\boldmath$\pi$}}}_{k}({\mathbf x})]=
i\delta_{ik}\tilde{\hat\rho}_{i}^{\prime\prime}({\mathbf x})
{\partial\over\partial{\mathbf x}}\delta({\mathbf x}-{\mathbf x}').
\]
Noting also that
\[
[\tilde{\hat\rho}_{i}^{\prime}({\mathbf
x}'),\tilde{\hat{\mbox{\boldmath$\pi$}}}_{k}({\mathbf
x})]=i\delta_{ik}\tilde{\hat\rho}_{i}^{\prime}({\mathbf
x}){\partial\over\partial{\mathbf x}}\delta({\mathbf x}-{\mathbf
x}'),
\]
we get finally
\[
[\tilde{\hat\rho}_{i}({\mathbf
x}'),\tilde{\hat{\mbox{\boldmath$\pi$}}}_{k}({\mathbf x})]=
i\delta_{ik}\tilde{\hat\rho}_{i}({\mathbf x}){\partial\over\partial
{\mathbf x}}\delta({\mathbf x}-{\mathbf x}').
\]
These commutation relations along with (\ref{65'}) result in
\[
[\tilde{\hat{\mathbf j}}_{0}({\mathbf
x}),\tilde{\hat\sigma}({\mathbf x}')]=-i\sum_{i=1}^{2}{e_{i}\over
m_{i}}\tilde{\hat\sigma}_{i}({\mathbf x}'){\partial\over\partial
{\mathbf x}}\delta({\mathbf x}-{\mathbf x}').
\]
Thus, according to (\ref{69}), (\ref{76}), (\ref{77}) we have
\[
\dot{\tilde{\hat\sigma}}({\mathbf x})=i[\tilde{\hat{\mathcal
H}}(t),\tilde{\hat\sigma}({\mathbf x})]={\rm div}\,
\tilde{\hat{\mathbf J}}({\mathbf x},t).
\]
The definitions (\ref{69}), (\ref{70}) of the longitudinal
component of the electric field along with the above formula give
\begin{equation}
\dot{\hat{\mathbf e}}^{l}({\mathbf x})\equiv i[\tilde{\hat{\mathcal
H}}(t),\hat{\mathbf e}^{l}({\mathbf
x})]={\partial\over\partial{\mathbf x}} {\rm div}\,\int d{\mathbf
x}'{\tilde{\hat {\mathbf J}}({\mathbf x}^{\prime},t)\over |{\mathbf
x}-{\mathbf x}'|}. \label{78}
\end{equation}

Let us find the equation of motion for $\hat{\mathbf e}^{t}$. The
use of commutation relations for creation $\hat{C}^{+}_{{\mathbf
k}\lambda}$ and annihilation $\hat{C}_{{\mathbf k}\lambda}$
operators of photons,
\[
[\hat{C}_{{\mathbf k}\lambda},\hat{C}^{+}_{{\mathbf
k}'\lambda^{\prime}}]=\delta_{{\mathbf k}{\mathbf
k}'}\delta_{\lambda\lambda^{\prime}}
\]
lead to
\begin{equation}
\hat{\mathbf e}^{t}({\mathbf x})=-i[\tilde{\hat{\mathcal
H}}(t),\hat{\mathbf a}({\mathbf x})] =i\sum_{{\mathbf
k}}\sum_{\lambda=1}^{2}\biggl({2\pi\omega_{k}\over {\mathcal
V}}\biggr)^{1/2}\biggl({\mathbf e}^{(\lambda)}_{\mathbf
k}\hat{C}_{{\mathbf k}\lambda}e^{i{\mathbf k}{\mathbf
x}}-h.c.\biggr), \label{79}
\end{equation}
\[
\dot{\hat{\mathbf e}}^{t}({\mathbf x})=-i[\tilde{\hat{\mathcal
H}}(t),\hat{\mathbf e}^{t}({\mathbf x})]={\rm rot}\,\hat{\mathbf
h}({\mathbf x})+i[\tilde{\hat V}(t),\hat{\mathbf e}^{t}({\mathbf
x})],
\]
and also
\[
[\hat{a}_{i}({\mathbf x}),\hat{e}_{s}^{t}({\mathbf x}')]=-4\pi
i\delta_{is}\delta({\mathbf x}-{\mathbf
x}')-i{\partial^{2}\over{\partial x_{i}}{\partial x_{s}}}{1\over
|{\mathbf x}-{\mathbf x}'|}.
\]
This formula and the definition of $\tilde{\hat V}(t)$ allows us
to find the commutator:
\[
i[\tilde{\hat V}(t),\hat{\mathbf e}^{t}({\mathbf
x})]=-4\pi\tilde{\hat{\mathbf J}}({\mathbf x},t) -
{\partial\over\partial{\mathbf x}} {\rm div}\, \int d{\mathbf
x}'{\tilde{\hat {\mathbf J}}({\mathbf x}^{\prime},t)\over |{\mathbf
x}-{\mathbf x}'|}. \label{80}
\]
Therefore, it follows from (\ref{78}), (\ref{79}) that
\begin{equation}
\dot{\hat{\mathbf e}}({\mathbf x})={\rm rot}\, \hat{\mathbf
h}({\mathbf x})-4\pi\tilde{\hat{\mathbf J}}({\mathbf x},t), \qquad
\dot{\hat{\mathbf h}}({\mathbf x})\equiv i[\tilde{\hat{\mathcal
H}}(t),\hat{\mathbf h}({\mathbf x})]=i[\hat{\mathcal
H}_{f},\hat{\mathbf h}({\mathbf x})].\label{81}
\end{equation}
Upon calculating this commutator with the use of (\ref{67}), one
obtains
\begin{equation}
\dot{\hat{\mathbf h}}({\mathbf x})=-{\rm rot}\, \hat{\mathbf
e}({\mathbf x}). \label{82}
\end{equation}
In addition, (\ref{67}), (\ref{69}) give ${\rm div}\, \hat{\mathbf
h}({\mathbf x})=0$, ${\rm div}\, \hat{\mathbf e}({\mathbf
x})=4\pi\tilde{\hat\sigma}({\mathbf x})$. These equations along with
(\ref{81}), (\ref{82}) represent the Maxwell--Lorentz equations for
the operators of electromagnetic field
\begin{gather*}{2}
\dot{\hat{\mathbf e}}({\mathbf x})={\rm rot}\, \hat{\mathbf
h}({\mathbf x})-4\pi\tilde{\hat{\mathbf J}}({\mathbf x},t), \qquad
{\rm div}\,
\hat{\mathbf e}({\mathbf x})=4\pi\tilde{\hat\sigma}({\mathbf x}), \\
\dot{\hat{\mathbf h}}({\mathbf x})=-{\rm rot}\, \hat{\mathbf
e}({\mathbf x}),\qquad {\rm div}\, \hat{\mathbf h}({\mathbf x})=0.
\label{83}
\end{gather*}
Notice that (\ref{64}), (\ref{67}), (\ref{79}) result in
\[
\hat{\mathcal H}_{f} = \sum_{\mathbf k}\sum_{\lambda =
1}^{2}\omega_{k}{\hat C}_{{\mathbf k}\lambda}^{+}{\hat C}_{{\mathbf
k}\lambda} = {1\over 8\pi}\int\,d{\mathbf x}\biggl( {{\mathbf
e}^{t}}^{2} + {\mathbf h}^{2}\biggr),
\]
where we neglect the unessential term $\sum_{{\mathbf
k}\lambda}{1\over 2}\omega_{k}$.

Up to now we used the Schroedinger representation, in which the
Schroedinger equation has the form
\[
i{\partial\Phi(t)\over\partial t}=\tilde{\hat{\mathcal H}}(t)\Phi(t).
\]
In this representation a dot above the certain operator $\dot{\hat
b}({\mathbf x})$ (see (\ref{69})) does not mean the differentiation
with respect to time. It means the following:
\begin{equation}
\dot{\hat b}({\mathbf x})=i[\tilde{\hat{\mathcal
H}}(\underline{\hat{\chi}}({\mathbf x}'),t),{\hat b}({\mathbf x})].
\label{84}
\end{equation}
Here the Hamiltonian $\tilde{\hat{\mathcal H}}$ depends on dynamic
variables $\underline{\hat{\chi}}({\mathbf x}')$ (these variables
unite $\hat{\chi}$ and $\hat{\eta}$) and time $t$. Introducing the
unitary operator $\hat{S}(t,0)$,
\[
i{\partial\hat{S}(t,0)\over\partial t}=\tilde{\hat{\mathcal
H}}(\underline{\hat{\chi}}({\mathbf x}'),t)\hat{S}(t,0), \qquad
\hat{S}(0,0)=1,
\]
we can write the solution of the equation for $\Phi(t)$ in the
form,
\[
\Phi(t)=\hat{S}(t,0)\Phi(0).
\]

In the Heisenberg representation the vector $\Phi(0)=\Psi$ is
taken as a time-independent state vector. Then, the time dependent
Heisenberg operators
\[
\underline{\hat{\chi}}({\mathbf x},t)\equiv
\underline{\hat{\chi}}(x)=\hat{S}^{+}(t,0)\underline{\hat{\chi}}({\mathbf
x})\hat{S}(t,0)
\]
satisfy the equation
\begin{equation}
i{\partial\underline{\hat{\chi}}(x)\over \partial
t}=[\underline{\hat{\chi}}(x),\tilde{\hat{\mathcal
H}}(\underline{\hat{\chi}}(x),t)]. \label{85}
\end{equation}

Since the external fields ${\mathbf H}^{(e)}, {\mathbf E}^{(e)}$
satisfy the Maxwell equations,
\begin{gather*}
{\partial {\mathbf H}^{(e)}\over \partial t}=-{\rm rot}\, {\mathbf
E}^{(e)},\qquad {\rm div}\, {\mathbf H}^{(e)}=0,\\
{\partial {\mathbf E}^{(e)}\over \partial t}={\rm rot}\, {\mathbf
H}^{(e)}-4\pi{\mathbf J}^{(e)}, \qquad {\rm div}\, {\mathbf
E}^{(e)}=4\pi\sigma^{(e)},
\end{gather*}
the operators of total fields $\hat{\mathbf E}$, $\hat{\mathbf H}$
in the Heisenberg representation satisfy, in accordance with
(\ref{85}), the equations
\begin{gather}
{\partial \hat{\mathbf H}\over \partial t}=-{\rm rot}\, \hat{\mathbf
E}, \qquad
{\rm div}\, \hat{\mathbf H}=0,\nonumber\\
{\partial \hat{\mathbf E}\over \partial t}={\rm rot}\, \hat{\mathbf
H}-4\pi(\tilde{\hat{\mathbf J}}+{\mathbf J}^{(e)}), \qquad {\rm
div}\, \hat{\mathbf E}=4\pi(\tilde{\hat\sigma}+\sigma^{(e)}),
\label{86}
\end{gather}
where ${\mathbf J}^{(e)}$ and $\sigma^{(e)}$ are the extrinsic
current and charge densities.

Let us write out the expressions for the charge density operator
$\tilde{\hat\sigma}$ and current density operator
$\tilde{\hat{\mathbf J}}$ in terms of creation and annihilation
operators of particles and bound states. Bearing in mind (\ref{66})
we have
\begin{gather}
\tilde{\hat\sigma}({\mathbf
x})=\sum_{i=1}^{2}e_{i}\tilde{\hat\rho}_{i}({\mathbf x}),\nonumber\\
\tilde{\hat{\mathbf J}}({\mathbf x})=-\hat{\mathbf A}({\mathbf
x})\sum_{i=1}^{2}{e_{i}^{2}\over m_{i}}\tilde{\hat\rho}_{i}({\mathbf
x})+\sum_{i=1}^{2}{e_{i}\over
m_{i}}\tilde{\hat{\mbox{\boldmath$\pi$}}}_{i}({\mathbf x}),
\label{87}
\end{gather}
where the particle density operators $\tilde{\hat\rho}_{i}({\mathbf
x})$ are defined by (\ref{40}) and the momentum density operators
$\tilde{\hat{\mbox{\boldmath$\pi$}}}_{i}({\mathbf x})$ by
(\ref{39}). The first terms in expressions for
$\tilde{\hat\rho}_{i}({\mathbf x})$ and
$\tilde{\hat{\mbox{\boldmath$\pi$}}}_{i}({\mathbf x})$ define the
contribution of particles, whereas the second terms define the
contribution of bound states (see (\ref{39}), (\ref{40})).

Let us find now the equations of motion for the field operators of
particles $\hat{\chi}_{i}({\mathbf x})$ and bound states
$\hat{\varphi}({\mathbf x},{\mathbf y})$. In the Heisenberg
representation these equations are obtained from the general
formulas (\ref{85}). Indeed, it follows from (\ref{54})--(\ref{56})
that
\begin{gather*}
[\hat{\chi}_{1}({\mathbf x}),\tilde{\hat{\mathcal
H}}_{int}(\hat{\chi}({\mathbf x}'))]=\biggl\{\int d{\mathbf
x}_{2}\nu_{11}({\mathbf x}-{\mathbf
x}_{2})\tilde{\hat\rho}_{1}({\mathbf x}_{2})+\int d{\mathbf
x}_{2}\nu_{12}({\mathbf x}-{\mathbf
x}_{2})\tilde{\hat\rho}_{2}({\mathbf
x}_{2})\biggr\}\hat{\chi}_{1}({\mathbf x}),\\
{[\hat{\chi}_{2}({\mathbf x}),\tilde{\hat{\mathcal
H}}_{int}(\hat{\chi}({\mathbf x}'))]}=\biggl\{\int d{\mathbf
x}_{1}\nu_{22}({\mathbf x}-{\mathbf
x}_{1})\tilde{\hat\rho}_{2}({\mathbf x}_{1})+\int d{\mathbf
x}_{2}\nu_{21}({\mathbf x}-{\mathbf
x}_{1})\tilde{\hat\rho}_{1}({\mathbf
x}_{1})\biggr\}\hat{\chi}_{2}({\mathbf x}), \\
{[\hat{\varphi}({\mathbf x},{\mathbf y}),\tilde{\hat{\mathcal
H}}_{int}(\hat{\chi}({\mathbf x}'))]}=\biggl\{\int d{\mathbf
x}_{2}\biggl(\nu_{11}({\mathbf x}-{\mathbf x}_{2})+\nu_{12}({\mathbf
y}-{\mathbf
x}_{2})\biggr)\tilde{\hat\rho}_{1}({\mathbf x}_{2})+\\
\int d{\mathbf x}_{2}\biggl(\nu_{22}({\mathbf y}-{\mathbf
x}_{2})+\nu_{21}({\mathbf x}-{\mathbf
x}_{2})\biggr)\tilde{\hat\rho}_{2}({\mathbf
x}_{2})\biggr\}\hat{\varphi}({\mathbf x},{\mathbf y}).
\end{gather*}
These formulas allow to find the equations of motion (\ref{85}) for
$\hat{\chi}_{1}({\mathbf x})$, $\hat{\chi}_{2}({\mathbf x})$ in the
Heisenberg representation,
\begin{gather*}
i\dot{\hat{\chi}}_{1}({\mathbf x},t)=\biggl\{-{1\over
2m_{1}}\biggl({\partial\over\partial{\mathbf x}}-ie_{1}\hat{\mathbf
A}({\mathbf x},t)\biggr)^{2}+e_{1}\varphi^{(e)}({\mathbf
x},t)+\hat{\upsilon}_{1}({\mathbf
x},t)\biggr\}\hat{\chi}_{1}({\mathbf x},t),\\
i\dot{\hat{\chi}}_{2}({\mathbf x},t)=\biggl\{-{1\over
2m_{2}}\biggl({\partial\over\partial{\mathbf x}}-ie_{2}\hat{\mathbf
A}({\mathbf x},t)\biggr)^{2}+e_{2}\varphi^{(e)}({\mathbf x},t)+
\hat{\upsilon}_{2}({\mathbf x},t)\biggr\}\hat{\chi}_{2}({\mathbf
x},t).
\end{gather*}
It is easy to find also the equation of motion for the field
operators $\hat{\varphi}({\mathbf x},{\mathbf y})$ of bound states,
\begin{equation}
\begin{split}
i\dot{\hat{\varphi}}({\mathbf x}_{1},{\mathbf x}_{2},t)&=
\sum_{\alpha}\varepsilon_{\alpha}\hat{\eta}_{\alpha}({\mathbf
X},t)\varphi_{\alpha}({\mathbf x})+\biggl\{-{1\over
2m_{1}}\biggl({\partial\over\partial{\mathbf
x}_{1}}-ie_{1}\hat{\mathbf A}({\mathbf x}_{1},t)\biggr)^{2}-{1\over
2m_{2}}\biggl({\partial\over\partial{\mathbf
x}_{2}}-ie_{2}\hat{\mathbf A}({\mathbf x}_{2},t)\biggr)^{2}\\
&+e_{1}\varphi^{(e)}({\mathbf x}_{1},t)+e_{2}\varphi^{(e)}({\mathbf
x}_{2},t)+\hat{\upsilon}({\mathbf x}_{1},{\mathbf
x}_{2},t)\biggr\}\hat{\varphi}({\mathbf x}_{1},{\mathbf x}_{2},t).
\label{89}
\end{split}
\end{equation}
Here
\begin{gather*}
\hat{\upsilon}_{1}({\mathbf x},t)=\int d{\mathbf
x}_{2}\nu_{11}({\mathbf x}-{\mathbf
x}_{2})\tilde{\hat\rho}_{1}({\mathbf x}_{2},t)+\int d{\mathbf
x}_{2}\nu_{12}({\mathbf x}-{\mathbf
x}_{2})\tilde{\hat\rho}_{2}({\mathbf
x}_{2},t), \\
{\hat{\upsilon}_{2}({\mathbf x},t)}=\int d{\mathbf
x}_{1}\nu_{22}({\mathbf x}-{\mathbf
x}_{1})\tilde{\hat\rho}_{2}({\mathbf x}_{1},t)+\int d{\mathbf
x}_{1}\nu_{21}({\mathbf x}-{\mathbf
x}_{1})\tilde{\hat\rho}_{1}({\mathbf x}_{1},t),\\
{\hat{\upsilon}({\mathbf x}_{1},{\mathbf x}_{2},t)}=\int d{\mathbf
x}'\biggl(\nu_{11}({\mathbf x}_{1}-{\mathbf x}')+\nu_{12}({\mathbf
x}_{2}-{\mathbf x}')\biggr)\tilde{\hat\rho}_{1}({\mathbf x}',t)+\int
d{\mathbf x}'\biggl(\nu_{22}({\mathbf x}_{2}-{\mathbf
x}')+\nu_{21}({\mathbf x}_{1}-{\mathbf
x}')\biggr)\tilde{\hat\rho}_{2}({\mathbf x}',t).
\end{gather*}
The operators $\hat{\eta}_{\alpha}({\mathbf X},t)$ related to
$\hat{\varphi}({\mathbf x}_{1},{\mathbf x}_{2},t)$ by
$\hat{\varphi}({\mathbf x}_{1},{\mathbf
x}_{2},t)=\varphi_{\beta}({\mathbf x})\hat{\eta}_{\beta}({\mathbf
X},t)$.

The equations of motion (\ref{86}), (\ref{89}) are
gauge-invariant. Indeed, under the following transformations of
fields $\hat{\chi}_{i}$, $\hat{\varphi}$:
\begin{gather}
\hat{\chi}_{1}({\mathbf
x},t)\rightarrow\hat{\chi}_{1}^{\prime}({\mathbf
x},t)=e^{ie_{1}\alpha({\mathbf x},t)}\hat{\chi}_{1}({\mathbf x},t),
\qquad \hat{\chi}_{2}({\mathbf
x},t)\rightarrow\hat{\chi}_{2}^{\prime}({\mathbf
x},t)=e^{ie_{2}\alpha({\mathbf x},t)}\hat{\chi}_{2}({\mathbf
x},t), \nonumber \\
\hat{\varphi}({\mathbf x},{\mathbf
y},t)\rightarrow\hat{\varphi}^{\prime}({\mathbf x},{\mathbf
y},t)=e^{ie_{1}\alpha({\mathbf x},t)+ie_{2}\alpha({\mathbf
y},t)}\hat{\varphi}({\mathbf x},{\mathbf y},t),  \label{91}
\end{gather}
and electromagnetic field,
\[
\hat{\mathbf A}({\mathbf x},t)\rightarrow\hat{\mathbf
A}^{\prime}({\mathbf x},t)=\hat{\mathbf A}({\mathbf
x},t)-{\partial\over\partial{\mathbf x}}\alpha({\mathbf x},t),
\qquad \varphi^{(e)}({\mathbf
x},t)\rightarrow{\varphi^{(e)}}^{\prime}({\mathbf
x},t)=\varphi^{(e)}({\mathbf x})-{\partial \alpha({\mathbf
x},t)\over\partial t}
\]
the operators $\tilde{\hat\rho}_{i}({\mathbf x})$ are invariant,
\[
\tilde{\hat\rho}_{i}({\mathbf
x},t)\rightarrow\tilde{\hat\rho}_{i}^{\prime}({\mathbf
x},t)=\tilde{\hat\rho}_{i}({\mathbf x},t), \label{92}
\]
whereas $\tilde{\hat{\mbox{\boldmath$\pi$}}}_{i}({\mathbf x})$
transform (see (\ref{89})), by the law
\[
\tilde{\hat{\mbox{\boldmath$\pi$}}}_{i}({\mathbf x},t)\rightarrow
\tilde{\hat{\mbox{\boldmath$\pi$}}}_{i}^{'}({\mathbf
x},t)=\tilde{\hat{\mbox{\boldmath$\pi$}}}_{i}({\mathbf
x},t)+e_{i}\tilde {\hat\rho}_{i}({\mathbf x},t)
{\partial\over\partial{\mathbf x}}\alpha({\mathbf x},t). \label{93}
\]
As a result we have (see (\ref{87}))
\[
\tilde{\hat{\mathbf J}}({\mathbf x},t)\rightarrow\tilde{\hat{\mathbf
J}}^{\prime}({\mathbf x},t)=\tilde{\hat{\mathbf J}}({\mathbf
x},t),\qquad \tilde{\hat{\mathcal H}}\rightarrow\tilde{\hat{\mathcal
H}}^{\prime}=\tilde{\hat{\mathcal H}} \label{94}
\]
that provides a gauge invariance of the Maxwell-Lorentz equations
(\ref{86}). Next noting that
\[
\biggl({\partial\over\partial{\mathbf x}}-ie_{i} \hat{\mathbf
A}({\mathbf x},t)\biggr)\hat{\chi}_{i}({\mathbf
x},t)\rightarrow\biggl({\partial\over\partial{\mathbf x}}-
ie_{i}\hat{\mathbf A}^{\prime}({\mathbf
x},t)\biggr)\hat{\chi}_{i}^{\prime}({\mathbf x},t)=
e^{ie_{i}\alpha({\mathbf x},t)}\biggl({\partial\over\partial{\mathbf
x}}- ie_{i}\hat{\mathbf A}({\mathbf
x},t)\biggr)\hat{\chi}_{i}({\mathbf x},t),
\]
we make sure of gauge-invariance of the equations of motion for the
field operators of particles $\hat{\chi}_{i}({\mathbf x})$. Taking
into account the transformation law for the field operators of bound
states $\hat{\varphi}({\mathbf x},{\mathbf y})$, it is easy to prove
a gauge invariance of the equation of motion for
$\hat{\varphi}({\mathbf x},{\mathbf y})$.

Since $\hat{\varphi}({\mathbf x}_{1},{\mathbf
x}_{2},t)=\varphi_{\alpha}({\mathbf x})\hat{\eta}_{\alpha}({\mathbf
X},t)$, we have under the gauge transformation (\ref{91}),
\[
\hat{\varphi}^{\prime}({\mathbf x}_{1},{\mathbf
x}_{2},t)=\varphi_{\alpha}({\mathbf
x})\hat{\eta}_{\alpha}^{\prime}({\mathbf X},t),
\]
where ${\mathbf x}_{1}={\mathbf X}+(m_{2}/M){\mathbf x}$, ${\mathbf
x}_{2}={\mathbf X}-(m_{1}/M){\mathbf x}$. Hence, we come to the
transformation law for the operators $\hat{\eta}_{\alpha}({\mathbf
X})$:
\[
\hat{\eta}_{\alpha}({\mathbf X},t)\rightarrow
\hat{\eta}_{\alpha}^{\prime}({\mathbf X},t)=
\sum_{\beta}L_{\alpha\beta}({\mathbf
X},t)\hat{\eta}_{\beta}({\mathbf X},t), \qquad
L_{\alpha\beta}({\mathbf X},t)=\int d{\mathbf
x}\varphi_{\alpha}^{*}({\mathbf x})e^{ie_{1}\alpha[{\mathbf
X}+(m_{2}/M){\mathbf x},t]+ie_{2}\alpha[{\mathbf X}-(m_{1}/
M){\mathbf x},t] }\varphi_{\beta}({\mathbf x}).
\]
In conclusion of this section let us note that
$\tilde{\hat\psi}_{i}(\mathbf x)$ transform under the gauge
transformation as well as $\hat{\chi}_{i}({\mathbf x})$:
\[
\tilde{\hat\psi}_{i}({\mathbf x},t) \rightarrow
\tilde{\hat\psi}_{i}^{\prime}({\mathbf x},t) =
e^{ie_{i}\alpha({\mathbf x},t)}\tilde{\hat\psi}_{i}({\mathbf x},t).
\]

\section{The Process of Spontaneous Radiation by Atom}

In this section we employ the obtained formulas in order to find
the probability distribution of spontaneous radiation of an
excited atom. To this end we use the general formulas for the
scattering matrix $\hat{S}$ in terms of the T-operator:
\begin{gather}
{\hat S}=1-2\pi\,i\int_{-\infty}^{\infty}dE\delta(E-{\hat {\mathcal
H}}_{0}){\hat T}^{(+)}(E)\delta(E-{\hat {\mathcal H}}_{0}),\nonumber\\
{\hat T}^{(+)}(E)=\lim_{\eta\to +0}{\hat T}(E+i\eta),\label{95}
\end{gather}
where ${\hat T}(z)$ is found from the equation
\begin{equation}
{\hat T}(z) = {\hat V}+{\hat V}{\hat R}_{0}{\hat T}(z), \qquad
{\hat R}_{0} = {1\over z-{\hat {\mathcal H}}_{0}}. \label{96}
\end{equation}
In the first order of the perturbative approach over charge we
have
\[
\langle f|{\hat T}(z)|i\rangle\approx\langle f|{\hat V}|i\rangle,
\]
where
\[
|i\rangle = {\hat \eta}^{+}_{\alpha {\mathbf p}}|0\rangle , \qquad
|f\rangle = {\hat C}_{{\mathbf k}\lambda}^{+}{\hat \eta}^{+}_{\alpha
{\mathbf p}}|0\rangle .
\]
Here $|i\rangle$ is the initial state (in which atom is in the state
$\alpha, {\mathbf p}$) and $|f\rangle$ is the final state (in which
atom is in the state $\alpha^{\prime}, {\mathbf p}^{\prime}$ and
photon is in the state $\lambda, {\mathbf k}$). According to
(\ref{65}), (\ref{66}) we can take ${\hat V}$ in the following form:
\[
\tilde{\hat V}=-\int\,d{\mathbf x} {\hat {\mathbf a}}({\mathbf
x})\tilde{\hat {\mathbf j}}({\mathbf x}),
\]
where the current of bound states is defined by
\begin{equation}
\tilde{\hat {\mathbf j}}({\mathbf x})={e_{1}\over
m_{1}}\tilde{\hat{\mbox{\boldmath$\pi$}}}_{1}({\mathbf x}) +
{e_{2}\over m_{2}}\tilde{\hat{\mbox{\boldmath$\pi$}}}_{2}({\mathbf
x}). \label{97}
\end{equation}
By using of (\ref{65}), (\ref{66}) we find
\[
\langle f|\tilde{\hat V}|i\rangle = -\biggl({2\pi\over {\mathcal
V}\omega_{k}} \biggr)^{1/2}{{\mathbf e}_{\mathbf
k}^{(\lambda)}}^{*}\int\,d{\mathbf x} e^{-i{\mathbf k}{\mathbf
x}}\langle{\alpha}^{\prime}{\mathbf p}^{\prime}|\tilde{\hat {\mathbf
j}}({\mathbf x})|\alpha{\mathbf p}\rangle .
\]
The formulas (\ref{97}), (\ref{39}) lead to
\[
\tilde{\hat {\mathbf j}}({\mathbf k})\equiv\int\,d{\mathbf x}
e^{-i{\mathbf k}{\mathbf x}}\tilde{\hat{\mathbf j}}({\mathbf
x})=\sum_{{\mathbf p}_{1},{\mathbf
p}_{2}}\sum_{\alpha_{1},\alpha_{2}}{\hat\eta}^{+}_{\alpha_{1}
{\mathbf p}_{1}}{\hat \eta}_{\alpha_{2} {\mathbf
p}_{2}}\biggl({\mathbf g}_{\alpha_{1}\alpha_{2}}({\mathbf
k})+{{\mathbf p}_{1}+{\mathbf p}_{2}\over
2M}g_{\alpha_{1}\alpha_{2}}({\mathbf k})\biggr)\Delta({\mathbf
p}_{1}-{\mathbf p}_{2}+{\mathbf k})
\]
and, consequently,
\[
\langle f|\tilde{\hat V}|i\rangle =-\biggl({2\pi\over {\mathcal
V}\omega_{k}} \biggr)^{1/2}{{\mathbf e}_{\mathbf
k}^{(\lambda)}}^{*}\biggl({\mathbf
g}_{\alpha\alpha^{\prime}}({\mathbf k})+{{\mathbf p}+{\mathbf
p}^{\prime}\over 2M}g_{\alpha\alpha^{\prime}}({\mathbf
k})\biggr)\Delta({\mathbf p}-{\mathbf p}^{\prime}-{\mathbf k}),
\quad {\mathbf k}={\mathbf p}-{\mathbf p}^{\prime},
\]
where $\Delta({\mathbf p})=\delta_{{\mathbf p} 0}$, and
\begin{gather}
g_{\alpha\beta}({\mathbf k}) = \int d{\mathbf
y}\biggl(e_{1}e^{i{\mathbf k}{\mathbf
y}(m_{2}/M)}+e_{2}e^{-i{\mathbf k}{\mathbf y}(m_{1}/M)}\biggr)\varphi_{\alpha}^{*}({\mathbf y})
\varphi_{\beta}({\mathbf y}), \nonumber \\
{\mathbf g}_{\alpha\beta}({\mathbf k})=-{i\over 2}\int d{\mathbf
y}\biggl({e_{1}\over m_{1}}e^{i{\mathbf k}{\mathbf
y}(m_{2}/M)}-{e_{2}\over m_{2}}e^{-i{\mathbf k}{\mathbf y}(m_{1}/
M})\biggr) \biggl(\varphi^{*}_{\alpha}({\mathbf
y}){\partial\varphi_{\beta}({\mathbf y}) \over\partial {\mathbf
y}}-{\partial\varphi_{\alpha}^{*}({\mathbf y}) \over\partial
{\mathbf y}}\varphi_{\beta}({\mathbf y}) \biggr).\label{98}
\end{gather}
The matrix element of $\hat{S}$-matrix is defined, in accordance
with (\ref{95}), by
\[
\langle f|{\hat S}|i\rangle=-2\pi i\langle f|\tilde{\hat
V}|i\rangle\delta({\varepsilon}({\mathbf p})-{\varepsilon}({\mathbf
p}^{\prime})-\omega_{k}).
\]
Next using the standard method we obtain the following expression
for the probability of transition per unit time from the initial
state $\hat{\eta}^{+}_{\alpha{\mathbf p}}|0\rangle$ into the final
state $\hat{\eta}^{+}_{\beta{\mathbf p}'}\hat{C}^{+}_{{\mathbf
k}\lambda}|0\rangle$:
\[
dw_{i\rightarrow f}={2\pi\over\omega_{k}}\bigg\vert {\mathbf
e}^{(\lambda)}_{\mathbf k}\biggl({\mathbf g}_{\alpha\beta}({\mathbf
k})+{{\mathbf p}+{\mathbf p}'\over 2M}g_{\alpha\beta}({\mathbf k})
\biggr)\bigg\vert^{2}\delta(\omega_{k}+\varepsilon_{\beta}({\mathbf
p}')-\varepsilon_{\alpha}({\mathbf p}))d{\mathbf k} . \label{99}
\]
In the domain of small ${\mathbf k}$ ($kr_{0} << 1$, see section II)
the Fourier-components of ${\mathbf g}_{\alpha\beta}({\mathbf x})$,
$g_{\alpha\beta}({\mathbf x})$ can be written in the form
\begin{equation}
{\mathbf g}_{\alpha\beta}(0)=\biggl({e_{1}\over m_{1}}-{e_{2}\over
m_{2}}\biggr)\int d{\mathbf y}\varphi^{*}_{\alpha}({\mathbf y}){\hat
{\mathbf p}}\varphi_{\beta}({\mathbf y})= \biggl({e_{1}\over
m_{1}}-{e_{2}\over m_{2}}\biggr)\langle\alpha |{\hat {\mathbf
p}}|\beta\rangle ,\quad
g_{\alpha\beta}(0)=(e_{1}+e_{2})\delta_{\alpha\beta}. \label{100}
\end{equation}

If $e_{2}=-e_{1}=-e$ (this case corresponds to the hydrogen atom),
then $g_{\alpha\beta}(0)=0$, ${\mathbf g}_{\alpha\beta}(0)=(e/
\mu)\langle\alpha |{\hat {\mathbf p}}|\beta\rangle$. The
Schroedinger equation for the wave function
${\varphi}_{\alpha}({\mathbf y})$ gives
\[
\langle\alpha |{\hat {\mathbf p}}|\beta\rangle =
i\mu({\varepsilon}_{\alpha} -{\varepsilon}_{\beta})\int\,d{\mathbf
y}{\varphi}_{\alpha}^{*}({\mathbf y}) {\mathbf y}
{\varphi}_{\beta}({\mathbf y}). \label{101}
\]
Therefore,
\[
{\mathbf g}_{\alpha\beta}(0)=i({\varepsilon}_{\alpha} -
{\varepsilon}_{\beta})\langle\alpha |{\hat {\mathbf d}}|\beta\rangle
,
\]
where ${\hat{\mathbf d}}=e{\hat{\mathbf y}}$ is the dipole moment of
the atom. Hence,
\[
dw_{i\rightarrow f}=2\pi\omega_{k}\vert \langle\alpha|{\mathbf
e}^{(\lambda)}_{\mathbf k}{\hat {\mathbf
d}}|\beta\rangle\vert^{2}\delta(\omega_{k}+\varepsilon_{\beta}({\mathbf
p}')-\varepsilon_{\alpha}({\mathbf p}))d{\mathbf k}. \label{102}
\]
Since $\varepsilon_{\alpha}({\mathbf p})=
\varepsilon_{\alpha}+p^{2}/2M$, we have for the atom of infinite
mass ($m_{2}\rightarrow\infty$, $M\rightarrow \infty$)${}^{5}$:
\[
dw_{i\rightarrow f}={2\pi\over\omega_{k}}\vert {\mathbf
e}^{(\lambda)}_{\mathbf k}{\mathbf g}_{\alpha\beta}({\mathbf
k})\vert^{2} \delta(\omega_{k}+
\varepsilon_{\beta}-\varepsilon_{\alpha})d{\mathbf k},
\]
where
\[
{\mathbf g}_{\alpha\beta}({\mathbf k})=-{ie\over
2m_{1}}\int\,d{\mathbf y}e^{{i\mathbf k}{\mathbf
y}}\biggl(\varphi_{\alpha}^{*}({\mathbf y}){\partial
\varphi_{\beta}({\mathbf y})\over\partial {\mathbf y}} - {\partial
\varphi_{\alpha}^{*}({\mathbf y})\over\partial {\mathbf
y}}\varphi_{\beta}({\mathbf y}) \biggr)
\]
($m_{1}$ is the mass of light fermion).

\section{Scattering of Photons and Fermions by Atoms}

Here we study the scattering process of long-wave photons by
atoms. For this purpose let us use the general formulas
(\ref{95}), (\ref{96}), where the Hamiltonian of free particles is
defined by
\[
\tilde{\hat{\mathcal H}}_{0}=\sum_{i=1}^{2}\int\,d{\mathbf
x}\nabla{\hat \chi}_{i}^{+}({\mathbf x})\nabla{\hat
\chi}_{i}({\mathbf x})+\sum_{\mathbf
k}\sum_{\lambda=1}^{2}\omega_{k}{\hat C}_{{\mathbf
k}\lambda}^{+}{\hat C}_{{\mathbf
k}\lambda}+\sum_{\alpha}\int\,d{\mathbf x}\biggl({1\over
2m}\nabla{\hat \eta}_{\alpha}^{+}({\mathbf x})\nabla{\hat
\eta}_{\alpha}({\mathbf x}) + {\varepsilon}_{\alpha}{\hat
\eta}_{\alpha}^{+}({\mathbf x}){\hat \eta}_{\alpha}({\mathbf
x})\biggr), \label{103}
\]
and the Hamiltonian associated with electromagnetic interaction by
\[
\tilde{\hat V}=\tilde{\hat V}^{\prime}+\tilde{\hat
V}^{\prime\prime}.
\]
In accordance with (\ref{65}), (\ref{66}),
\begin{gather}
\tilde{\hat V}^{\prime}={1\over 2}\int\,d{\mathbf x} {\hat {\mathbf
a}}^{2}({\mathbf x})\biggl({e^{2}\over m_{1}}\tilde
{\hat{\rho}}_{1}({\mathbf x}) + {e^{2}\over
m_{2}}\tilde{\hat{\rho}}_{2}({\mathbf x})\biggr), \qquad \tilde{\hat
V}^{\prime\prime}=-\int\,d{\mathbf x}{\hat {\mathbf a}}({\mathbf
x})\tilde{\hat {\mathbf j}}({\mathbf x}), \nonumber  \\
\tilde{\hat{\mathbf j}}({\mathbf x})= {e\over
m_{1}}{\tilde{\hat{\mbox{\boldmath$\pi$}}}}_{1}({\mathbf x}) +
{e\over m_{2}}{\tilde{\hat{\mbox{\boldmath$\pi$}}}}_{2}({\mathbf x})
\label{104}
\end{gather}
(further we assume $e_{1}=-e_{2}=e$). Since $\tilde{\hat
V}^{\prime}\sim e^{2}$, $\tilde{\hat V}^{\prime\prime}\sim e$, in
the second order in charge $e$, the $T$-matrix can be written as
\[
{\hat T}(z)=\tilde{\hat V}^{\prime}+\tilde{\hat
V}^{\prime\prime}{\hat R}_{0}(z)\tilde{\hat V}^{\prime\prime},
\qquad {\hat R}_{0} = {1\over z-\tilde{\hat{\mathcal H}}_{0}}
\]
(within the considered approximation it is not necessary to take
into account the fields ${\hat \chi}_{i}({\mathbf x})$). Bearing in
mind (\ref{104}), (\ref{40}), the operator $\tilde{\hat V}^{\prime}$
can be written as
\[
\tilde{\hat V}^{\prime}={1\over 2{\mathcal V}}\sum_{{\mathbf
p}_{1},{\mathbf p}_{2}} {\hat\eta}^{+}_{\alpha}({\mathbf
p}_{1}){\hat \eta}_{\beta}({\mathbf p}_{2})q_{\alpha\beta}({\mathbf
p}_{1} - {\mathbf p}_{2}) \int\,d{\mathbf x}{\hat {\mathbf
a}}^{2}({\mathbf x})e^{i{\mathbf x}({\mathbf p}_{2}-{\mathbf
p}_{1})} ,
\]
where
\[
q_{\alpha\beta}({\mathbf k})=\int\,d{\mathbf y}\biggl\{{e^{2}\over
m_{1}}e^{i{\mathbf k}{\mathbf y}(m_{2}/ M)}+{e^{2}\over
m_{2}}e^{-i{\mathbf k}{\mathbf
y}(m_{1}/M)}\biggr\}{\varphi}_{\alpha}^{*}({\mathbf
y}){\varphi}_{\beta}({\mathbf y}).
\]
Similarly,
\[
\tilde{\hat V}^{\prime\prime}=-{1\over {\mathcal V}}\sum_{{\mathbf
p}_{1}, {\mathbf p}_{2}}{\hat\eta}^{+}_{\alpha}({\mathbf
p}_{1}){\hat \eta}_{\beta}({\mathbf p}_{2}){\mathbf
I}_{\alpha\beta}({\mathbf p}_{1}+{\mathbf p}_{2}, {\mathbf
p}_{1}-{\mathbf p}_{2})\int\,d{\mathbf x}{\hat{\mathbf a}}({\mathbf
x})e^{i{\mathbf x}({\mathbf p}_{2} - {\mathbf p}_{1})},
\]
where
\[
{\mathbf I}_{\alpha\beta}({\mathbf p}_{1} + {\mathbf p}_{2},
{\mathbf p}_{1} - {\mathbf p}_{2})={\mathbf g}_{\alpha
\beta}({\mathbf p}_{1} - {\mathbf p}_{2}) +{{\mathbf p}_{1} +
{\mathbf p}_{2}\over 2M} g_{\alpha \beta}({\mathbf p}_{1} - {\mathbf
p}_{2}), \qquad {\mathbf p}_{1}-{\mathbf p}_{2}={\mathbf k}, \quad
{\mathbf p}_{1}+{\mathbf p}_{2}={\mathbf p},
\]
moreover, ${\mathbf g}_{\alpha \beta}({\mathbf k})$, $g_{\alpha
\beta}({\mathbf k})$ are defined by (\ref{98}).

Let ${\hat C}_{{\mathbf k}\lambda}^{+}{\hat
\eta}_{\alpha}^{+}({\mathbf p})|0\rangle$ and ${\hat C}_{{\mathbf
k}^{\prime}\lambda^{\prime}}^{+}{\hat
\eta}_{\alpha^{\prime}}^{+}({\mathbf p}^{\prime})|0\rangle$ be the
initial and final states of the system,
\[
|i\rangle = {\hat C}_{{\mathbf k}\lambda}^{+}{\hat
\eta}_{\alpha}^{+}({\mathbf p})|0\rangle, \qquad |i\rangle = {\hat
C}_{{\mathbf k}^{\prime}\lambda^{\prime}}^{+}{\hat
\eta}_{\alpha^{\prime}}^{+}({\mathbf p}^{\prime})|0\rangle.
\]
Then the matrix element $\langle f|{\hat T}(z)|i\rangle$ can be
represented in the form
\[
\langle f|{\hat T}(z)|i\rangle=\langle f|\tilde{\hat
V}^{\prime}|i\rangle+\langle f|\tilde{\hat V}^{\prime\prime}{\hat
R}_{0}(z)\tilde{\hat V}^{\prime\prime}|i\rangle,
\]
where
\begin{gather*}
\langle f|\tilde{\hat V}^{\prime}|i\rangle=-{4\pi^{2}i\over
{\mathcal V}}{1\over\sqrt{\omega\omega^{\prime}}}
R_{\alpha^{\prime}\alpha}^{\prime}\Delta({\mathbf p} + {\mathbf k} -
{\mathbf
p}^{\prime} - {\mathbf k}^{\prime}), \\
\langle f|\tilde{\hat V}^{\prime\prime}{\hat R}_{0}(z)\tilde{\hat
V}^{\prime\prime}|i\rangle=-{4\pi^{2}i\over {\mathcal
V}}{1\over\sqrt{\omega\omega^{\prime}}}
R_{\alpha^{\prime}\alpha}^{\prime\prime}\Delta({\mathbf p} +
{\mathbf k} - {\mathbf p}^{\prime} - {\mathbf k}^{\prime}), \qquad
\Delta({\mathbf k})=\delta_{{\mathbf k}0},
\end{gather*}
and the matrices $R_{\alpha^{\prime}\alpha}^{\prime}$,
$R_{\alpha^{\prime}\alpha}^{\prime\prime}$ are given by
\begin{gather*}
R_{\alpha^{\prime}\alpha}^{\prime}={\mathbf e}^{(\lambda)}_{{\mathbf
k}}{{\mathbf e}^{(\lambda)}_{{\mathbf
k}}}^{*}q_{\alpha\alpha^{\prime}}({\mathbf k}), \qquad {\mathbf
k}={\mathbf
p}^{\prime} - {\mathbf p}, \\
R_{\alpha^{\prime}\alpha}^{\prime\prime}=\sum_{\beta}\Biggl\{{
({\mathbf e}^{\prime}{\mathbf I}_{\beta\alpha^{\prime}}({\mathbf
p}^{\prime},{\mathbf k}^{\prime}))^{*}({\mathbf e}{\mathbf
I}_{\beta\alpha}({\mathbf p},{\mathbf k}))\over\omega +
{\varepsilon}_{\alpha}({\mathbf p})-{\varepsilon}_{\beta}({\mathbf
p} + {\mathbf k})} +{({\mathbf e}^{\prime}{\mathbf
I}_{\alpha\beta}({\mathbf p},-{\mathbf k}^{\prime}))^{*}({\mathbf
e}{\mathbf I}_{\alpha^{\prime}\beta}({\mathbf p}^{\prime},-{\mathbf
k}))\over -\omega^{\prime} + {\varepsilon}_{\alpha}({\mathbf p}) -
{\varepsilon}_{\beta}({\mathbf p} - {\mathbf k}^{\prime})} \Biggr\}.
\end{gather*}
Here we have taken into account that $\omega +
{\varepsilon}_{\alpha}({\mathbf p}) = \omega^{\prime} +
{\varepsilon}_{\beta}({\mathbf p}^{\prime})$ is a consequence of
(\ref{95}) and assumed $E=\omega + {\varepsilon}_{\alpha}({\mathbf
p})$. According to our approximation $kr_{0}<<1$ ($\lambda >>
r_{0}$) we have
\begin{gather*}
q_{\alpha\beta}({\mathbf k})\approx {e^{2}\over
\mu}\delta_{\alpha\beta}, \quad g_{\alpha\beta}\approx 0,\\
{\mathbf g}_{\alpha\beta}({\mathbf k})\approx -{ie\over\mu}
\int\,d{\mathbf y}{\varphi}_{\alpha}^{*}({\mathbf y}){\partial
{\varphi}_{\beta}({\mathbf y})\over \partial {\mathbf y}} =
{e\over\mu}\langle\alpha|{\hat {\mathbf p}}|\beta\rangle,
\end{gather*}
whence
\begin{equation}
R_{\alpha^{\prime}\alpha}^{\prime} \approx {e^{2}\over \mu}{\mathbf
e}^{(\lambda)}_{\mathbf k}{{\mathbf
e}^{(\lambda^{\prime})}_{{\mathbf
k}^{\prime}}}^{*}\delta_{\alpha\alpha^{\prime}}, \qquad {\mathbf
I}_{\alpha\beta}({\mathbf p},{\mathbf k})\approx {\mathbf
g}_{\alpha\beta}({\mathbf k})\approx{e\over\mu}\langle\alpha|{\hat
{\mathbf p}}|\beta\rangle, \label{108}
\end{equation}
where ${\hat {\mathbf p}}=-i\nabla$ is the momentum operator. Noting
that ${\varepsilon}_{\alpha}({\mathbf p})={\varepsilon}_{\alpha}+
(p^{2}/2M)$ we neglect the kinetic energy in denominators of
$R_{\alpha^{\prime}\alpha}^{\prime\prime}$ (the kinetic energy is
small in comparison to binding energy of bound states; see section
II). The result is
\[
R_{\alpha^{\prime}\alpha}^{\prime\prime}={e^{2}\over \mu^{2}}
\sum_{\beta}\left\{{\langle\alpha^{\prime}|{\mathbf e}^{\prime
*}{\hat {\mathbf p}}|\beta\rangle\langle\beta|{\mathbf e}{\hat {\mathbf
p}}|\alpha\rangle \over \omega + {\varepsilon}_{\alpha} -
{\varepsilon}_{\beta}} + {\langle\alpha^{\prime}|{\mathbf e}{\hat
{\mathbf p}}|\beta\rangle\langle\beta|{\mathbf e}^{\prime *}{\hat
{\mathbf p}}|\alpha\rangle \over -\omega^{\prime} +
{\varepsilon}_{\alpha} - {\varepsilon}_{\beta}} \right\}.
\]
Let us express this formula through the matrix elements of the
dipole moment of atom. It is easy to verify (by using of
(\ref{100}) that
\begin{gather*}
{\langle\alpha^{\prime}|{\mathbf e}^{\prime *}{\hat {\mathbf
p}}|\beta\rangle\langle\beta|{\mathbf e}{\hat {\mathbf
p}}|\alpha\rangle \over \omega + {\varepsilon}_{\alpha} -
{\varepsilon}_{\beta}} = {\mu}^{2}\left({\varepsilon}_{\alpha} -
{\varepsilon}_{\beta} - \omega^{\prime} +
{\omega\omega^{\prime}\over \omega + {\varepsilon}_{\alpha} -
{\varepsilon}_{\beta}}\right)\langle\alpha^{\prime}|{\mathbf
e}^{\prime
*}{\hat {\mathbf y}}|\beta\rangle\langle\beta|{\mathbf e}{\hat
{\mathbf y}}|\alpha\rangle,\\
{\langle\alpha^{\prime}|{\mathbf e}{\hat {\mathbf
p}}|\beta\rangle\langle\beta|{\mathbf e}^{\prime *}{\hat {\mathbf
p}}|\alpha\rangle \over -\omega^{\prime} + {\varepsilon}_{\alpha} -
{\varepsilon}_{\beta}}=
{\mu}^{2}\left({\varepsilon}_{\alpha^{\prime}} -
{\varepsilon}_{\beta} + \omega^{\prime} +
{\omega\omega^{\prime}\over -\omega^{\prime} +
{\varepsilon}_{\alpha} -
{\varepsilon}_{\beta}}\right)\langle\alpha^{\prime}|{\mathbf e}{\hat
{\mathbf y}}|\beta\rangle\langle\beta|{\mathbf e}^{\prime *}{\hat
{\mathbf y}}|\alpha\rangle.
\end{gather*}
Therefore, the quantity $R_{\alpha^{\prime}\alpha}^{\prime\prime}$
is of the form
\[
R_{\alpha^{\prime}\alpha}^{\prime\prime} = Q_{\alpha^{\prime}\alpha}
+\omega\omega^{\prime}\sum_{\beta}\left\{{\langle\alpha^{\prime}|{\mathbf
e}{\hat {\mathbf d}}|\beta\rangle\langle\beta|{\mathbf e}^{\prime
*}{\hat {\mathbf d}}|\alpha\rangle \over -\omega^{\prime} +
{\varepsilon}_{\alpha} - {\varepsilon}_{\beta}} +
{\langle\alpha^{\prime}|{\mathbf e}^{\prime *}{\hat {\mathbf
d}}|\beta\rangle\langle\beta|{\mathbf e}{\hat {\mathbf
d}}|\alpha\rangle \over \omega + {\varepsilon}_{\alpha} -
{\varepsilon}_{\beta}} \right\},
\]
where ${\hat {\mathbf d}} = e{\hat {\mathbf y}}$ is the dipole
moment of atom, and
\[
Q_{\alpha^{\prime}\alpha}=
\mu^{2}\sum_{\beta}\Biggl\{({\varepsilon}_{\alpha} -
{\varepsilon}_{\beta} - \omega^{\prime})
\langle\alpha^{\prime}|{\mathbf e}^{\prime *}{\hat {\mathbf
y}}|\beta\rangle\langle\beta|{\mathbf e}{\hat {\mathbf
y}}|\alpha\rangle +(\omega^{\prime}+{\varepsilon}_{\alpha^{\prime}}
- {\varepsilon}_{\beta})\langle\alpha^{\prime}|{\mathbf e}{\hat
{\mathbf y}}|\beta\rangle\langle\beta|{\mathbf e}^{\prime *}{\hat
{\mathbf y}}|\alpha\rangle \Biggr\}.
\]
Next, using that
\[
\left[{\hat {\mathbf y}}{\mathbf e}, {p^{2}\over 2\mu} + {\hat
V}\right] = \left[{\hat {\mathbf y}}{\mathbf e}, {p^{2}\over
2\mu}\right]={i\over\mu}{\mathbf e}{\hat {\mathbf p}},
\]
one finds
\[
Q_{\alpha^{\prime}\alpha} = -{e^{2}\over\mu}{\mathbf e}^{\prime
*}{\mathbf e}\delta_{\alpha\alpha^{\prime}}.
\]
According to (\ref{108}), the quantity
$R^{\prime}_{\alpha^{\prime}\alpha}$ is equal to
$R^{\prime}_{\alpha^{\prime}\alpha} = {e^{2}\over\mu}{\mathbf
e}^{\prime *}{\mathbf e}\delta_{\alpha\alpha^{\prime}}$. Thus, the
matrix $R=R^{\prime}+R^{\prime\prime}$ takes the form
\[
R_{\alpha^{\prime}\alpha}=\omega\omega^{\prime}\sum_{\beta}
\left\{{\langle\alpha^{\prime}|{\mathbf e}{\hat {\mathbf
d}}|\beta\rangle\langle\beta|{\mathbf e}^{\prime *}{\hat {\mathbf
d}}|\alpha\rangle \over -\omega^{\prime} + {\varepsilon}_{\alpha} -
{\varepsilon}_{\beta}} + {\langle\alpha^{\prime}|{\mathbf e}^{\prime
*}{\hat {\mathbf d}}|\beta\rangle\langle\beta|{\mathbf e}{\hat {\mathbf
d}}|\alpha\rangle \over \omega + {\varepsilon}_{\alpha} -
{\varepsilon}_{\beta}}\right\}
\]
and the amplitude of transition from the initial state into the
final state is given by (see${}^{5}$)
\[
S_{i\rightarrow f}=-{4\pi^{2}i\over {\mathcal V}} {1\over
\sqrt{\omega\omega^{\prime}}} R_{\alpha^{\prime}\alpha} \delta\left(
{\varepsilon}_{\alpha}({\mathbf p}) + \omega -
{\varepsilon}_{\alpha^{\prime}}({\mathbf p}^{\prime}) -
\omega^{\prime}\right)\Delta\left({\mathbf p}+{\mathbf k}-{\mathbf
p}^{\prime}-{\mathbf k}^{\prime}\right).
\]
The differential scattering cross-section is of the form
\[
d\sigma_{i\rightarrow f} = {1\over
\omega\omega^{\prime}}|R_{\alpha^{\prime}\alpha}|^{2}\delta\left(
{\varepsilon}_{\alpha}({\mathbf p}) + \omega -
{\varepsilon}_{\alpha^{\prime}}({\mathbf p}^{\prime}) -
\omega^{\prime}\right)d{\mathbf k}^{\prime}.
\]

Let us consider the scattering of an elementary particle by
compound particle (the scattering of an electron by atom). Due to
the structure of the Hamiltonian of interaction (\ref{54}) it is
sufficient to study the scattering of particle of the first kind
by bound state. The matrix element of $\hat{S}$-matrix in the
first nonvanishing approximation over interaction is defined by
\[
S_{i\rightarrow f}=-2\pi i\langle f|\tilde{\hat{\mathcal
H}}_{int}^{1}|i\rangle\delta({\varepsilon}_{i}-{\varepsilon}_{f}),
\]
where $\tilde{\hat{\mathcal H}}_{int}^{1}$ is given by (\ref{54}) and
consequently,
\[
\langle f|\tilde{\hat{\mathcal H}}_{int}^{1}|i\rangle={1\over
{\mathcal V}^{2}}\int d{\mathbf x}_{1} d{\mathbf x}_{2} d{\mathbf
y}{\varphi}_{\alpha}({\mathbf x}_{2}-{\mathbf y})e^{i{\mathbf
p}{\mathbf X}}{\varphi}_{\alpha^{\prime}}({\mathbf x}_{2}-{\mathbf
y})e^{-i{{\mathbf p}^{\prime}}{\mathbf X}}\biggl(\nu_{11}({\mathbf
x}_{1}-{\mathbf x}_{2}) - \nu_{21}({\mathbf x}_{1}-{\mathbf
y})\biggr)e^{i{\mathbf x}_{1}({\mathbf k}_{1} - {\mathbf
k}_{1}^{\prime})}.
\]
Here $\langle f|=\langle 0|{\hat \eta}_{\alpha^{\prime}}({\mathbf
p}^{\prime}) {\hat{\chi}}_{1}({\mathbf k}_{1}^{\prime})$ and
$|i\rangle ={\hat \chi}_{1}^{+}({\mathbf k}_{1}){\hat
\eta}_{\alpha}^{+}({\mathbf p})|0\rangle$. The simple calculations
lead to
\[
\langle f|\tilde{\hat{\mathcal H}}_{int}^{1}|i\rangle={1\over
{\mathcal V}}\Delta({\mathbf p} + {\mathbf k} - {\mathbf p}^{\prime}
- {\mathbf k}^{\prime})\int\,d{\mathbf
x}{\varphi_{\alpha^{\prime}}^{*}}({\mathbf
x}){\varphi}_{\alpha}({\mathbf x})\left\{\nu_{11}({\mathbf
q})e^{-i{\mathbf q}{\mathbf x}(m_{2}/M)} + \nu_{21}({\mathbf
q})e^{i{\mathbf q}{\mathbf x}(m_{1}/M)}\right\}, \quad {\mathbf
q}={\mathbf k}^{\prime}-{\mathbf k}.
\]

For Coulomb's interaction of particles we have
\[
\nu_{11}({\mathbf q})=-\nu_{21}({\mathbf q})\equiv \nu({\mathbf
q})={4\pi e^{2}\over q^{2}}.
\]
Therefore, in the long-wave approximation ($qr_{0}<<1$), one gets
\begin{equation}
\langle f|\tilde{\hat{\mathcal H}}_{int}^{1}|i\rangle=4\pi i
e{1\over {\mathcal V}}\Delta({\mathbf p} + {\mathbf k} - {\mathbf
p}^{\prime} - {\mathbf k}^{\prime}){1\over q^{2}}{\mathbf q}{\mathbf
d}_{\alpha^{\prime}\alpha}, \label{109}
\end{equation}
where ${\hat {\mathbf d}}=e{\hat {\mathbf x}}$ is the dipole moment
of atom (${\mathbf d}_{\alpha^{\prime}\alpha}=\int\,d{\mathbf
x}{\varphi}^{*}_{\alpha^{\prime}}{\mathbf x}({\mathbf x})
{\varphi}_{\alpha}({\mathbf x})$). Note that the Hamiltonian
\[
\tilde{\hat V}=\int d{\mathbf x}_{1}d{\mathbf x}_{2} {\hat
\chi}^{+}({\mathbf x}_{1}){\hat \chi}({\mathbf
x}_{1})v_{\alpha^{\prime}\alpha}({\mathbf x}_{1}-{\mathbf
x}_{2}){\hat \eta}_{\alpha^{\prime}}^{+}({\mathbf x}_{2}) {\hat
\eta}_{\alpha}({\mathbf x}_{2}),
\]
which describes the interaction between particles and bound states
leads to the same result. Here $\hat{\chi}$, $\hat{\eta}_{\alpha}$
are the field operators of particles and bound states respectively,
and $v_{\alpha^{\prime}\alpha}({\mathbf x})=(e/x^{3}){\mathbf
x}{\mathbf d}_{\alpha^{\prime}\alpha}$. The relation (\ref{109})
defines the interaction energy of the dipole moment ${\mathbf d}$
and electric charge $e$, which are at the distance ${\mathbf x}$
from each other.

\section{The Van der Waals Forces}

In this section  we investigate the forces acting between neutral
atoms being in the ground state (the van der Waals forces) on the
basis of the developed formalism. In order to solve this problem
let us address to the Schroedinger equation that determines the
energy spectrum of the system,
\begin{equation}
\tilde{\hat{\mathcal H}}\Phi=E\Phi, \qquad \tilde{\hat{\mathcal
H}}=\tilde{\hat{\mathcal H}}_{0}+\tilde{\hat{V}}. \label{110}
\end{equation}
Here $\tilde{\hat{\mathcal H}}_{0}$ and
$\tilde{\hat{V}}=\tilde{\hat{\mathcal H}}_{int}^{1}+\tilde{\hat{\mathcal
H}}_{int}^{2}+\tilde{\hat{\mathcal H}}_{int}^{3}$ are defined by
(\ref{53}), (\ref{54})-(\ref{56}). Since the system in question
consists of two atoms, we should seek the solution of (\ref{110})
in the form
\begin{equation}
\Phi_{\alpha\beta}({\mathbf X},{\mathbf X}')=\sum_{\lambda\rho}\int
d{\mathbf Y}d{\mathbf Y}' K_{\alpha\beta;\lambda\rho}({\mathbf
X},{\mathbf X}';{\mathbf Y},{\mathbf
Y}')\hat{\eta}^{+}_{\lambda}({\mathbf
Y})\hat{\eta}^{+}_{\rho}({\mathbf Y}')\Phi_{0}. \label{111}
\end{equation}
The Hamiltonian of interaction $\tilde{\hat{V}}$ is equal, in
accordance with (\ref{54})--(\ref{56}), to
\begin{equation}
\tilde{\hat{V}}={1\over 2}\int d{\mathbf X}d{\mathbf
Y}\hat{\eta}^{+}_{\alpha}({\mathbf
X})\hat{\eta}^{+}_{\beta}({\mathbf Y})\hat{\eta}_{\gamma}({\mathbf
Y})\hat{\eta}_{\delta}({\mathbf
X})G_{\delta\gamma;\alpha\beta}({\mathbf X}-{\mathbf Y}),
\label{112}
\end{equation}
where
\begin{equation*}
\begin{split}
G_{\delta\gamma;\alpha\beta}({\mathbf X}-{\mathbf Y})=\int d{\mathbf
x}d{\mathbf y}\varphi^{*}_{\alpha}({\mathbf
x})\varphi^{*}_{\beta}({\mathbf y})\varphi_{\gamma}({\mathbf
y})\varphi_{\delta}({\mathbf x})&\biggl\{\nu_{12}\left({\mathbf
X}-{\mathbf Y}-{m_{1}{\mathbf x}+m_{2}{\mathbf y}\over
M}\right)+\nu_{21}\left({\mathbf X}-{\mathbf Y}+{m_{1}{\mathbf
y}+m_{2}{\mathbf x}\over M}\right)
\\&+\nu_{11}\left({\mathbf X}-{\mathbf Y}+{m_{2}\over M}({\mathbf x}-{\mathbf
y})\right)+\nu_{22}\left({\mathbf X}-{\mathbf Y}-{m_{1}\over
M}({\mathbf x}-{\mathbf y})\right)\biggr\}.
\end{split}
\end{equation*}

We suppose that the kinetic energy of atoms is small in comparison
to the energy levels $|\varepsilon_{\alpha}|$
($\varepsilon_{\alpha}<0$). In this case, according to (\ref{53}),
the operator $\tilde{\hat{\mathcal H}}_{0}$ can be represented in the
form
\[
\tilde{\hat{\mathcal H}}_{0}=\sum_{\alpha}\int d{\mathbf X}
{\varepsilon}_{\alpha}\hat{\eta}^{+}_{\alpha}({\mathbf
X})\hat{\eta}_{\alpha}({\mathbf X}). \label{113}
\]
It can be easily seen that
\begin{gather}
\tilde{\hat{\mathcal H}}_{0}\hat{\eta}^{+}_{\lambda}({\mathbf
Z})\hat{\eta}_{\rho}^{+}({\mathbf
Z}')\Phi_{0}=({\varepsilon}_{\lambda}+{\varepsilon}_{\rho})
\hat{\eta}^{+}_{\lambda}({\mathbf
Z})\hat{\eta}_{\rho}^{+}({\mathbf Z}')\Phi_{0}, \label{114} \\
\tilde{\hat{V}}\hat{\eta}^{+}_{\lambda}({\mathbf
Z})\hat{\eta}_{\rho}^{+}({\mathbf
Z}')\Phi_{0}=\sum_{\alpha\beta}G_{\lambda\rho;\alpha\beta}({\mathbf
Z}-{\mathbf Z}')\hat{\eta}^{+}_{\alpha}({\mathbf
Z})\hat{\eta}_{\beta}^{+}({\mathbf Z}')\Phi_{0}. \nonumber
\end{gather}
These formulas show that we can seek the solution of (\ref{110})
in a more simple, rather than (\ref{111}), form
\[
\Phi_{\alpha\beta}({\mathbf X},{\mathbf
X}')=\sum_{\lambda\rho}K_{\alpha\beta;\lambda\rho}({\mathbf
X},{\mathbf X}')\hat{\eta}^{+}_{\lambda}({\mathbf
X})\hat{\eta}^{+}_{\rho}({\mathbf X}')\Phi_{0},
\]
so that the coordinates of atoms have the definite values in the
state $\Phi_{\alpha\beta}({\mathbf X},{\mathbf X}')$. Upon
substituting this expression into (\ref{110}) and using (\ref{114}),
one finds
\[
K_{\alpha\beta;\gamma\delta}({\varepsilon}_{\gamma}
+{\varepsilon}_{\delta})+
\sum_{\lambda\rho}K_{\alpha\beta;\lambda\rho}
G_{\lambda\rho;\gamma\delta}({\mathbf Z}-{\mathbf
Z}')=E_{\alpha\beta}K_{\alpha\beta;\gamma\delta}.
\]
A perturbative approach for this equation can be easily developed in
the domain of great $|{\mathbf Z}-{\mathbf Z}'|$ when the quantity
$G_{\lambda\rho;\gamma\delta}({\mathbf Z}-{\mathbf Z}')$ becomes
small (see (\ref{112})). Expanding $K_{\alpha\beta;\gamma\delta}$ in
$G$,
\begin{gather*}
K_{\alpha\beta;\gamma\delta}=K_{\alpha\beta;\gamma\delta}^{1} +
K_{\alpha\beta;\gamma\delta}^{1}+K_{\alpha\beta;\gamma\delta}^{2}+\ldots,\\
E_{\alpha\beta}=E_{\alpha\beta}^{0}+E_{\alpha\beta}^{1}+
E_{\alpha\beta}^{2}+\ldots,
\end{gather*}
one obtains in the zeroth order
\[
K_{\alpha\beta;\gamma\delta}^{0}({\varepsilon}_{\gamma}
+{\varepsilon}_{\delta})=
E_{\alpha\beta}^{0}K_{\alpha\beta;\gamma\delta}^{0},
\]
whence
\[
K_{\alpha\beta;\gamma\delta}^{0}=K_{\alpha\beta}^{0}
\delta_{\alpha\gamma} \delta_{\beta\delta}, \qquad
E_{\alpha\beta}^{0}={\varepsilon}_{\alpha}+{\varepsilon}_{\beta}.
\label{116}
\]
Taking into account this result we have in the first
approximation,
\[
K_{\alpha\beta;\gamma\delta}^{1}({\varepsilon}_{\gamma}
+{\varepsilon}_{\delta}) +
K_{\alpha\beta}^{0}G_{\alpha\beta;\gamma\delta}({\mathbf Z}-{\mathbf
Z}')= ({\varepsilon}_{\alpha}+{\varepsilon}_{\beta})
K_{\alpha\beta;\gamma\delta}^{1} +
E_{\alpha\beta}^{1}K_{\alpha\beta}^{0}\delta_{\alpha\gamma}
\delta_{\beta\delta}.
\]
By setting here $\alpha=\gamma$, $\beta=\delta$, one gets
\begin{equation}
E_{\alpha\beta}^{1}=G_{\alpha\beta;\alpha\beta}({\mathbf Z}-{\mathbf
Z}'), \label{117}
\end{equation}
and for $\alpha, \beta \ne \gamma, \delta$,
\begin{equation}
K_{\alpha\beta;\gamma\delta}^{1} =
K_{\alpha\beta}^{0}{G_{\alpha\beta;\gamma\delta}({\mathbf
Z}-{\mathbf Z}') \over
({\varepsilon}_{\alpha}+{\varepsilon}_{\beta}-{\varepsilon}_{\gamma}
- {\varepsilon}_{\delta})}, \qquad \alpha, \beta \ne \gamma, \delta.
\label{118}
\end{equation}
The second order of the perturbative approach give
\[
K_{\alpha\beta;\gamma\delta}^{2}({\varepsilon}_{\gamma}+{\varepsilon}_{\delta})
+ \sum_{\lambda\rho}K_{\alpha\beta;\lambda\rho}^{1}
G_{\lambda\rho;\gamma\delta}({\mathbf Z}-{\mathbf Z}')=
({\varepsilon}_{\alpha}+{\varepsilon}_{\beta})
K_{\alpha\beta;\gamma\delta}^{2} +
E_{\alpha\beta}^{1}K_{\alpha\beta;\gamma\delta}^{1} +
E_{\alpha\beta}^{2}K_{\alpha\beta;\gamma\delta}^{0}.
\]
Taking here $\alpha=\gamma$, $\beta=\delta$, one finds
\[
\sum_{\lambda\rho}K_{\alpha\beta;\lambda\rho}^{1}
G_{\lambda\rho;\alpha\beta}({\mathbf Z}-{\mathbf
Z}')=E_{\alpha\beta}^{1}K_{\alpha\beta;\alpha\beta} +
K_{\alpha\beta}^{0}E_{\alpha\beta}^{2},
\]
whence, according to (\ref{117}), (\ref{118}), we have
\begin{equation}
E_{\alpha\beta}^{2}={\sum_{\lambda\rho}}^{\prime}{
G_{\alpha\beta;\lambda\rho}({\mathbf Z}-{\mathbf Z}')
G_{\lambda\rho;\alpha\beta}({\mathbf Z}-{\mathbf Z}')\over
({\varepsilon}_{\alpha}+{\varepsilon}_{\beta} -
{\varepsilon}_{\lambda}-{\varepsilon}_{\rho})}. \label{119}
\end{equation}
The prime above the sum means that the terms with $\lambda=\alpha$,
$\rho=\beta$ are omitted. The state vector
$\Phi_{\alpha\beta}({\mathbf X},{\mathbf X}')$ in the main
approximation of the perturbative approach is determined by
\[
\Phi_{\alpha\beta}({\mathbf X},{\mathbf
X}')=K_{\alpha\beta}^{0}\hat{\eta}_{\alpha}^{+}({\mathbf
X})\hat{\eta}_{\beta}^{+}({\mathbf X}')\Phi_{0}+\ldots \label{120}
\]
(the constant $K_{\alpha\beta}^{0}$ may be found from the
normalization relation,
$(\Phi_{\alpha\beta},\Phi_{\alpha\beta})=1$). Formulas
(\ref{117}), (\ref{119}) for $E_{\alpha\beta}^{1}$ and
$E_{\alpha\beta}^{2}$ give us the corrections to the energy levels
$E_{\alpha\beta}^{0}={\varepsilon}_{\alpha}+{\varepsilon}_{\beta}$.
It follows from the obtained formulas that the energy of two atoms
being in the ground state $\alpha$ and spaced apart for
sufficiently long distances is defined by
\begin{equation}
E_{\alpha\alpha}=2\varepsilon_{\alpha}+
G_{\alpha\alpha;\alpha\alpha}({\mathbf Z}-{\mathbf
Z}')+{\sum_{\lambda\rho}}^{\prime}{
G_{\alpha\alpha;\lambda\rho}({\mathbf Z}-{\mathbf Z}')
G_{\lambda\rho;\alpha\alpha}({\mathbf Z}-{\mathbf Z}')\over
(2{\varepsilon}_{\alpha}-{\varepsilon}_{\lambda}-
{\varepsilon}_{\rho})}+\ldots \label{121}
\end{equation}

Let us prove now that $G_{\alpha\alpha;\alpha\alpha}({\mathbf
Z}-{\mathbf Z}')\equiv 0$. In doing so we use the following
formula${}^{9}$:
\begin{gather*}
{1\over\sqrt{R^{2}-2R\rho x +\rho^{2}}}={1\over|{\mathbf
R}-\mbox{\boldmath$\rho$}|}={1\over
R}+\sum_{n=1}^{\infty}\biggl({\rho\over R}\biggr)^{n}{\mathcal
P}_{n}(x),\\
\end{gather*}
where $x=\cos\vartheta$, $\vartheta$ is the angle between vectors
${\mathbf R}$ and $\mbox{\boldmath$\rho$}$, and ${\mathcal
P}_{n}(x)$ are the Legendre polynomials. Noting that (see
(\ref{112}))
\begin{multline*}
\begin{split}
G_{\alpha\alpha;\alpha\alpha}({\mathbf Z}-{\mathbf Z}')=\int
d{\mathbf x} d{\mathbf y}\vert\varphi_{\alpha}({\mathbf
x})\vert^{2}\vert\varphi_{\alpha}({\mathbf y})\vert^{2}
&\biggl\{\nu_{12}\left({\mathbf Z}-{\mathbf z}'-{m_{1}{\mathbf
x}+m_{2}{\mathbf y}\over M}\right)+ \nu_{21}\left({\mathbf
Z}-{\mathbf z}' + {m_{1}{\mathbf y}+m_{2}{\mathbf x}\over M}\right)
\\ &+ \nu_{11}\left({\mathbf Z}-{\mathbf z}' + {m_{2}\over M}({\mathbf x}-{\mathbf
y})\right)+\nu_{22}\left({\mathbf Z}-{\mathbf z}' - {m_{1}\over
M}({\mathbf x}-{\mathbf y})\right)\biggr\},
\end{split}
\end{multline*}
we come, taking into account the spherical symmetry of
$\vert\varphi_{\alpha}({\mathbf x})\vert^{2}$, to
\begin{equation}
G_{\alpha\alpha;\alpha\alpha}({\mathbf Z}-{\mathbf Z}')=0.
\label{122}
\end{equation}
We have also employed here that according to (\ref{61})
\[
\nu_{ab}({\mathbf Z}-{\mathbf Z}')={e_{a}e_{b}\over |{\mathbf
Z}-{\mathbf Z}'|}, \qquad e_{1}=-e_{2}=e.
\]
Let us find now $G_{\alpha\alpha;\lambda\rho}({\mathbf Z}-{\mathbf
Z}')$,
\begin{multline*}
\begin{split}
G_{\alpha\alpha;\lambda\rho}({\mathbf Z}-{\mathbf Z}')=\int
d{\mathbf x}d{\mathbf y}\varphi^{*}_{\rho}({\mathbf
x})\varphi^{*}_{\lambda}({\mathbf y})\varphi_{\alpha}({\mathbf
y})\varphi_{\alpha}({\mathbf x})&\biggl\{\nu_{12}\left({\mathbf
Z}-{\mathbf z}' - {m_{1}{\mathbf x}+m_{2}{\mathbf y}\over M}\right)+
\nu_{21}\left({\mathbf Z}-{\mathbf z}' + {m_{1}{\mathbf
y}+m_{2}{\mathbf x}\over M}\right)
\\ &+\nu_{11}\left({\mathbf Z}-{\mathbf z}' + {m_{2}\over M}({\mathbf x}-{\mathbf
y})\right)+\nu_{22}\left({\mathbf Z}-{\mathbf z}' - {m_{1}\over
M}({\mathbf x}-{\mathbf y})\right)\biggr\}.
\end{split}
\end{multline*}
The presence of multipliers $\varphi_{\alpha}({\mathbf y})$ and
$\varphi_{\alpha}({\mathbf x})$ makes it possible to expand the
expression in braces in powers of ${\mathbf x}$, ${\mathbf y}$. As a
result we have
\[
G_{\alpha\alpha;\beta\lambda}({\mathbf Z}-{\mathbf Z}')={1\over
|{\mathbf Z}-{\mathbf Z}'|^{3}}\biggl(-3({\mathbf n}{\mathbf
d}_{\beta\alpha})({\mathbf n}{\mathbf d}_{\lambda\alpha})+({\mathbf
d}_{\beta\alpha}{\mathbf d}_{\lambda\alpha}) \biggr),
\]
where
\[
{\mathbf n}={{\mathbf Z}-{\mathbf Z}'\over |{\mathbf Z}-{\mathbf
Z}'|}, \qquad {\mathbf d}_{\beta\alpha}=e\int d{\mathbf x}{\mathbf
x}\varphi_{\beta}^{*}({\mathbf x})\varphi_{\alpha}({\mathbf x}).
\]
(${\mathbf d}_{\beta\alpha}$ are the matrix elements of the dipole
moment of atom). Thus, the potential energy of interaction between
atoms, $V({\mathbf Z}-{\mathbf Z}')$ is defined by${}^{6}$
\[
V({\mathbf Z}-{\mathbf Z}')\equiv
E_{\alpha\alpha}^{2}-2\varepsilon_{\alpha}={1\over |{\mathbf
Z}-{\mathbf Z}'|^{6}} {\sum_{\beta\lambda}}^{\prime}{|-3({\mathbf
n}{\mathbf d}_{\beta\alpha})({\mathbf n}{\mathbf
d}_{\lambda\alpha})+({\mathbf d}_{\beta\alpha}{\mathbf
d}_{\lambda\alpha})|^{2}\over
2{\varepsilon}_{\alpha}-{\varepsilon}_{\beta}-{\varepsilon}_{\lambda}}<0.
\]
Since $\varepsilon_{\beta},
\varepsilon_{\lambda}>\varepsilon_{\alpha}$, $V({\mathbf Z}-{\mathbf
Z}')<0$. Therefore, the attractive forces (the van der Waals forces)
act between the neutral atoms at long distances.

Consider now the scattering of a compound particle by compound
particle,  which are in the ground state. We start from the
scattering matrix representation in terms of $T$-operators
(\ref{95}). Accurate to the terms, which are of the second order
in interaction $\tilde{\hat V}$ between the atoms (see
(\ref{96})), one gets
\begin{equation}
{\hat S}=1-2\pi
i\int_{-\infty}^{\infty}\,dE\delta(E-\tilde{\hat{\mathcal
H}}_{0})\tilde{\hat V}\delta(E-\tilde{\hat{\mathcal H}}_{0})-2\pi
i\int_{-\infty}^{\infty}\,dE\delta(E-\tilde{\hat{\mathcal
H}}_{0}){\hat V}{1\over E-\tilde{\hat{\mathcal
H}}_{0}+i\eta}\tilde{\hat V}\delta(E-\tilde{\hat{\mathcal H}}_{0}).
\label{124}
\end{equation}
Neglecting the kinetic energy of atoms we replace in the resolvent
$(E-\tilde{\hat{\mathcal H}}_{0}+i\eta)^{-1}$ the Hamiltonian of free
particles $\tilde{\hat{\mathcal H}}_{0}$ by the Hamiltonian $\tilde
{\hat H}_{0}$ of atoms in the rest (see (\ref{53}))
\[
\tilde{\hat{\mathcal H}}_{0}\rightarrow \tilde{\hat H}_{0} =
\sum_{\alpha}\int\,d{\mathbf x} {\varepsilon}_{\alpha}{\hat
\eta}_{\alpha}^{+}({\mathbf x}){\hat \eta}_{\alpha}({\mathbf x})
\]
(in the low-energy limit we can not do this under the argument of
$\delta$-function).

Since we study the collision of two atoms and the number of atoms
is conserved during the process of collision, we can restrict
ourselves by considering of this process in the two-particle
subspace with the following completeness condition:
\begin{equation}
{1\over 2}\int\,d{\mathbf x}_{1}d{\mathbf
x}_{2}\sum_{\alpha_{1},\alpha_{2}} {\hat
\eta}_{\alpha_{1}}^{+}({\mathbf x}_{1}){\hat
\eta}_{\alpha_{2}}^{+}({\mathbf x}_{2})|0\rangle\langle 0|{\hat
\eta}_{\alpha_{2}}({\mathbf x}_{2}){\hat \eta}_{\alpha_{1}}({\mathbf
x}_{1}) = 1. \label{125}
\end{equation}
In this case formulas (\ref{114}) are true. Since the atoms are in
the ground state, (\ref{122}) is also valid and, therefore, the
matrix element of $\hat{S}$-matrix turns to zero in the first
approximation of the perturbation theory.

 The second approximation of the perturbation theory is defined, according to (\ref{124}),
(\ref{125}), by the formula
\begin{gather*}
\begin{split}
\tilde{\hat V}{1\over E-\tilde{\hat {H}}_{0}+i\eta}\tilde{\hat
V}&={1\over 2}\sum_{\alpha_{1},\alpha_{2}}\sum_{\beta_{1},\beta_{2}}
\sum_{\gamma_{1},\gamma_{2}} \int\,d{\mathbf x}_{1}d{\mathbf x}_{2}
{1\over
E-{\varepsilon_{\alpha_{1}}}-{\varepsilon_{\alpha_{2}}}+i\eta}
G_{\alpha_{1}\alpha_{2};\beta_{1}\beta_{2}}({\mathbf x}_{1}-{\mathbf
x}_{2})
\\ &\times{\hat \eta}_{\beta_{1}}^{+}({\mathbf x}_{1}){\hat
\eta}_{\beta_{2}}^{+}({\mathbf x}_{2})|0\rangle\langle 0| {\hat
\eta}_{\gamma_{1}}({\mathbf x}_{1}){\hat \eta}_{\gamma_{2}}({\mathbf
x}_{2}) G^{+}_{\gamma_{1}\gamma_{2};\alpha_{1}\alpha_{2}}({\mathbf
x}_{1}-{\mathbf x}_{2}).
\end{split}
\end{gather*}
In the two-particle subspace this formula is equivalent to
\[
\tilde{\hat V}{1\over E-\tilde{\hat {H}}_{0}+i\eta}\tilde{\hat V}=
\int\,d{\mathbf x}_{1}d{\mathbf x}_{2} \sum_{\beta_{1},\beta_{2}}
\sum_{\gamma_{1},\gamma_{2}} {\hat \eta}_{\beta_{1}}^{+}({\mathbf
x}_{1}){\hat \eta}_{\beta_{2}}^{+}({\mathbf
x}_{2})V_{\beta_{1}\beta_{2};\gamma_{1}\gamma_{2}}({\mathbf
x}_{1}-{\mathbf x}_{2};E){\hat \eta}_{\gamma_{1}}({\mathbf
x}_{1}){\hat \eta}_{\gamma_{2}}({\mathbf x}_{2}),
\]
where
\[
V_{\beta_{1}\beta_{2};\gamma_{1}\gamma_{2}}({\mathbf x}_{1}-{\mathbf
x}_{2};E)=\sum_{\alpha_{1}\alpha_{2}}{1\over
E-{\varepsilon_{\alpha_{1}}}-{\varepsilon_{\alpha_{2}}}+i\eta}
G^{+}_{\gamma_{1}\gamma_{2};\alpha_{1}\alpha_{2}}({\mathbf
x}_{1}-{\mathbf x}_{2})
G_{\alpha_{1}\alpha_{2};\beta_{1}\beta_{2}}({\mathbf x}_{1}-{\mathbf
x}_{2}).
\]
Since the initial and final states of a particle are the ground
states, the scattering matrix in the two-particle subspace can be
represented in the form
\[
{\hat S} = -2\pi
i\int_{-\infty}^{\infty}\,dE\delta(E-\tilde{\hat{\mathcal
H}}_{0}){\hat V}_{eff}\delta(E-\tilde{\hat{\mathcal H}}_{0}),
\]
where
\begin{gather*}
{\hat V}_{eff}={1\over 2} \int\,d{\mathbf x}_{1}\int\,d{\mathbf
x}_{2}{\hat \eta}_{\alpha_{1}}^{+}({\mathbf x}_{1}){\hat
\eta}_{\alpha_{2}}^{+}({\mathbf x}_{2})V_{eff}({\mathbf
x}_{1}-{\mathbf x}_{2}){\hat \eta}_{\alpha_{2}}({\mathbf
x}_{2}){\hat
\eta}_{\alpha_{1}}({\mathbf x}_{1}), \\
V_{eff}({\mathbf x}_{1}-{\mathbf x}_{2})=\sum_{\beta_{1}\beta_{2}}
{G^{+}_{\alpha\alpha;\beta_{1}\beta_{2}}({\mathbf x}_{1}-{\mathbf
x}_{2})G_{\beta_{1}\beta_{2};\alpha\alpha}({\mathbf x}_{1}-{\mathbf
x}_{2}) \over 2{\varepsilon}_{\alpha} - {\varepsilon}_{\alpha_{1}} -
{\varepsilon}_{\alpha_{2}}}.
\end{gather*}
We can see that the second order in $\tilde{\hat{V}}$ of
perturbation theory for $\hat{S}$-matrix is equivalent to the
first order of perturbation theory over effective interaction
$\hat{V}_{eff}$. This effective interaction is determined by van
der Waals forces (see formula (\ref{121})).

It follows from (\ref{55}), that the Hamiltonian of interaction of
atoms at low-energies is defined by
\[
\tilde{\hat V}=\int d{\mathbf x}_{1}d{\mathbf
x}_{2}v_{\alpha\beta;\gamma\delta}({\mathbf x}_{1}-{\mathbf
x}_{2}){\hat\eta}_{\alpha}^{+}({\mathbf x}_{1})
{\hat\eta}_{\beta}^{+}({\mathbf x}_{2}){\hat\eta}_{\gamma}({\mathbf
x}_{2}){\hat\eta}_{\delta}({\mathbf x}_1),
\]
where
\[
v_{\alpha\beta;\gamma\delta}({\mathbf x}) = {1\over
x^{5}}\left(x^{2}({\mathbf d}_{\alpha\delta}{\mathbf
d}_{\beta\gamma}) - 3({\mathbf x}{\mathbf
d}_{\alpha\delta})({\mathbf x}{\mathbf d}_{\delta\gamma})\right).
\]
This Hamiltonian corresponds to the dipole-dipole interaction of
atoms.

\section{Conclusion}

The goal of this paper was to develop a microscopic approach for
describing the physical processes in many-particle systems in the
presence of bound states of particles. To achieve this goal we
developed an approximate second quantization method for systems
with bound states of particles.

The basic results obtained in this paper are the following:\\
\noindent 1. The Fock space was introduced in the second
quantization formalism. In this space the creation and
annihilation operators of elementary particles $\hat{\chi}^{+}$,
$\hat{\chi}$ and their bound states $\hat{\eta}^{+}$, $\hat{\eta}$
were introduced on an equal status.\\
\noindent 2. The operators of basic physical quantities acting in
this space were constructed. These operators include the
Hamiltonians of interaction between elementary particles and their bound states.\\
\noindent 3. It was shown that in the approximation when the
radius of interaction is small the above mentioned Hamiltonians
transform into the well-known Hamiltonians for Coulomb's and
dipole interactions between the particles of various kinds.\\
\noindent 4. The non-relativistic quantum electrodynamics of
charged and neutral (the bound states) particles was
constructed.\\
\noindent 5. The various physical effects including the theory of
the van der Waals forces acting between atoms were considered as
the approbation of the developed formalism. The description of
such effects within the usual formalism requires more considerable
efforts associated with introduction of the interaction between
neutral currents of bound states and electromagnetic field.\\
\noindent 6. More detailed calculations, which fall out the limits
of our approximation should result in appearance of three- and
many-body interactions between unbound particles and two-body
bound states. These many-body interactions should be consistent
with pair interaction of original particles. The appearance of
these many-body terms is analogous to the origin of the terms with
photon-photon scattering in transition from the usual standard
Lagrangian to the effective low-energy Euler-Heisenberg Lagrangian
of quantum electrodynamics.

Especially we would like to emphasize a role of the obtained
Hamiltonians, which describe the interaction of quasi-neutral
particles (the bound states of charged fermions) with
electromagnetic field, the elementary particles with bound states,
and also the bound states with bound states. On the basis of these
Hamiltonians one can study such phenomena as Bose-Einstein
condensation in a gas of excited atoms, the interaction of
condensates with an electromagnetic field in Bose and Fermi
systems. These Hamiltonians can be also the basis of the kinetic
theory for systems with bound states of particles.

Finally we would like to stress that the developed method can be
easily generalized to the case of bound states consisting of more
than two particles. The generalization of the offered method for
describing the systems with bound states of bosons and also of
bosons and fermions taking into account the spin of particles can
be also performed without principal difficulties.

\begin{acknowledgments}
The authors would like to acknowledge the financial support of
INTAS (project 00-00577) and the Russian Fund of Fundamental
Researches (project 03-02-17695). The authors thank A.S.
Peletminskii for useful discussions and for his help in preparing
of the paper.

\end{acknowledgments}

\vspace{0.5cm}
\noindent
${}^{1}$S.~Chu, Rev. Mod. Phys. \textbf{70}, 685 (1998).\\
${}^{2}$C.N.~Cohen-Tannoudji, Rev. Mod. Phys. \textbf{70}, 707
(1998).\\
${}^{3}$W.D.~Phillips, Rev. Mod. Phys. \textbf{70}, 721 (1998).\\
${}^{4}$M.H.~Anderson et al., Science, \textbf{269}, 198 (1995).\\
${}^{5}$A.I.~Akhiezer, V.B.~Berestetskii, {\it{Quantum
Electrodynamics}} (Interscience Publisher, New York,1965).\\
${}^{6}$L.~Landau, E.~Lifshitz, {\it{Quantum Mechanics}} (Pergamon
Press, New York,1967).\\
${}^{7}$A.S.~Davydov, {\it{Quantum Mechanics}} (Pergamon Press,
New York,1976).\\
${}^{8}$A.I.~Akhiezer, S.V.~Peletminskii, {\it{Methods of
Statistical Physics}} (Pergamon Press, Oxford, 1981).\\
${}^{9}$D.~Jackson, {\it{Fourier series and orthogonal
polinomials}} (The Carus Mathematical Monographs, Number Six,
Minnesota, 1941).\\
${}^{10}$For Coulomb's interaction the inequality $R>>r_{0}$ is
equivalent to ${\mathcal E}<<\varepsilon_{0}$ because of
$r_{0}={n^{2}{\bar h}^{2}\over me^{2}}$,
$\varepsilon_{0}={me^{4}\over 2n^{2}{\bar h}^{2}}$, $n\sim 1$
\end{document}